\documentclass[preprint,3p]{elsarticle}
\pdfoutput=1
\usepackage{amsmath,subeqnarray,multirow,hhline}
\biboptions{sort&compress}

\newcommand{\diff}[3][]{\dfrac{\mathrm{d}^{#1}#2}{\mathrm{d}{#3}^{#1}}}

\newcommand{\pdiff}[3][]{\dfrac{\partial^{#1} #2}{\partial {#3}^{#1}}}

\def \bd {\boldsymbol}

\begin{document}

\begin{frontmatter}

\title{Finite Element Framework for Describing Dynamic Wetting Phenomena}

\author{J.E. Sprittles\footnote{Corresponding author. Tel.:+44 121 4142916; fax: +44 121 4143389}}

\ead{sprittlj@maths.bham.ac.uk}

\author{Y.D. Shikhmurzaev}

\ead{yulii@for.mat.bham.ac.uk}
\address{School of Mathematics, University of Birmingham,
Edgbaston, Birmingham, B15 2TT, U.K.}

\begin{abstract}

The finite element simulation of dynamic wetting phenomena, requiring the computation of flow in a domain confined by intersecting a liquid-fluid free surface and a liquid-solid interface, with the three-phase contact line moving across the solid, is considered.  For this class of flows, different finite element method (FEM) implementations have been proposed in the literature and these are seen to produce apparently contradictory results.  The purpose of this paper is to develop a robust framework for the FEM simulation of these flows and then, by performing numerical experiments,  provide guidelines for future investigations.  In the new framework, the boundary conditions on the liquid-solid interface are implemented in a methodologically similar way to those on the free surface so that the equations at the contact line, where the interfaces meet, are applied without any ad-hoc alterations.  The new implementation removes the need for complex rotations of the momentum equations, usually required to apply boundary conditions normal and tangent to the solid surface.  The developed code allows the convergence of the solution to be studied as the spatial resolution of the computational mesh is varied over many orders of magnitude.  This makes it possible to provide practical recommendations on the spatial resolution required by a numerical scheme for a given set of non-dimensional similarity parameters. Furthermore, one can examine various implementations used in the literature and evaluate their performance. Finally, it is shown how the framework may be generalized to account for additional physical effects, such as gradients in surface tensions.  A user-friendly step-by-step guide specifying the entire implementation and allowing the reader to easily reproduce all presented results is provided in the Appendix.

\end{abstract}


\end{frontmatter}

\section{Introduction}

\begin{figure}
\centering
\includegraphics[scale=0.80]{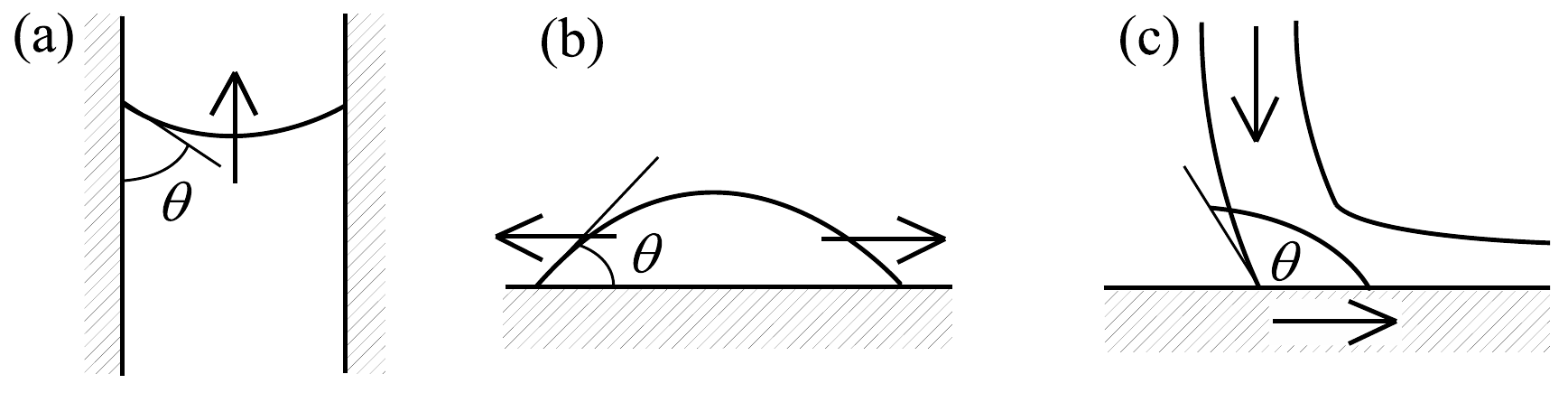}
\caption{Examples of dynamic wetting flows: a) propagation of a meniscus through a capillary tube, b) drop spreading over a solid surface, c) liquid curtain falling and coating a moving solid substrate.} \label{F:wetting}
\end{figure}

Processes in which a liquid wets or dewets a solid surface are ubiquitous throughout industry and nature. They are known as dynamic wetting flows \cite{berg93,shik07}. Flows from this class occur, for example, when a liquid-gas meniscus propagates through a tube (Figure~\ref{F:wetting}a) and during the spreading of liquid drops over solid surfaces (Figure~\ref{F:wetting}b).  In industrial applications, dynamic wetting flows are usually referred to as `coating flows', emphasizing the technological objective to `coat' the solid substrate with a liquid.  One can distinguish between discrete wetting, as for ink-jet printing \cite{calvert01,singh10}, and continuous coating, where the liquid is deposited as a continuous film \cite{ruschak85,weinstein04}, e.g.\ using a liquid curtain falling onto a moving solid substrate (Figure~\ref{F:wetting}c).  It is well established that controlling the dynamics of wetting is critical to the manufacture of a defect-free coating at an optimal speed \cite{blake94,blake06}. Manipulation of the wetting dynamics is often achieved by introducing additional physical effects which either alter the surface tension of an interface such as, for example, surfactant transport \cite{pearson09,hammond09} or influence both the interfaces and the bulk \cite{yamamura07}.

A defining feature of flows involving dynamic wetting is that, as confirmed by numerous experiments, e.g.\ \cite{hoffman75,chen97,blake99}, the free surface meets the solid boundary at an angle $\theta$, known as the `contact angle', which is less than $180^\circ$ (Figure~\ref{F:wetting}). As a result, mathematically, one faces a problem of finding a solution in a domain with a non-smooth boundary with different boundary conditions for the bulk equations of fluid mechanics on the two boundaries that form the contact angle \cite{huh71,dussan76}.

From the modeling viewpoint, the difficulties are two-fold. Firstly, regardless of the contact
angle value, there is no solution to the Navier-Stokes equations that would simultaneously satisfy
all classical boundary conditions on the free surface and the solid boundary \cite{shik07,shik06}.  This
nonexistence of what would be a `classical solution' for dynamic wetting problems (in a weaker
formulation also described in terms of a non-integrable shear-stress singularity) came to be known
as the `moving contact-line problem' \cite{shik07,huh71,dussan74}.

The second difficulty in the modelling of dynamic wetting phenomena is that the value of the
contact angle, which provides a boundary condition for the equation describing the free-surface
shape and hence is vital for finding the configuration of the flow domain, depends not only on the
speed at which the liquid-fluid-solid contact line moves across the solid surface but, as experiments show, also on the
flow field in the vicinity of this contact line \cite{blake99,clarke06}. Therefore, in order to model
dynamic wetting and be able to incorporate this effect, one has to accurately describe the flow near the moving contact line.

Theoretical works that aim at modelling the process of dynamic wetting as a solution to a clear-cut
closed mathematical problem (see \cite{shik07} for a review) invariably preserve the contact angle,
and hence a piecewise smooth boundary of the flow domain, and focus on formulating boundary
conditions that would resolve the moving contact-line problem, i.e.\ allow for the existence of a
solution (or, in a weaker formulation, make the shear stress integrable). In the vicinity of the moving contact line, the `extra' physics that
kicks in and has to be incorporated into the boundary conditions to resolve the problem makes it necessary to generalize the no-slip condition on the
solid surface to allow for slip, i.e.\ a difference between the tangential to the solid component of the liquid's velocity and that of the solid. The simplest example is the Navier slip condition \cite{navier23} in which slip is proportional to the tangential stress acting from the liquid on the liquid-solid interface.

The complexity of free-boundary problems with moving contact lines makes it almost inevitable that numerical methods are required. If the free surface position and evolution are to be tracked explicitly and with high accuracy, it is advantageous to use an arbitrary Lagrangian-Eulerian (ALE) finite element method (FEM) in which computational nodes on the free surface evolve with it whilst those in the bulk move in some other specified manner that prevents the elements from degeneration. This method easily handles complex geometries, allows for a natural incorporation of boundary conditions formulated in terms of stresses and has been successfully used to describe a variety of free-boundary flows, e.g.\ \cite{kistler84,christodoulou92,cairncross00,hadji00,heil04,wilson06,lukyanov07,bajaj08,madasu09}.

In applying numerical methods, including the FEM, to dynamic wetting flows, one encounters two main difficulties: (a) the characteristic length scale over which the classical boundary conditions have to be altered, i.e.\ in which the no-slip condition is violated, is much smaller than the scale associated with the bulk flow and (b) the boundary conditions that are applied on the two parts of the boundary that form the contact angle, i.e.\ on the free surface and the solid, are of different type.

The disparity of length scales in dynamic wetting problems is usually so large that any attempt to resolve both scales using a numerical method with a fixed spatial resolution will inevitably produce a computationally intractable problem. Consequently, previous numerical implementations of dynamic wetting models split into two groups: those that explicitly attempted to resolve the smaller length scale using specially designed meshes, e.g.\ \cite{zhou90,wilson06,sprittles10,lowndes80,bazhlekov96}, and, the majority, where no such attempt was made.

The consequence of each interface having different types of boundary conditions is that the contact line, where the two interfaces meet and the contact angle must be specified, does not fit naturally into the numerical scheme. In the FEM literature many different schemes have been proposed for the contact line implementation and, as discussed in detail in this work, many of them give different results for the same problem.

A priori, it is unclear, both qualitatively and quantitatively,  how the aforementioned difficulties affect the accuracy of a numerical
scheme and how localized to the contact line any errors will be. To resolve these issues, we develop a computational framework which both fully captures the dynamics of slip and
allows the contact line to be incorporated in a natural way. Then, this numerical tool is used to
examine previous approaches and provide guidelines for future investigations.

There is a large body of literature on the use of the FEM to simulate fluid flow \cite{gresho2}, and it is well known how to accurately incorporate liquid-fluid free surfaces into this method, e.g.\ \cite{kistler84,tezduyar92,christodoulou92,cairncross00}.  In this paper, we complement these works by providing a detailed exposition, including practical recommendations, of a framework for incorporating dynamic wetting phenomena in a regular way into a FEM scheme.

In \S\ref{equations}, we give the equations of a basic, representative, model of dynamic wetting.  The standard FEM approach to solving the governing equations is then outlined in \S\ref{FEM}. It is then described how previous works have attempted a number of different computational approaches to incorporate the dynamic wetting process.  In \S\ref{new1}, we present a new framework which, by implementing the boundary conditions on the liquid-solid interface in a methodologically similar way to that used on the free surface, allows the contact line to be incorporated into the finite element procedure in a natural way.  In the main body of the text the new approach is constructed so that it may be applied in an arbitrary coordinate system and is not dependent on the specific features of the FEM scheme, e.g.\ mesh design,  which have been chosen.  In the Appendix, we describe, in a step-by-step user-friendly guide, our FEM implementation in detail for the cases of two-dimensional and three-dimensional axisymmetric flow.

In \S\ref{example} we consider a representative example: the steady propagation of a liquid-gas meniscus through a capillary tube. This allows us to validate the new implementation against analytic results for specific flow regimes and check the convergence of the code as the spatial resolution of the FEM mesh is increased. The code allows one to critically analyze previous numerical implementations of dynamic wetting models, compare them to the new one and provide practical quantitative recommendations for future code development.

In \S\ref{surf}, we use the new framework to show how `additional' surface physics can be naturally incorporated into the FEM procedure and the effect which the dynamics has on the flow.  Finally, in \S\ref{conclusions}, we summarize our findings and suggest future directions of research.

\section{A representative model of dynamic wetting flows}\label{equations}

Consider the steady spreading of a Newtonian liquid, with constant density $\rho$ and viscosity $\mu$, over a chemically homogeneous smooth solid surface. The liquid is surrounded by a gas which, for simplicity, is assumed to be inviscid and dynamically passive, of a constant pressure $p_g$.  Let the flow be characterized by scales for length $L$, velocity $U$, pressure $\mu U/L$ and external body force $F_{0}$. In the dimensionless form, the continuity and momentum balance equations are then given by
\begin{equation}\label{ns}
 \nabla\cdot\mathbf{u} = 0,\qquad Re~\mathbf{u}\cdot\nabla\mathbf{u} = \nabla\cdot\mathbf{P} + St~\mathbf{F},
 \end{equation}
 where
\begin{equation}\label{stens}
\mathbf{P} = -p\mathbf{I}+\left[\nabla\mathbf{u}+\left(\nabla\mathbf{u}\right)^{T}\right],
\end{equation}
is the stress tensor, $\mathbf{u}$ and $p$ are the liquid's velocity and pressure, $\mathbf{F}$ is the external force density and $\mathbf{I}$ is the metric tensor of the coordinate system.   The non-dimensional parameters are the Reynolds number $Re=\rho U L/\mu$ and the Stokes number $St=\rho F_{0}L^{2}/(\mu U)$.

Boundary conditions to the bulk equations are required at the liquid-gas free surface $\mathbf{x}=\mathbf{x}_1(s,t)$, whose position is found as part of the solution, and at the liquid-solid interface $\mathbf{x}=\mathbf{x}_2(s,t)$, whose position is known; $(s,t)$ are the coordinates that parameterize the surfaces.

On the free surface the kinematic condition of impermeability is
\begin{equation}\label{kin}
\mathbf{u}\cdot\mathbf{n}_1 = 0.
\end{equation}
Hereafter, $\mathbf{n}$ is the unit normal to a surface pointing into the liquid, and subscripts $1$ and $2$ refer to the free surface and solid surface, respectively.

In what follows, it is convenient to introduce a (symmetric) tensor $\mathbf{I}-\mathbf{n}\mathbf{n}$, which is essentially a metric tensor on the surface.  If an arbitrary vector $\mathbf{a}$ is decomposed into a scalar normal component $a_n=\mathbf{a}\cdot\mathbf{n}$ and a vector tangential part $\mathbf{a}_{||}$ (the subscript $||$ henceforth denotes the tangential component of a vector), so that $\mathbf{a}=\mathbf{a}_{||} + a_n\mathbf{n}$, we can see that, because $\mathbf{n}\cdot\left(\mathbf{I}-\mathbf{n}\mathbf{n}\right)=\mathbf{0}$, the tensor $\left(\mathbf{I}-\mathbf{n}\mathbf{n}\right)$  extracts the component of a vector which is tangential to the surface, i.e.\ $\mathbf{a}\cdot\left(\mathbf{I}-\mathbf{n}\mathbf{n}\right)=\mathbf{a}_{||}$. Then,  $\nabla^s = \left(\mathbf{I}-\mathbf{n}\mathbf{n}\right)\cdot\nabla$ is the two-dimensional gradient operator playing the same role on the surface as $\nabla$ does in the bulk.

The classical boundary conditions for the tangential and normal stress on the free surface can now be written down as
\begin{equation}\label{fsa}
\mathbf{n}_1\cdot\left[\nabla\mathbf{u}+\left(\nabla\mathbf{u}\right)^T\right]\cdot
\left(\mathbf{I}-\mathbf{n}_1\mathbf{n}_1\right) +\frac{\nabla^{s}\sigma_1}{Ca} =\mathbf{0},\qquad
-p+ \mathbf{n}_1\cdot\left[\nabla\mathbf{u}+\left(\nabla\mathbf{u}\right)^T\right]\cdot
\mathbf{n}_1=\frac{\sigma_1\nabla^s\cdot\mathbf{n}_1}{Ca}.
\end{equation}
Here, the superscript $T$ denotes the transposition, $\sigma_1$ is the surface tension of the interface and the capillary number $Ca=\mu U/\sigma_{1e}$ is based on its equilibrium value $\sigma_{1e}$. The pressure in the liquid is considered relative to its value $p_g$ in the gas.

The solid is considered rigid, impermeable and moving at velocity $\mathbf{U}$, so that at the liquid-solid interface the standard form of the Navier-slip boundary condition, which in the present formulation replaces the condition of no-slip, is
\begin{equation}\label{ss_alt}
\pdiff{\mathbf{u}_{||}}{\mathbf{n}_2} = \bar{\beta}\left(\mathbf{u}_{||}-\mathbf{U}_{||}\right), \qquad
\left(\mathbf{u}-\mathbf{U}\right)\cdot\mathbf{n}_2=0,
\end{equation}
where $\bar{\beta}$ is the non-dimensional coefficient of slip\footnote{Note that in some works $\bar{\beta}$ is defined as the inverse of what we have chosen, e.g.\ \cite{kistler84,baer00}}.  The no-slip condition is recovered in the limit $\bar{\beta}^{-1}\rightarrow 0$.  The first term in (\ref{ss_alt}) represents the tangential stress acting on the interface and may be rewritten, to keep a consistent notation, in terms of $\mathbf{P}$, so that conditions (\ref{ss_alt}) take the form
\begin{equation}\label{ss}
\mathbf{n}_2\cdot\mathbf{P}\cdot\left(\mathbf{I}-\mathbf{n}_2\mathbf{n}_2\right)
=\bar{\beta}\left(\mathbf{u}_{||}-\mathbf{U}_{||}\right), \qquad
\left(\mathbf{u}-\mathbf{U}\right)\cdot\mathbf{n}_2=0.
\end{equation}

The differential term $\sigma_1\nabla\cdot\mathbf{n}_1$ in the normal stress equation (\ref{fsa}) indicates that this equation requires its own boundary condition where the free surface terminates, i.e.\ at the contact line, where $\mathbf{n}_1$ is prescribed provided that one specifies the contact angle $\theta$ at which the free surface meets the solid surface:
\begin{equation}\label{angle}
\mathbf{n}_{1}\cdot\mathbf{n}_2 = -\cos\theta.
\end{equation}
Here, $\mathbf{n}_2$ is determined by the (known) shape of the solid, the contact angle $\theta$ is specified by a particular model (or simply prescribed) and consequently $\mathbf{n}_1$ may be determined and provides a boundary condition for the normal stress equation (\ref{fsa}).

Equations (\ref{ns}) with boundary conditions (\ref{kin}), (\ref{fsa}), (\ref{ss}) and (\ref{angle}) are the `conventional' ones to describe flow near a moving contact line.  Although the shortcomings of these conventional boundary conditions as a mathematical model for dynamic wetting are well known (see, for example \cite{shik07}), as far as numerics is concerned, the described formulation is a necessary intermediate step towards implementing more sophisticated models.

Next, we shall show how the FEM converts the outlined continuous problem into an approximate discrete one which may be solved computationally to a desired degree of accuracy.

\section{Conventional FEM simulation of dynamic wetting flows}\label{FEM}

In this section, we will briefly recapitulate some of the main elements of the FEM, which is required in what follows as well as in the Appendix where a step-by-step implementation of the developed procedure is given.

\subsection{Weak form of the governing equations}

The defining feature of the FEM is that the computational domain $V$ is tessellated into a finite number of non-overlapping elements, each containing a fixed number of nodes at which the functions' values are determined.  Between these nodes the functions are approximated using interpolating functions whose functional dependence on position is chosen. Let $N_p$ be the total number of nodes in $V$ at which the pressure is determined, $N_u$ be the number of nodes at which the velocity components are to be found and $N_1$ be the number of nodes on the free surface $A_1$.  The functions are approximated as a linear combination of interpolating functions each weighted by the corresponding nodal value. This gives a trial solution of the form
\begin{equation}\label{approxn_global}
p^{\ast}               = \sum^{N_p}_{\textrm{j}=1} p_{\textrm{j}}\psi_{\textrm{j}}(\mathbf{x}),       \qquad
\mathbf{u}^{\ast}     = \sum^{N_u}_{\textrm{j}=1}\mathbf{u}_{\textrm{j}}\phi_{\textrm{j}}(\mathbf{x}),    \qquad
\mathbf{x}^{\ast}_1 = \sum^{N_1}_{\textrm{j}=1}\mathbf{x}_{1{\textrm{j}}} \phi_{1,{\textrm{j}}}(s,t), \qquad (\mathbf{x}\in V,~~\mathbf{x}_{1}\in A_1),
\end{equation}
where $\psi_{\textrm{j}}$, $\phi_{\textrm{j}}$ and $\phi_{1,\textrm{j}}$ are the interpolating functions and the asterisk is used to distinguish the approximate representations from the exact solution ($p,\mathbf{u},\mathbf{x}_1$). The interpolating functions are constructed so that they take the value one at the node with which they are associated $\textrm{j}$ and zero at all the other nodes to ensure that ($p_\textrm{j},\mathbf{u}_\textrm{j},\mathbf{x}_{1\textrm{j}}$) are nodal values.  Note that here we use Roman letters for the indices ($\textrm{i},\textrm{j}$, etc) to refer to the nodal values and approximating functions spanning the whole domain (globally); in the Appendix, where all the numerical details are given, these indices will be used alongside italicized ones ($i,j$, etc), which will refer to local, element-based, values and functions. It is not necessary to use the same level of approximation for the shape of the free surface and the velocity on it, but it is assumed here.  In practice, each interpolating function is constructed in an element-by-element fashion and will only be non-zero over a handful of elements: this procedure is described in the Appendix but is not required for the following exposition.

The representation $(p^{\ast},\mathbf{u}^{\ast},\mathbf{x}_1^\ast)$ of the actual solution $(p,\mathbf{u},\mathbf{x}_1)$ is not exact and hence the substitution of (\ref{approxn_global}) into the governing equations (\ref{ns})--(\ref{stens}) and the kinematic boundary condition (\ref{kin}) will result in errors (the $r$'s):
\begin{equation}\label{goveqn1}
\nabla\cdot\mathbf{u}^{\ast} = r^{C},
\end{equation}
\begin{equation}\label{goveqn2}
\mathbf{e}_\alpha\cdot\left[Re~\mathbf{u}^{\ast}\cdot\nabla\mathbf{u}^{\ast} - \nabla\cdot\mathbf{P}^{\ast} - St~\mathbf{F}\right] = r^{M,\alpha}, \qquad \mathbf{P}^{\ast} = -p^{\ast}\mathbf{I}+\left[\nabla\mathbf{u}^{\ast}+\left(\nabla\mathbf{u}^{\ast}\right)^{T}\right],
\end{equation}
\begin{equation}\label{goveqn3}
\mathbf{u}^{\ast}\cdot\mathbf{n}^{\ast}_1 = r^K  \qquad\hbox{at}\qquad \mathbf{x} = \mathbf{x}_1^\ast,
\end{equation}
where the unit basis vectors of the coordinate system $\mathbf{e}_{\alpha}~(\alpha=1,2,3)$ have been used to obtain scalar equations. Hereafter, the superscript $C$ will be used for quantities associated with the continuity of mass equation, $M$ for the momentum equations and $K$ for the kinematic boundary condition.    We have chosen to use the kinematic condition on the free surface (\ref{kin}) as the equation which determines the surface's position, as opposed to the normal stress condition (\ref{fsa}) which is shown in \S\ref{weakform} to be incorporated into the weak form of the momentum residuals.  Then, given these three equations (\ref{goveqn1})--(\ref{goveqn3}) the problem becomes symmetric with respect to the boundary conditions: on each interface, i.e.\ the free surface and the solid boundary, two boundary conditions are applied.  Henceforth, we shall drop the asterisks and realize that all functions to be considered represent approximate solutions to the problem.

The aim is to minimize the errors spatially over the entire domain.  In the method of weighted residuals \cite{finlayson66} on which the FEM is based, this is achieved by minimizing a weighted spatial average of the error.  Specifically, we multiply a given error by an arbitrary test function (the weight) $W$ and integrate the equation over the corresponding domain.  Applying this procedure to all errors creates the so-called residual equations.  For the bulk equations this requires integration over $V$ whilst for the kinematic equation integration is over the entire free surface $A_1$, so that the residuals to be considered are
\begin{equation}\label{weakeqn1}
R^{C} = \int_{V} W^{C}(\mathbf{x}) r^{C}~dV,\qquad R^{M,\alpha} = \int_{V} W^{M}(\mathbf{x}) r^{M,\alpha}~dV,\qquad R^{K} = \int_{A_1} W^{K}(s,t) r^{K}~dA_{1} .
\end{equation}
Insisting that these integrals vanish
\begin{equation}\label{weakeqn}
R^{C}=R^{M,\alpha}=R^{K}=0,
\end{equation}
for \emph{all} test functions, of which there are infinitely many, creates the so-called \emph{weak form} of the (strong form) governing equations of \S\ref{equations}. A link between the weak and strong forms is provided by the fundamental lemma of the calculus of variations which states that if certain equation-specific constraints on the test functions are satisfied, such as continuity and boundedness, and the relations above hold for all such test functions, then an approximate solution which is continuous will be exact, i.e.\ a solution of the strong form, see \cite[p.~58]{cuvelier86} for specific details.  In practice we choose a finite set of test functions to generate a discrete set of equations for the nodal values of the functions, which, it is expected, will approximate the continuous problem as closely as required.

\subsection{The Galerkin method}

It is useful to think of an equation associated with each node as `determining' the nodal value of a function there. Mathematically, this is crude, but numerically it is helpful, and we henceforth use the word `determine' in this sense.  Specifically, the $N_p$ continuity of mass equations can be thought of as determining the $N_p$ pressure nodal unknowns, the $N_u$ momentum equations in each coordinate direction determine the $N_u$ velocity nodal unknowns in that direction and the $N_1$ kinematic equations on the free surface determine the $N_1$ nodal unknowns which specify the free surface shape (see the Appendix for a specific free surface representation).

To discretize the equations, one has to replace the infinite set of test functions, the $W$'s, by a finite set $W\rightarrow W_{\textrm{i}}~~(\textrm{i}=1\hbox{--}N)$ which spans the solution space as completely as possible.  In the Galerkin method this is achieved by choosing the test functions to be the same interpolating functions used for our variables.  Specifically, the interpolating function of a given variable is the same as the test function for the equation which determines that variable's value, so that
\begin{equation}\label{approxn_weight}
W^C_{\textrm{i}}(\mathbf{x})   = \psi_{\textrm{i}}(\mathbf{x}) \quad (\textrm{i}=1\hbox{--}N_p),\qquad
W^{M}_{\textrm{i}}(\mathbf{x}) = \phi_{\textrm{i}}(\mathbf{x}) \quad (\textrm{i}=1\hbox{--}N_u),\qquad
W^{K}_{\textrm{i}}(s,t) = \phi_{1,\textrm{i}}(s,t) \quad (\textrm{i}=1\hbox{--}N_1).
\end{equation}
After substituting the interpolating functions in (\ref{approxn_weight}) into the expressions for residuals (\ref{weakeqn1}), equations (\ref{weakeqn}) are transformed into a finite set of algebraic equations
\begin{equation}\label{weakish}
R^{C}_{\textrm{i}}=0~~~(\textrm{i}=1\hbox{--}N_p),\qquad R^{M,\alpha}_{\textrm{i}}=0~~~(\textrm{i}=1\hbox{--}N_u),\qquad R^{K}_{\textrm{i}}=0~~~(\textrm{i}=1\hbox{--}N_1).
\end{equation}
In \S\ref{weakform}, we give specific expressions for these and show how they can be manipulated to allow some of the boundary conditions to be easily incorporated. As one increases the number of nodes in the domain, the error from approximating a continuous problem with a discrete one should decrease, and this will be checked for the problem in question in \S\ref{convergence}.

Here, \emph{global} residuals and functions, i.e.\ quantities considered over the entire domain, have been considered. In practice, the global interpolating functions are constructed piecewise over elements, i.e.\ \emph{locally}, as low order polynomials.  Then, the global residuals are obtained by summing up contributions from each element. In the velocity-pressure formulation of the incompressible Navier-Stokes equations, mixed interpolation \cite{fortin81}, with velocity approximated by a polynomial of degree at least one more than pressure, ensures that the Ladyzhenskaya-Babu\u{s}ka-Brezzi \cite{babuska72} condition is satisfied\footnote{Equal-order methods may also be used with stabilization, e.g.\ \cite{hughes86,codina00}, but these issues lie beyond the scope of this paper.}. Instead of considering the elements of varying shapes in the global domain, it is useful to map each element onto a \emph{master element} over which analysis can be systematically performed: this ensures that the solution to problems in complex domains pose no additional difficulty.  This procedure is described in detail in the Appendix, but will not be required in what follows.

\subsection{Discrete problem formulation}\label{weakform}
In this section, using (\ref{weakeqn1}) and (\ref{approxn_weight}) the discretized weak form of the governing equations (\ref{weakish}) is presented and it is shown how the boundary conditions are applied. Functions appearing in the residuals are approximated using (\ref{approxn_global}).

The continuity of mass (incompressibility of the fluid) residuals $R^{C}_{\textrm{i}}$ are
\begin{equation}\label{incom}
R^{C}_{\textrm{i}} = \int_{V} \psi_{\textrm{i}}\nabla\cdot\mathbf{u}~dV\qquad (\textrm{i}=1\hbox{--}N_p) ,
\end{equation}

Momentum equations are
\begin{equation}\label{ns1}
R^{M,\alpha}_{\textrm{i}} = \int_{V} \phi_{\textrm{i}}\mathbf{e}_{\alpha} \cdot \left[Re~\mathbf{u}\cdot\nabla\mathbf{u} - \nabla\cdot\mathbf{P} -
St~\mathbf{F}\right]~dV \qquad (\textrm{i}=1\hbox{--}N_u).
\end{equation}
Assuming that the interpolating function is at least once differentiable, the stress tensor's contribution can be represented as
\begin{equation}
-\phi_{\textrm{i}} \mathbf{e}_\alpha\cdot\nabla\cdot\mathbf{P} = -\nabla\cdot\left(\phi_{\textrm{i}}\mathbf{e}_\alpha\cdot\mathbf{P}\right) + \nabla\left(\phi_{\textrm{i}} \mathbf{e}_\alpha\right):\mathbf{P},
\end{equation}
so that, after applying the divergence theorem, one has
\begin{equation}\label{ns2}
R^{M,\alpha}_{\textrm{i}} = \int_{V}  \left[\phi_{\textrm{i}}\mathbf{e}_\alpha\cdot\left(Re~ \mathbf{u}\cdot\nabla\mathbf{u} - St~\mathbf{F}\right)+
\nabla(\phi_{\textrm{i}}\mathbf{e}_\alpha):\mathbf{P}\right]~dV +
\int_{A} \phi_{\textrm{i}}\mathbf{e}_\alpha\cdot\mathbf{P}\cdot\mathbf{n}~dA ,
\end{equation}
where, as before, $\mathbf{n}$ is the inward normal to the surface $A$ which forms the boundary to $V$.  Therefore, the volume
contribution to the momentum residuals is
\begin{equation}\label{vol}
\left(R^{M,\alpha}_{\textrm{i}}\right)_V = \int_{V}
\left[\phi_{\textrm{i}}\mathbf{e}_\alpha\cdot\left(Re~ \mathbf{u}\cdot\nabla\mathbf{u} - St~\cdot\mathbf{F}\right)+
\nabla(\phi_{\textrm{i}}\mathbf{e}_\alpha):\mathbf{P}\right]~dV,
\end{equation}
whilst the contribution to the momentum residual from the surface is
\begin{equation}\label{stress}
\left(R^{M,\alpha}_{\textrm{i}}\right)_A = \int_{A} \phi_{\textrm{i}}\,\mathbf{e}_\alpha\cdot\mathbf{P}\cdot\mathbf{n}~dA.
\end{equation}
In the last expression, only when node $\textrm{i}$ is on the surface $A$ will $\phi_{\textrm{i}}$ be non-zero, i.e.\ only nodes on the surface of $V$ will give contributions to the momentum residuals via (\ref{stress}).

Previously, it was described how for each nodal unknown there is an equation which is assumed to determine its value. In all numerical methods, to maintain the equality between the number of unknowns and equations, in applying the boundary conditions we must either (a) replace a bulk equation at boundary nodes by an equation representing the discretized form of the boundary condition, or (b) manipulate the bulk equations to allow the boundary condition to be incorporated into them. We refer to (a) as an \emph{essential} imposition of the boundary condition; the bulk equation which is replaced is referred to as the \emph{dropped} equation. The (b)-type implementation of boundary conditions will be referred to as \emph{natural} imposition, which, for stress conditions, is made possible by the contribution (\ref{stress}) that is replaced by the corresponding expression for the stress on a given surface (the FEM is therefore referred to as allowing a natural incorporation of stress-type boundary conditions).

\begin{figure}
\centering
\includegraphics[scale=0.60]{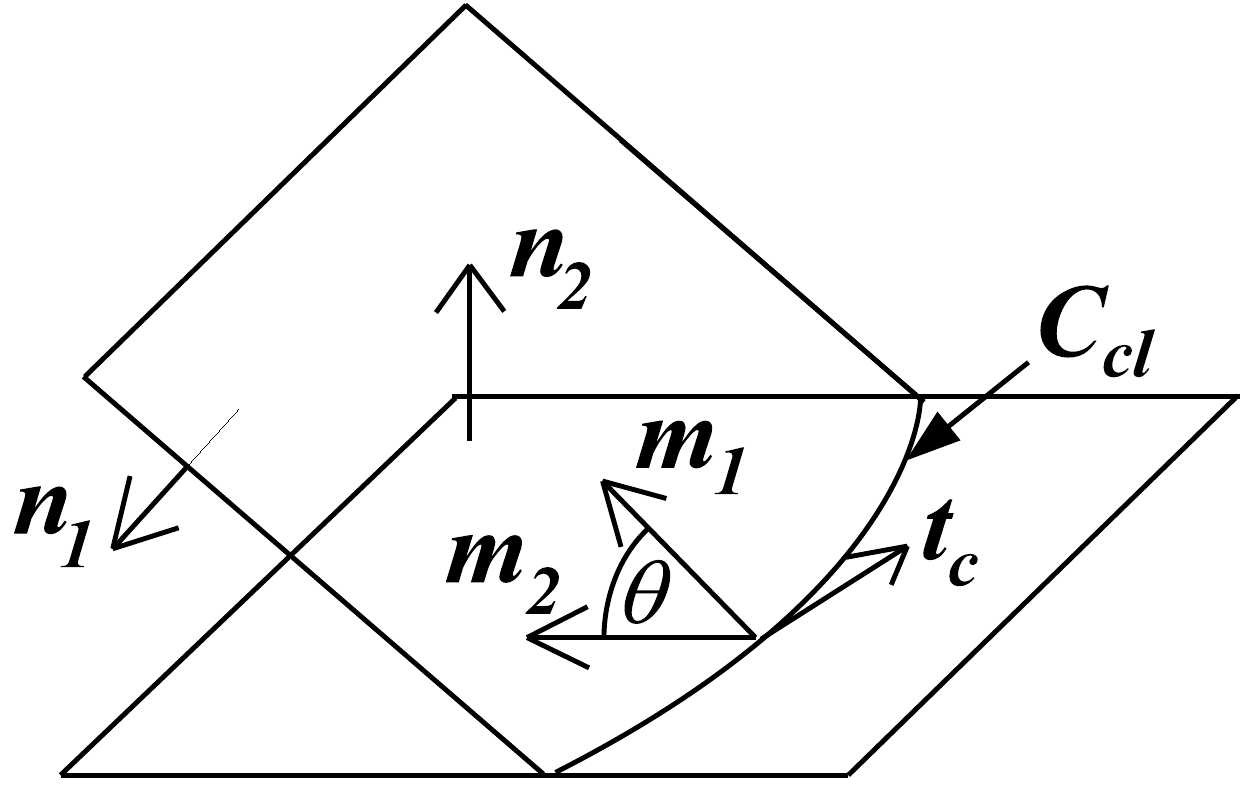}
\caption{Illustration showing the vectors associated with the liquid-gas free surface $A_1$ and the liquid-solid surface $A_2$ in the vicinity of the contact line $C_{cl}$.}
\label{F:angles}
\end{figure}

On the free surface, boundary conditions (\ref{fsa}) can be combined into a more computationally favourable form
\begin{equation}\label{fs} 	
Ca~\mathbf{n}_1\cdot\mathbf{P}
+\nabla^s\cdot\left[\sigma_{1}\left(\mathbf{I}-\mathbf{n}_1\mathbf{n}_1\right)\right]=\mathbf{0},
\end{equation}
where $\sigma_{1}\left(\mathbf{I}-\mathbf{n}_1\mathbf{n}_1\right)$ is the surface stress, playing the same role on the surface as $\mathbf{P}$ does in the bulk, so that (\ref{fs}) is actually a balance of stress acting on the interface (first term) and that acting in the interface (second term).  Then, using (\ref{fs}), one can rewrite (\ref{stress}) on the free surface as
\begin{equation}\label{fs2}
\int_{A_1} \phi_{\textrm{i}}\mathbf{e}_\alpha\cdot\mathbf{P}\cdot\mathbf{n}_1~dA = -\frac{1}{Ca}\int_{{A}_1}
\phi_{1,\textrm{i}} \mathbf{e}_\alpha\cdot\nabla^s\cdot\left[\sigma_{1}\left(\mathbf{I}-\mathbf{n}_1\mathbf{n}_1\right)\right]~dA_1.
\end{equation}
Direct calculation of the integrand in (\ref{fs2}) would require the evaluation of the derivatives
of the vector normal to the surface. It was initially suggested by Ruschak \cite{ruschak80}, and generalized for
three-dimensional problems in \cite{ho91,cairncross00}, that by using the surface divergence
theorem one could obtain an expression in which no such derivatives are required.  Lowering the highest derivatives reduces the constraints on the differentiability of the interpolating functions which are used to approximate the free surface shape, i.e.\ $\phi_{1,{\textrm{i}}}$, and hence allows lower-order polynomials to be used. The required transformation of (\ref{fs2}) is given by
\begin{equation}\label{fs3}
\int_{A_1} \phi_{\textrm{i}}\mathbf{e}_\alpha\cdot\mathbf{P}\cdot\mathbf{n}_1~dA_1 = -\frac{1}{Ca}\int_{{A}_1}\left[
\nabla^s\cdot\left(\phi_{1,\textrm{i}}\sigma_1\mathbf{e}_\alpha\cdot\left(\mathbf{I}-\mathbf{n}_1\mathbf{n}_1\right)\right)-
\sigma_1\nabla^s\cdot\left(\phi_{1,\textrm{i}}\mathbf{e}_\alpha\right)\right]~dA_1.
\end{equation}
Then, using the surface divergence theorem \cite[p.~224]{aris62}, which for a surface vector $\mathbf{a}$
with no normal component $\mathbf{a}=\mathbf{a}_{||}$ is
\begin{equation}
\int_{A}\nabla^s\cdot\mathbf{a}_{||}~dA = -\int_{C}\mathbf{a}_{||}\cdot\mathbf{m}~dC,
\end{equation}
where the unit vector $\mathbf{m}$ is the inwardly facing normal to the contour $C$ that confines $A$ (Figure~\ref{F:angles}), with $\mathbf{a}_{||}=\phi_{1,{\textrm{i}}}\sigma_1\mathbf{e}_\alpha\cdot\left(\mathbf{I}-\mathbf{n}_1\mathbf{n}_1\right)$ we obtain
\begin{equation}\label{fs4}
\int_{A_1} \phi_{\textrm{i}}\mathbf{e}_\alpha\cdot\mathbf{P}\cdot\mathbf{n}_1~dA_1 =
\frac{1}{Ca}\int_{{A}_1}\sigma_1\nabla^s\cdot\left(\phi_{1,\textrm{i}}\mathbf{e}_\alpha\right)~dA_1 +
\frac{1}{Ca}\int_{C_1}\phi_{1,\textrm{i}}\sigma_1\mathbf{e}_\alpha\cdot\mathbf{m}_{1}~dC_1.
\end{equation}
Thus, on the free surface, the term (\ref{stress}) in the momentum residual is now split into surface and line contributions
\begin{equation}\label{fs5}
\left(R^{M,\alpha}_{\textrm{i}}\right)_{A_{1}} =
\frac{1}{Ca}\int_{{A}_1}\sigma_1\nabla^s\cdot\left(\phi_{1,\textrm{i}}\mathbf{e}_\alpha\right)~dA_1, \qquad
\left(R^{M,\alpha}_{1,\textrm{i}}\right)_{C_{1}} = \frac{1}{Ca}\int_{C_1}\phi_{1,\textrm{i}}\sigma_1\mathbf{e}_\alpha\cdot\mathbf{m}_{1}~dC_1,
\end{equation}
with the stress boundary conditions (\ref{fs}) incorporated.

Notably, the same procedure of integrating by parts and using the divergence theorem has been used on both the surface stress term $\nabla^s\cdot\left[\sigma_{1}\left(\mathbf{I}-\mathbf{n}\mathbf{n}\right)\right]$  and the bulk stress term $\nabla\cdot\mathbf{P}$.  In both cases, this has created contributions from the boundary of that term's domain, i.e.\ the confining contour and surface, respectively. Consequently, the momentum residual now contains a cascade of scales
\begin{equation}\label{casc}
R^{M,\alpha}_{\textrm{i}}=\left(R^{M,\alpha}_{\textrm{i}}\right)_{V}+\left(R^{M,\alpha}_{\textrm{i}}\right)_{A}+\left(R^{M,\alpha}_{\textrm{i}}\right)_{C},
\end{equation}
which represent, respectively, the volume, surface and line contributions. In particular, part of the contour which bounds the free surface is the contact line $C_{cl}$ where the liquid-gas interface meets the solid. Other boundaries to the free surface further away from the contact line, for example an axis of symmetry, are not considered here.

At the contact line, it is useful to rearrange the term in the integrand of the contour integral in (\ref{fs5}) by representing the vector $\mathbf{m}_1$ in terms of a linear combination of its components parallel to $\mathbf{n}_2$ and $\mathbf{m}_2$ (Figure~\ref{F:angles}):
\begin{equation}\label{decomp}
\mathbf{m}_1  =
\left(\mathbf{m}_{1}\cdot\mathbf{m}_2\right)\mathbf{m}_2 +
\left(\mathbf{m}_{1}\cdot\mathbf{n}_2\right)\mathbf{n}_2.
\end{equation}
Using this to impose the contact angle (\ref{angle}), which is a boundary condition for the free surface shape, results in
\begin{equation}\label{fscl}
\left(R^{M,\alpha}_{\textrm{i}}\right)_{C_{cl}} =
\frac{1}{Ca}\int_{C_{cl}}\sigma_1\phi_{\textrm{i}}\mathbf{e}_\alpha\cdot\left(\mathbf{m}_2\cos\theta +
\mathbf{n}_2\sin\theta\right)~dC_{cl},
\end{equation}
where the normal and tangent vectors on the solid surface are known a-priori. If contact line's tangent vector is $\mathbf{t}_c$, then $\mathbf{m}_2=\pm\mathbf{t}_c\times\mathbf{n}_2$, with the sign chosen to ensure the inward facing vector $\mathbf{m}_2$ is selected.

Thus, the contact angle equation (\ref{angle}) can be applied in a \emph{natural} way, that is without needing to drop another equation in order to impose the contact angle condition and thus fix the shape of the free surface at the contact line.

The kinematic condition residuals $R^{K}_{\textrm{i}}$ complete the residuals on the free surface:
\begin{equation}\label{w_kin}
R^{K}_{\textrm{i}} = \int_{A_1}\phi_{1,\textrm{i}}
\mathbf{u}\cdot\mathbf{n}_1~dA_1\qquad (\textrm{i}=1\hbox{--}N_1).
\end{equation}

Thus far, the implementation has been straightforward: the stress conditions are applied in the weak form of the momentum equation, and their integration by parts allows specification of the angle at which the free surface meets the solid boundary.  This means that momentum equations on the free surface do not need to be replaced in order to apply essential surface or line conditions.  In other words, all of the bulk equations are applied at every node, to determine the velocity components, with both the free surface stress conditions and the specification of the contact angle fitting naturally into the weak form of the momentum equations. The kinematic condition is applied as a separate equation to determine the free surface position. Notably, the implementation is independent of the free surface shape, that is the curved nature of the free surface is as easy to handle as, say, a planar surface which aligns with coordinate axes.

On the rigid solid surface, where we have $N_2$ nodes, it is the Navier condition and the condition of impermeability (\ref{ss}) that have to be imposed.  The
Navier-slip condition is of stress type and hence contributes to the weak form of the momentum
equations in a natural way, giving
\begin{equation}\label{ss2}
\left(R^{M,\alpha}_{\textrm{i}}\right)_{A_{2}} = \bar{\beta}\int_{{A}_2} \phi_{2,\textrm{i}}
\mathbf{e}_\alpha\cdot\left(\mathbf{u}_{||}-\mathbf{U}_{||}\right)~dA_2,
\end{equation}
where $\phi_{2,\textrm{i}}$ is an interpolating function for the liquid-solid interface corresponding to the $\textrm{i}$--th node. However, the impermeability condition is not of stress type and hence does not fit naturally into the weak form of the momentum equations. Consequently, it is usually applied as an essential condition which creates a new residual equation at each of the nodes on the liquid-solid interface:
\begin{equation}\label{ss3}
R^{I}_{\textrm{i}}=0 \qquad (\textrm{i}=1\hbox{--}N_{2})\qquad\hbox{where}\qquad R^{I}_{\textrm{i}} = \int_{{A}_2} \phi_{2,\textrm{i}}
\left(\mathbf{u}-\mathbf{U}\right)\cdot\mathbf{n}_2~dA_2.
\end{equation}
Then, as previously described, in order to retain equality between the number of unknowns and number of equations in the numerical scheme, an equation must be dropped to make room for (\ref{ss3}). As (\ref{ss3}) is an equation which determines a component of velocity, and it is the momentum equations that are regarded as determining these, it is a momentum equation that should be dropped.  The dropping of a momentum equation in order to make space for an essential condition is achieved more formally, see e.g.\ \cite{gresho2,cuvelier86}, by defining the interpolating functions such that they are zero at nodes at which an essential condition is to be applied. Either way, the end result is exactly the same.

The procedure outlined above is straightforwardly achieved when the solid surface is parallel to a coordinate plane, so that one can drop the momentum equation normal to this plane and apply (\ref{ss3}) instead. The momentum equations tangential to the boundary only require an expression for the tangential stress and this is provided by the Navier-slip condition.  If the solid surface is not parallel to a coordinate plane, then, so as not to favour one of the axes, before applying the boundary conditions one must project the momentum equations at each node on the surface so that they are tangent and normal to it \cite{engelman82}. This is not a complex procedure, but it is considerably more cumbersome, and hence more time-consuming, than that encountered on the free surface, where curved shapes do not cause any additional difficulties. One can imagine that such a procedure will be computationally intricate when considering flows over topographically complex surfaces where the momentum residuals will have to be projected on different axes at each node.

So far, the weak formulation of the bulk equations has been derived and it has been shown how the boundary conditions on the liquid-gas free surface and the liquid-solid interface are implemented.  Additionally, it was shown that the contact angle may be naturally imposed by specifying a contribution to the momentum equations from the contact line.  So, on the liquid-gas free surface all momentum equations are applied, with stress-type boundary conditions naturally imposed, as well as the kinematic condition, whereas on the liquid-sold interface one momentum equation was dropped in order to apply the essential condition of impermeability.  Next, we consider what happens when we reach the nodes at which these interfaces, which, as described above, are treated in methodologically different ways, meet at the contact line.

\subsection{Conventional FEM implementation of the contact line}\label{c_ex}

Ideally, one would use an implementation which allows all the boundary conditions from both sides of the contact line to be applied in addition to the contact angle specification at the contact line.  However, it will be shown that a count of the number of equations and unknowns indicates that this is not possible.  Consequently, in the method as it stands, one has to favour certain boundary conditions over others to arrive at an equality between equations and unknowns.

To make the argument most transparent, consider as an example the free surface meeting a solid which is parallel to the $z$-axis and motion occurs in the $(r,z)$-plane of a Cartesian coordinate system with the liquid below, in terms of $z$, the free surface (e.g.\ see the geometry of Figure~\ref{F:sketch}).  In this geometry, the free surface can be represented, for example, as $z=h(s)$, where $s$ is the arclength along it, so that in the discretized form $h_\textrm{i}$ ($\textrm{i}=1\hbox{--}N_1$) are the unknowns that determine the free surface's position.

The incompressibility equation is applied at every node and determines the pressure. At every node of the liquid-gas free surface both momentum equations are used, with the stress boundary conditions applied naturally, for the components of velocity $(u_{\textrm{i}},w_{\textrm{i}})$, and one kinematic condition is applied for the free surface unknown $h_{\textrm{i}}$. On the liquid-solid interface the $z$-momentum equation is applied to determine $w_{\textrm{i}}$, with the Navier boundary condition applied naturally, whilst the $r$-momentum equation is dropped to allow the essential impermeability condition, $u_{\textrm{i}}=0$, to be satisfied.

At the contact line, node $\textrm{i}=1$, we have unknowns $u_1,w_1,h_1$, with $h_1$ corresponding to the $z$-coordinate of the contact line.  Application of the $z$-momentum equation, to determine $w_1$, allows both the Navier boundary condition along the solid and the $z$-projection of the stress conditions on the free surface to be imposed.  This is a good start, but there are now three equations left - the $r$-momentum equation, the kinematic condition on the free surface and the impermeability condition on the solid surface - for only two unknowns $u_1,h_1$.  The conventional solution to this difficulty is to favour the latter two equations and not apply the $r$-momentum equation at the contact line.  This is in keeping with the usual FEM approach of replacing bulk equations wherever essential boundary are applied. However, the equations at the contact line now enforce the $z$-projection of the free surface stress boundary condition but not the $r$-projection.  Furthermore, the contact angle specification is no longer applied into the $r$-momentum equation at the contact line. Using (\ref{fscl}) and noting that $\mathbf{m}_2=(0,-1)$ and $\mathbf{n}_2=(-1,0)$, the contribution from the contour integral in (\ref{fscl}) to the $z$-momentum equation at the contact line will be
\begin{equation}\label{contri1}
\left(R^{M,z}_1\right)_{C_{cl}} =-\phi_{1,1}\sigma_1\cos\theta/Ca.
\end{equation}

In \S\ref{previous},  implementations of the contact line in FEM codes known in the literature are discussed. In all these cases, as above, at least one momentum equation is not applied at the contact line.  This type of implementation, in which equations are dropped at the contact line, will be referred to as Type A and in this framework the following two different implementations of the contact angle specification have been developed:
\begin{itemize}\label{1or2}
  \item[(A-a)] The contact angle is imposed naturally, as described above, using (\ref{fscl}) (whilst one momentum equation is dropped already).
  \item[(A-b)] The contact angle is imposed as an essential condition (with one more momentum equation also dropped).
\end{itemize}
Exactly the same methodological issues arise in fully three-dimensional flow simulations.

\begin{figure}
\centering
\includegraphics[scale=0.650]{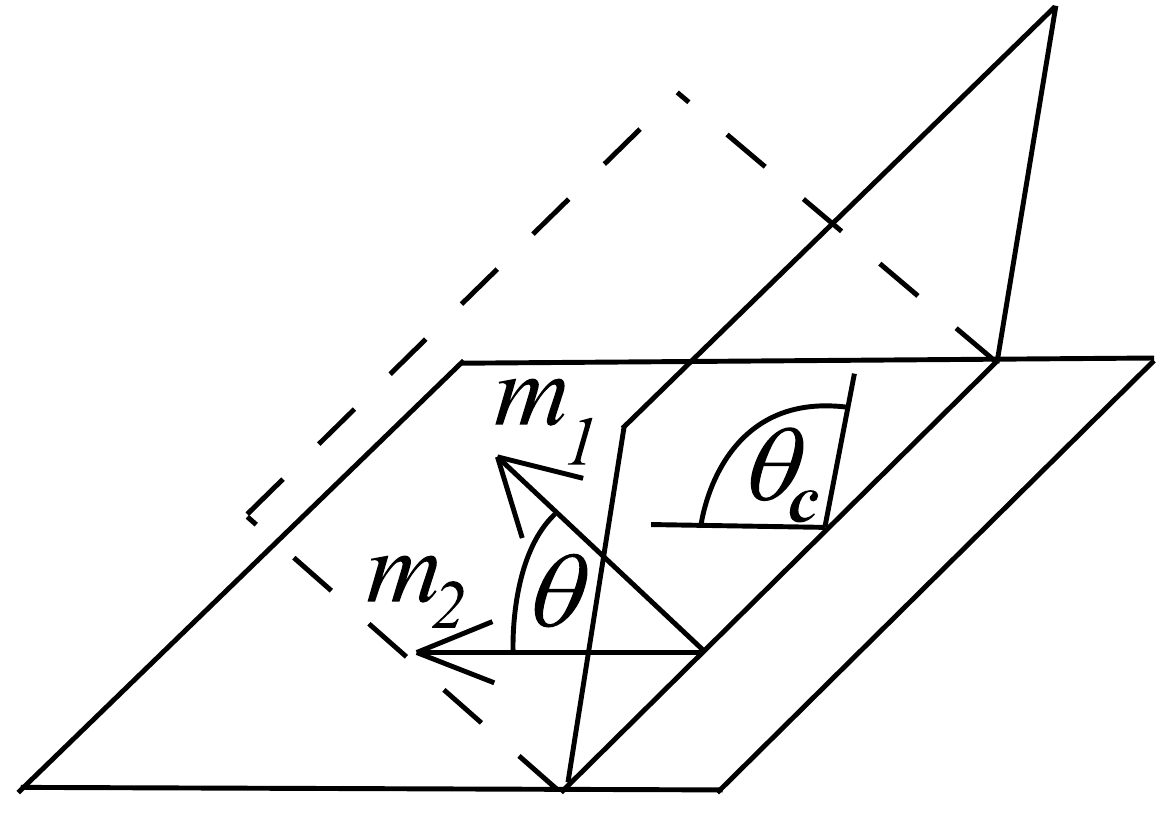}
\caption{Illustration of a solution in which the contact angle at which the computed free surface meets the solid $\theta_c$ differs from the applied angle $\theta$ which appears in the mathematical model.}
\label{F:computed}
\end{figure}

Before reviewing previous works that use the FEM to simulate dynamic wetting flows, with particular attention to the development of different implementations of the contact angle, it is necessary to make a remark concerning the different `contact angles' now involved in the modelling.  When the contact angle specified in the mathematical model is incorporated into the FEM code in a natural way, i.e.\ absorbed into the contribution to the residual from the contact line, as in the (A-a) implementation, there is no guarantee that the contact angle that emerges after computations are performed, i.e.\ the angle at which the computed free surface meets the solid boundary, will be equal to the specified one.  We will refer to the angle incorporated naturally into the code, and appearing in the mathematical model, as \emph{applied}, and denote it $\theta$, whereas the contact angle at which the computed free surface meets the solid will be referred to as \emph{computed} and denoted as $\theta_c$ (Figure~\ref{F:computed}).  In contrast, in the (A-b) implementation there is a single equation which ensures that $\theta_c = \theta$.

\subsection{Previous approaches}\label{previous}

We will now consider the development of the two different FEM implementations, (A-a) and (A-b) in the literature and, in particular, focus on how they appear to produce contradictory results. This is by no means supposed to be an exhaustive review, rather a pointer to some key papers and their conclusions.

Building on early articles considering the FEM simulation of free surface flows, for example \cite{nickell74}, pioneering research into the use of FEM to describe the dynamics of coating flows, and hence dynamic wetting phenomena, came from the University of Minnesota. The study of flows with contact lines began by computing static meniscus shapes in \cite{orr75,orr77,brown80} after which problems such as the die-swell phenomenon, with static contact lines, were simulated \cite{silliman78,silliman80,saito81},
and coating processes involving dynamic contact lines considered in \cite{kistler83,kistler84,christodoulou89,christodoulou92}. The conclusions of these investigations and the techniques, subsequently used in many works, are as follows.

Initially, the Minnesota group used the weak formulation to impose the shape of a free surface where it intersected a boundary, i.e.\ implementation (A-a). This appears to have worked well for imposing the contact angle of a static meniscus shape. Using this implementation to impose the contact angle in a dynamic wetting problem was discussed in \cite[p.~263]{kistler83}.  However, in \cite[p.~217]{kistler84}, it was found that implementation (A-a) used in \cite{kistler83} resulted in the computed angle $\theta_c$ differing significantly from the applied angle $\theta$, and this discrepancy, in the worst case, even prevented the numerical scheme from obtaining a solution.  To circumvent the problems observed in \cite{kistler84}, it was suggested that \emph{all} the momentum equations should be dropped at the contact line to allow the contact angle to be imposed as an essential condition, i.e.\ implementation (A-b) was proposed. Similar conclusions are discussed in \cite[p.~333]{christodoulou89} and \cite[p.~46]{christodoulou92}, and it appears that this method was henceforth adopted.  These papers appear to be the first in which it is recognized that implementation (A-a) does not guarantee that the applied angle will be equal to the computed one.  Their conclusion is that implementation (A-a) \emph{must} be utilized to guarantee that $\theta_c=\theta$ and allow seemingly satisfactory results to be obtained.


In \cite{bach85}, an FEM code is designed for the simulation of free surface flows and implementation (A-a) is used to specify the contact angle.  This paper is the first to suggest a new interpretation of the phenomenon which the Minnesota group had observed. The authors stress that (A-a) does not, and should not necessarily, make $\theta_c$ equal to $\theta$: the term (\ref{fscl}), where the contact angle is incorporated into the weak formulation, merely creates a net interfacial force on the contact line.  The computed angle $\theta_c$, which varies from the applied one $\theta$, is then referred to as the `dynamic' contact angle, which, for small capillary numbers, was found to be close to $\theta$.  This interpretation was adopted in \cite{fukai93,fukai95}, where implementation (A-a) was used in studies of drop impact and spreading phenomena. As the drop spreading process commences, with the initial computed contact angle at $\theta_c = 180^{\circ}$, for early times there is a huge, unavoidable, difference between $\theta$ (in the model it is assumed that $\theta$ is a constant which in all cases is less than $100^\circ$) and $\theta_c$.  From these publications the idea emerged that the dynamic behaviour of the \emph{computed} angle, which differs from the (constant) applied one, is a physical effect, despite the fact that there is no $\theta_c$ in the original mathematical formulation of the problem whose solution the code is supposed to describe.

In all of the aforementioned papers, the FEM mesh has been designed to capture the general characteristics of the bulk flow, and the region near the contact line is given no special treatment.  In particular, there is no attempt to capture the dynamics of slip, which approximately occurs on the (non-dimensional) slip length $\bar{\beta}^{-1}$, by placing a number of elements within this distance of the contact line sufficient to resolve this region. The articles use what is referred to as `numerical slip', which broadly refers to numerical schemes where slip is only prescribed, often in a couple of elements adjacent to the contact line, to ensure that the nodes at the contact line are not pinned to the solid by the no-slip condition.  In other words, slip is used as a numerical convenience and no attention is paid to the extra physics which it introduces that occurs on a length scale greatly smaller than the elements used.  Given that the region of numerical slip depends on the size of the elements adjacent to the contact line, the results of such computations are obviously mesh-dependent.

An alternative approach is to create a special mesh which allows small elements to be placed around the contact line and the slip dynamics to be captured.  This is achieved in \cite{lowndes80,bazhlekov96} where the propagation of a liquid-fluid meniscus through a capillary tube is considered.  In \cite{bazhlekov96}, implementation (A-a) is used, whilst in \cite{lowndes80} the normal-stress boundary condition is used to iterate the free surface position with the contact angle applied as an integral constraint on the pressure level, i.e.\ it is not imposed as an essential condition. In \cite{bazhlekov96}, it is reported that the computed angle $\theta_c$ was within $0.01^{\circ}$ of $\theta$ and in \cite{lowndes80} no significant deviation is noted.  These results appear to suggest that, if sufficiently small elements are used near the contact line, so that the dynamics of slip is well resolved, then the weak formulation can indeed be used to impose a specified contact angle.

Even in recent papers, there is still much disagreement over which approach is correct. For example, in \cite{ganesan09} approach (A-a) is used, with the computed angle, which differs from the applied one, interpreted as the dynamic contact angle; whereas, say, in \cite{baer00} the implementation (A-b) is taken, with all the momentum equations dropped at the contact line. Variance also exists in whether or not the dynamics of slip is resolved: some authors design special meshes to do so, e.g.\ in \cite{salamon97,wilson06,sprittles10}, whilst in many other publications the region of slip is under-resolved either because of constraints on computational power, as could be the case for three-dimensional simulations, or simply because this phenomenon is not considered important.

Previous investigations show that when a `standard' mesh is used, i.e.\ one which does not incorporate smaller elements near the contact line, implementations (A-a) and (A-b) certainly create different solutions: in case (A-a) it has been established that $\theta_c$ deviates from $\theta$ imposed via the weak formulation, whereas in (A-b) the contact angle is imposed as an essential condition at the expense of dropping another of the momentum equations at the contact line, so that it is unclear whether or not the imposition of $\theta_c=\theta$ creates problems with the accuracy elsewhere in the scheme. The deviation of $\theta_c$ from $\theta$ which was observed when using (A-a) has been attributed to the fact that the implementation only imposes a net interfacial force at the contact line, but, as far as we are aware, there has been no thorough convergence studies performed to see if the contact angle varies as the mesh is refined and whether additional mesh fineness in the immediate vicinity of the contact line has an influence on the flow further away.

What has been presented thus far broadly summarizes the popular implementations used in the literature. However, a closer look at previous works reveals even more sub-varieties of implementation. For example, in implementation (A-a) it is usually the projection of the momentum equation tangent to the solid surface which is retained; however, in \cite{kistler83} the projection of the momentum equation onto the free surface's normal is chosen.  Furthermore, in \cite{salamon97}, instead of applying the impermeability equation separately on the free surface and the solid surface at the contact line, i.e.\ creating two FEM residual equations, the two were combined to form one impermeability residual with contributions from the two surfaces. Encouragingly, this created room in the formulation to apply both momentum equations at the contact line alongside the (one) impermeability equation.  However, it was observed that using this approach resulted in an $O(1)$ velocity at the contact line, which violated the requirement that both the free surface and the solid surface were impermeable.  Implementation (A-b) was then adopted to overcome this issue. No doubt, other variations exist and clearly the implementation options at the contact line are plentiful.


By now, it is unclear as to which implementation of the above, if any, is correct, what effect the dominant parameters in the contact line implementation $\bar{\beta}$ and $Ca$ have on the seemingly contradictory outcomes and how the spatial resolution of the mesh influences the results.  In the next section, we describe an approach which ensures that both the liquid-solid and liquid-gas interfaces are treated in a methodologically similar way so that the contact line does not present an obstacle to the numerical implementation. This approach does not contain any loose-ends at the contact line and hence does not allow for creativity on the part of the user. After rigorously validating our numerical code we will use it to address the aforementioned issues and provide practical advice for future code development.

\section{New approach}\label{new1}

In the two implementations (A-a) and (A-b) that have been described and which are both frequently used in the FEM literature, at least one of the momentum equations is dropped at the contact line.  Consequently, some of the naturally imposed stress-type boundary conditions and line contributions are dropped.  By now, it is unclear whether these droppings at the contact line will cause the numerical scheme to produce a spurious solution or whether there will be an influence on the speed of convergence of the code towards a correct solution.  To see if this is the case, we develop a new implementation which requires no equations to be dropped at the contact line and thus allows us to examine how dropping equations at the contact line influences accuracy.  To achieve this, we will outline a known, but rarely used, method for incorporating boundary conditions on the liquid-solid interface and show, for the first time, how this gives a perfect framework for incorporating the contact line.

It was shown in \S\ref{weakform} that the stress boundary conditions on the free surface are incorporated naturally into the weak form of the momentum equations.  The kinematic condition (\ref{kin}) is then the extra equation to determine the free surface shape. However, because the solid surface's shape is prescribed, on the liquid-solid interface the normal stress is determined as part of the solution, as opposed to appearing in a boundary condition which would slot into the weak formulation.  Consequently, there appears one less unknown on the liquid-solid interface than on the liquid-gas interface and hence an equation needs to be dropped to enable the essential boundary condition of impermeability of the solid surface equation to be satisfied. In other words, when the surface's shape is prescribed, the normal stress condition can no longer be applied and the kinematic condition plays the role of the second boundary condition for the Navier-Stokes equations. Clearly then, because the free surface and the solid surface are handled in different ways, the contact line where they join will have to be specially treated.

In order to eliminate the asymmetry between the two interfaces, an extra unknown, denoted $\lambda$, is introduced into the scheme; this unknown will be equal to the normal stress at the liquid-solid interface
\begin{equation}\label{lambda}
\lambda=\mathbf{n}_2\cdot\mathbf{P}\cdot\mathbf{n}_2.
\end{equation}
The introduction of $\lambda$ as an extra unknown makes it possible to reinstate the momentum equation normal to the solid surface at all nodes on the liquid-solid interface: all momentum equations will then be applied everywhere.  In other words, on the \emph{free} surface all momentum equations are applied with the impermeability condition used to determine the `extra' unknowns specifying the \emph{position} of the free surface, and now we have that all the momentum equations are also applied on the \emph{solid} surface where the geometric constraint of the prescribed shape allows the impermeability condition to be used to determine the `extra' unknown $\lambda$, i.e.\ the \emph{normal stress}.  As a result, the free surface and the solid boundary appear to have been treated in the same way and hence, once the two boundaries meet at the contact line, the contact line conditions can be implemented naturally, without dropping any of the equations there.

The new unknown is non-dimensionalized using the characteristic scale $\mu U/L$ and then approximated using the velocity interpolating function on the liquid-solid interface, so that
\begin{equation}\label{lambda1}
\lambda = \sum^{N_{2}}_{\textrm{j}=1} \lambda_{\textrm{j}} \phi_{2,\textrm{j}}.
\end{equation}
Then, after splitting the stress into components normal and tangential to the solid
\begin{equation}\label{components}
\mathbf{e}_\alpha\cdot\mathbf{P} = \mathbf{e}_\alpha\cdot\left[
\left(\mathbf{n}_2\cdot\mathbf{P}\cdot\mathbf{n}_2\right)\mathbf{n}_2 +
\mathbf{n}_2\cdot\mathbf{P}\cdot\left(\mathbf{I}-\mathbf{n}_2\mathbf{n}_2\right)\right],
\end{equation}
and applying the Navier boundary condition ({\ref{ss}) as before, the contribution from the liquid-solid interface to
the momentum residuals (\ref{stress}) becomes
\begin{equation}\label{ns55}
\left(R^{M,\alpha}_{\textrm{i}}\right)_{A_{2}} =\int_{A_2}
\phi_{2,\textrm{i}}\left[\lambda~(\mathbf{e}_\alpha\cdot\mathbf{n}_2) +
\bar{\beta}~\mathbf{e}_\alpha\cdot\left(\mathbf{u}_{||}-\mathbf{U}_{||}\right)\right]~dA_2,
\end{equation}
with the first and second terms in the integrand associated with the normal and tangential
stresses, respectively.  Now, on both interfaces all momentum equations are applied to determine the velocity components and the extra kinematic-type equation determines the additional unknowns $\mathbf{x}_1$ or $\lambda$.

The new approach introduces a significant simplification, at the expense of relatively few additional unknowns, when the solid surface is not parallel to a coordinate axis: instead of having to project the momentum equations in the normal and tangential directions to the surface, to allow the impermeability condition to be applied, no additional work is required as all equations are applied everywhere in a coordinate-free manner.  This is particularly advantageous for the solid surface with complex topography. Furthermore, in contrast to implementations of Type A, all momentum equations are applied at the contact line, with contributions to the momentum residuals from each interface, so that the contact line equations no longer require special attention.  In other words, the new approach allows the contact line to naturally fuse the liquid-solid and liquid-gas interfacial equations so that there is no scope for options on how the equations are implemented at the contact line, i.e.\ there are no `loose ends'.  This will be referred to as an implementation of Type B:
\begin{itemize}\label{or3}
  \item[(B)] All momentum equations are applied at the contact line and the contact angle is imposed
  naturally into the weak form using (\ref{fscl}).
\end{itemize}

The realization that this particular implementation of the boundary conditions on the liquid-solid interface allows the contact line to be handled naturally is new; however, the idea of introducing additional unknowns on the liquid-solid interface has been well documented.  Building on the ideas of Babu\u{s}ka \cite{babuska73}, the method was considered by Verf\"{u}rth \cite{verfurth87} for implementing slip conditions, and it was shown that $\lambda$ is the Lagrange multiplier of the impermeability equation (\ref{ss}). This is analogous to the situation with pressure, which is well known to be the Lagrange multiplier of the incompressibility constraint, as, in both cases, modelling assumptions (zero Mach number/rigid solid) result in a variable ($p$/$\lambda$) being determined by an equation in which it does not appear (incompressibility/impermeability).  In \cite{bansch00}, it is shown, for a case in which there are no contact lines, that results obtained from the approach of dropping the normal momentum and replacing it with the impermeability condition converge to the same computational results as the Lagrange multiplier approach. Lagrange multipliers are also used in \cite{berghezan94} to impose both the liquid-liquid free surface and the far-field boundary conditions. In \cite{page97}, for the problem of flow over a backward facing step on which a slip boundary condition is applied, it is shown that the best results are produced when the Lagrange multiplier is approximated with the same basis functions as the velocity, as we have chosen to.

The method described above ensures that contact lines may be treated in a methodologically consistent way with no equations dropped and hence no loose ends that leave room for user intervention and can lead to the myriad of approaches described in \S\ref{previous}. Now we shall investigate the accuracy of the new implementation, and then compare its results to those obtained using previous implementations (A-a) and (A-b).

\section{Numerical experiments: meniscus in a capillary tube}\label{example}

The new implementation, as well as the ones considered in previous works, have been incorporated into the FEM scheme described in the Appendix.  First, results from the new implementation are compared to those obtained from asymptotic analysis. Having confirmed the accuracy of the scheme, a practical guide to the spatial resolution which numerical schemes modelling dynamic wetting phenomena require is given in \S\ref{convergence} after which the previous approaches are examined in \S\ref{exam}.

As a representative dynamic wetting flow, consider the steady propagation of a liquid-gas meniscus through a cylindrical capillary tube.  This flow configuration has the advantage that the geometry is simple and that published results, which agree with each other, exist from spatially well resolved FEM schemes \cite{lowndes80,bazhlekov96}.

Axisymmetric cylindrical polar coordinates $(r,z)$ are used with the $z$-axis along the centre of the tube and pointing towards the gas; azimuthal symmetry about this axis is then assumed
(Figure~\ref{F:sketch}). The characteristic length $L$ and velocity scale $U$ are provided by the radius of the capillary and the speed at which the meniscus propagates, respectively. The solution is computed in a frame in which the contact line is stationary, so that in this frame the solid, which is located at $r=1$, moves with velocity $\mathbf{U}=-\mathbf{e}_z$.

The bulk equations are (\ref{ns}), with the standard free-surface conditions (\ref{kin})--(\ref{fsa}), the Navier-slip and impermeability conditions (\ref{ss}) and new normal stress specification (\ref{lambda}) at the liquid-solid interface, and the contact angle condition (\ref{angle}), with a prescribed velocity-independent value of $\theta$, at the contact line whose location is (1,0). The surface tension of the free surface takes its constant equilibrium value of $\sigma_1 = 1$.  Additional boundary conditions at the axis of symmetry $r=0$ are
\begin{equation}\label{sym}
\mathbf{u}\cdot\mathbf{e}_r = \mathbf{e}_r\cdot\mathbf{P}\cdot\mathbf{e}_z=0,
\end{equation}
and the condition of smoothness on the free surface shape at $r=0$ that
$\mathbf{n}_1\cdot\mathbf{e}_r = 0$. Boundary conditions in the truncated far field at $z=-z_f$ are
provided from the assumption that the flow is fully developed
\begin{equation}\label{far}
u=0,\qquad w = a r^{2} + b;\quad a = -2/(4\bar{\beta}^{-1} +1),\quad b = -a/2;
\end{equation}
where $(u,w)$ are the components of velocity in the axisymmetric coordinate system, and if we have no-slip $\bar{\beta}^{-1}=0$ the usual Poiseuille profile is recovered.  For a similar problem, in \cite{bazhlekov96} it is noted that extending the truncated far-field further than two and a half radii away from the contact line did not alter their results.  As the main interest here lies in the computational aspects near the contact line anyway, the truncated far-field is placed at $z_f = 3$ and the issue is not considered further.  The implementation of the boundary conditions (\ref{sym}) and (\ref{far}) into the finite element procedure is carried out in the conventional way with the stress conditions incorporated naturally into the weak formulation and the Dirichlet conditions applied instead of momentum equations as essential conditions.  New variables could have been introduced on these boundaries to ensure the momentum equations are always applied, but the implementation of the conditions on these boundaries into the FEM is not the focus of this paper.
\begin{figure}
\centering
\includegraphics[scale=0.70]{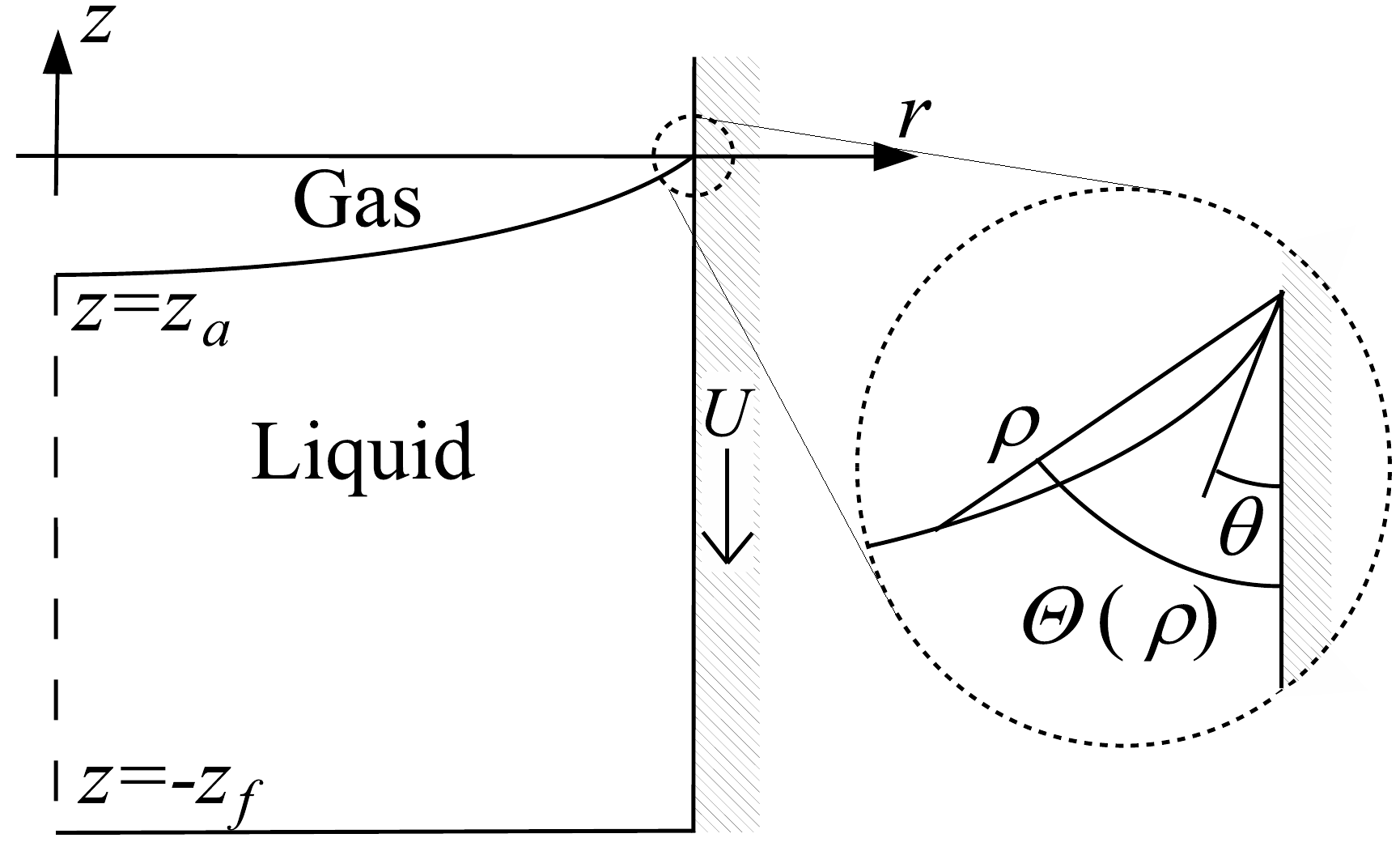}
\caption{The computational domain for a liquid-gas meniscus propagating through a capillary tube with a close-up of the contact line region showing the coordinate system used in the derivation of the local asymptotic results.} \label{F:sketch}
\end{figure}

\subsection{Parameter regime of interest}

To obtain estimates on the values of the similarity parameters, consider the displacement of air by water with viscosity $\mu=1$ mPa~s and density $\rho=10^{3}$ kg~m$^{-3}$ in a capillary of radius $100$ $\mu$m. The water-air meniscus advances at a speed $U=1$ m~s$^{-1}$ and has a surface tension of $\sigma_{1e} = 70$ mN~m$^{-1}$. Finally, an estimate for $\beta$ often used in the literature for Newtonian fluids \cite{lowndes80,bazhlekov96,blake02}, is that $\beta\sim 10^{9}\mu$. This gives that $Re~\approx 100$, $Ca\approx 10^{-2}$, $\bar{\beta}\approx10^{5}$ and external forces are neglected so that $St=0$.  For simplicity and transparency, we consider acute contact angles to avoid the spurious numerical behaviour encountered, and then resolved, in \cite{sprittles10b} for obtuse angles.  In particular,  $\theta=30^{\circ}$ is taken as a base state.

Initially, we shall consider Stokes flow, i.e.\ $Re=0$, to allow for a comparison with asymptotic results.  After confirming the accuracy of our approach, we consider more
numerically challenging flows where the Reynolds number is non-zero and the capillary number is
such that the free-surface becomes heavily deformed.

\subsection{Comparison to local asymptotics}\label{comp_loc}

To confirm the accuracy of our numerical results it will be useful to compare them to local asymptotics
derived in the limit $\rho\rightarrow 0$ where $(\rho,\Theta)$ are polar coordinates centred at
the contact line with the angular coordinate measured from the solid surface through the liquid (Figure~\ref{F:sketch}).  Such an
analysis can be found, for example, in \cite{shik06}. The leading-order term of the streamfunction, introduced using
\begin{equation}
u_\rho=\frac{1}{\rho}\pdiff{\psi}{\Theta},\qquad u_\Theta=-\pdiff{\psi}{\rho},
\end{equation}
is given by
\begin{equation}
\label{stream_2} \psi = \rho^{2}F(\Theta),\qquad F(\Theta) =
\left(B_{1}+B_{2}\Theta+B_{3}\sin2\Theta+B_{4}\cos2\Theta\right),
\end{equation}
where the constants of integration $B_i~(i=1\hbox{--}4)$ are: $B_1=-B_4=-\bar{\beta}/4$, $B_2=-B_1/\theta$, $B_3=B_1\cot2\theta$. For comparison with our numerical results, we will use the radial velocity $u_\rho$ and the normal stress $\lambda$ on the solid surface $\Theta=0$, which are given by
\begin{equation}\label{u_asymp}
u_\rho = \rho F'(\Theta=0);\qquad \lambda = -4B_2\ln \rho - 2(B_2+2B_3) + p_0,
\end{equation}
where $p_0$ is a constant of integration. Later, the free-surface shape will also be needed, which is parameterized (Figure~\ref{F:sketch}) as $\Theta = \Theta(\rho)$, and locally has the form
\begin{equation}\label{fs_asymp}
\Theta = \theta + Ca\bar{\beta}\left[A\rho\ln \rho + B\rho\right] + o(\rho),
\end{equation}
where $A,B$ are $O(1)$ functions of $\theta$.  Expressions in (\ref{u_asymp}) will be sufficient to validate the numerical results.  Note that to leading order the domain is wedge-shaped so that the radial velocity $u_\rho$ should closely approximate the velocity tangential to the free surface $u_s$ from the computed solution, which will be used for comparison.

Consider typical parameter values $Ca=0.01,~\bar{\beta}=10^{5},~\theta=30^{\circ}$. Figure~\ref{F:cap} shows the corresponding
streamlines and the mesh which was used for the computation.  It can be seen that the motion of the
solid causes fluid nearby to be dragged down and consequently, to ensure continuity of mass, this
is replenished by a flow of fluid up the centre of the capillary. The free surface is relatively
undeformed and appears to be close to its static shape, i.e.\ a spherical cap which meets the solid at
the given contact angle. As one zooms in on the contact line region, it can be seen in
Figure~\ref{F:cap_zoom} that the domain is approximately a wedge of angle $\theta=30^\circ$. It is also
apparent that the FEM mesh has significant spatial resolution, so that, the dynamics of
slip, on the (non-dimensional) length scale of $\bar{\beta}^{-1}=10^{-5}$, is well resolved as the
smallest element has size $l_{min} = O(10^{-9})$.
\begin{figure}
\centering
\includegraphics[scale=0.50]{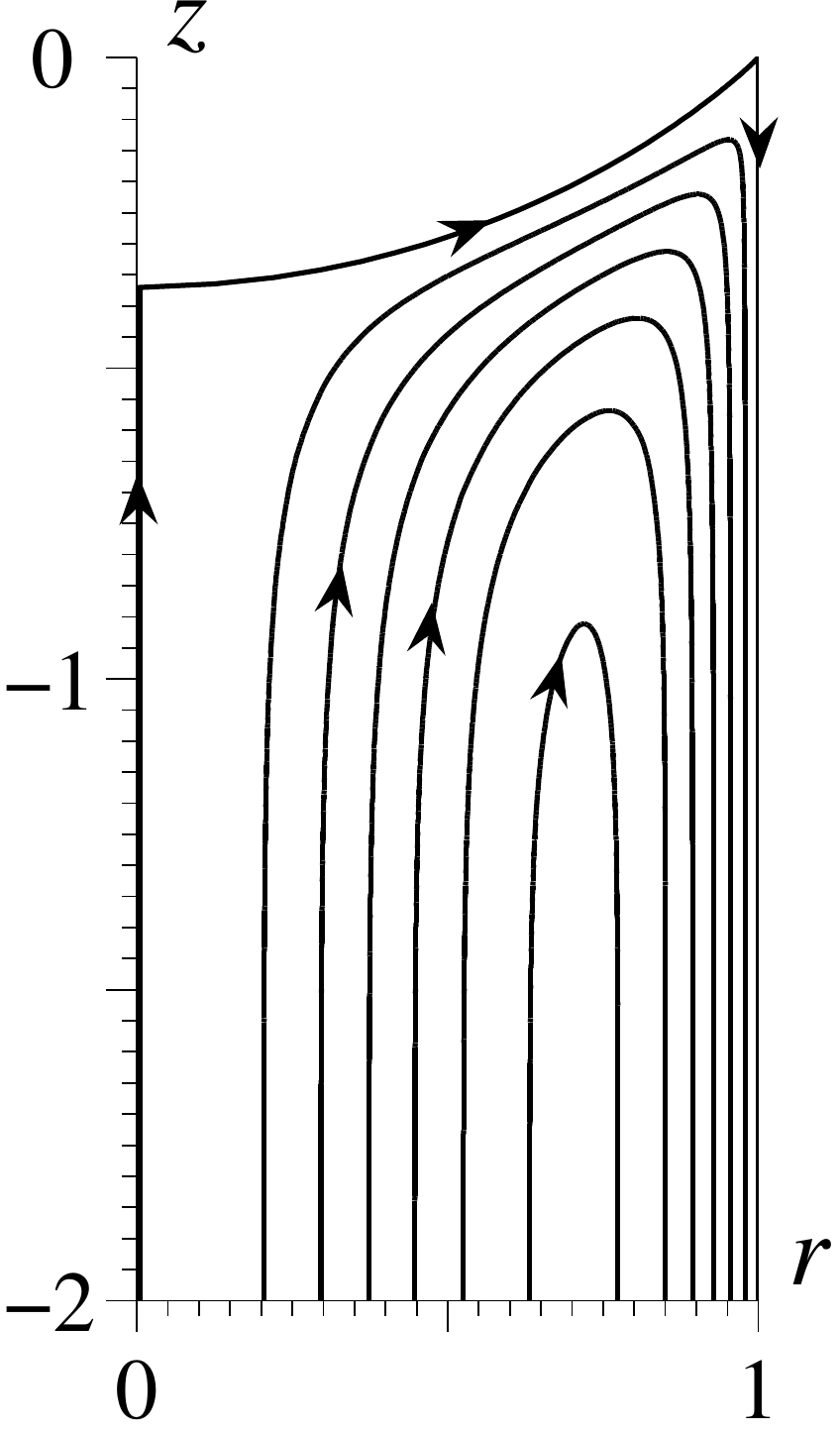}\hspace{3cm}
\includegraphics[scale=0.50]{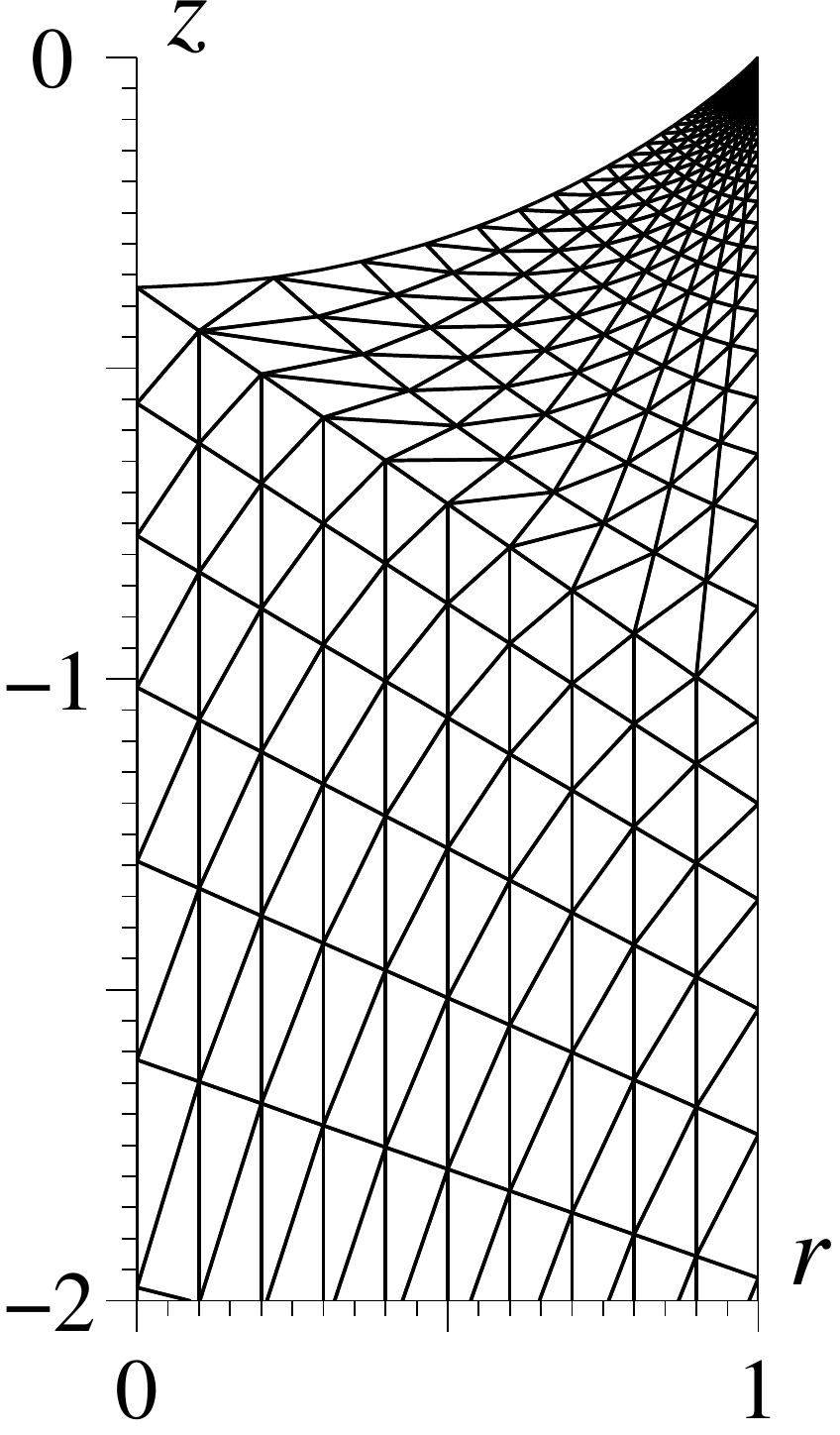}
\caption{Left: streamlines in increments of $\psi=-0.02$ computed for
$Ca=0.01,\bar{\beta}=10^{5},\theta=30^{\circ}$.  Right: corresponding mesh used in this region.}
\label{F:cap}
\end{figure}
\begin{figure}
\centering
\includegraphics[scale=0.380]{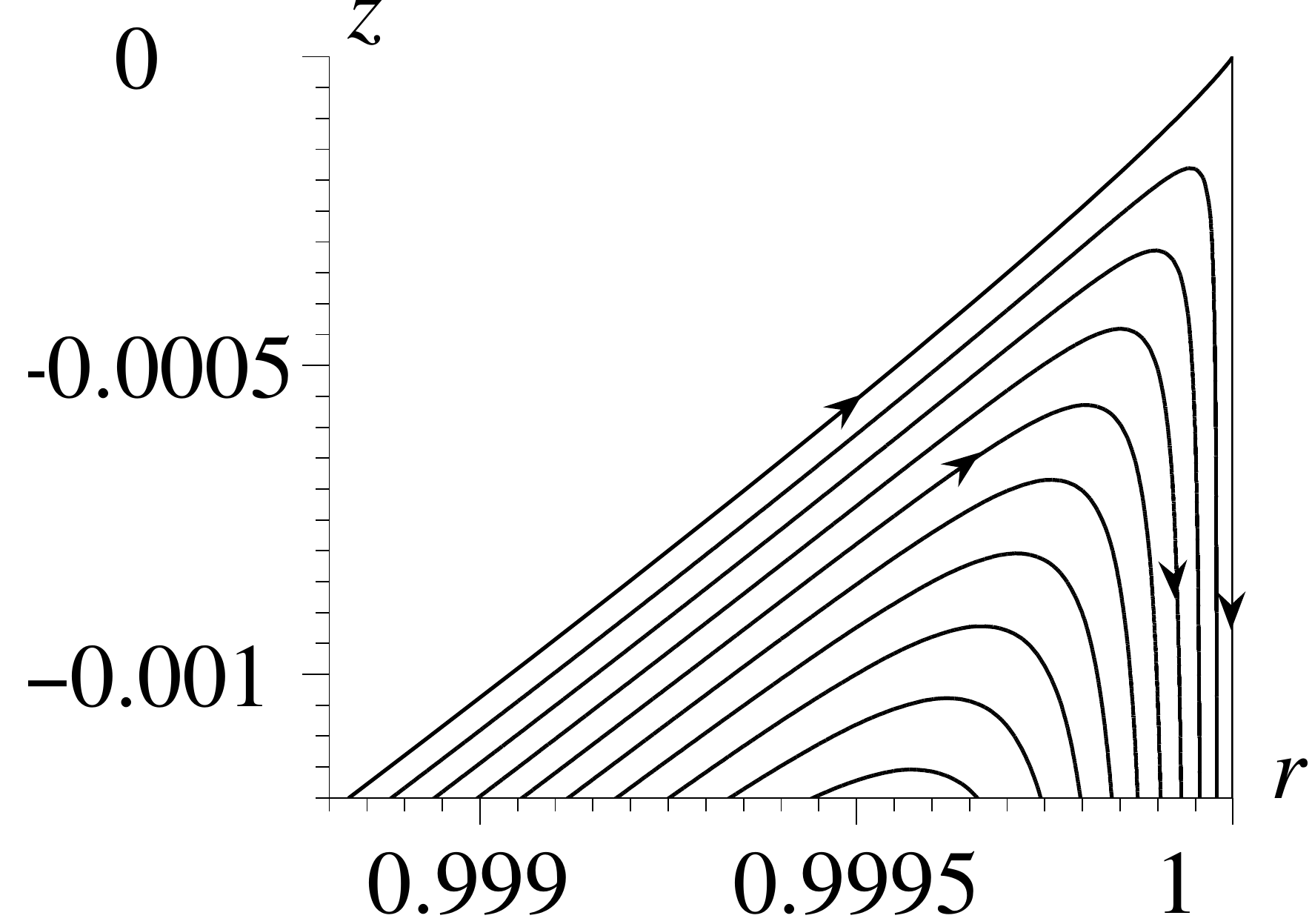}\hspace{2cm}
\includegraphics[scale=0.380]{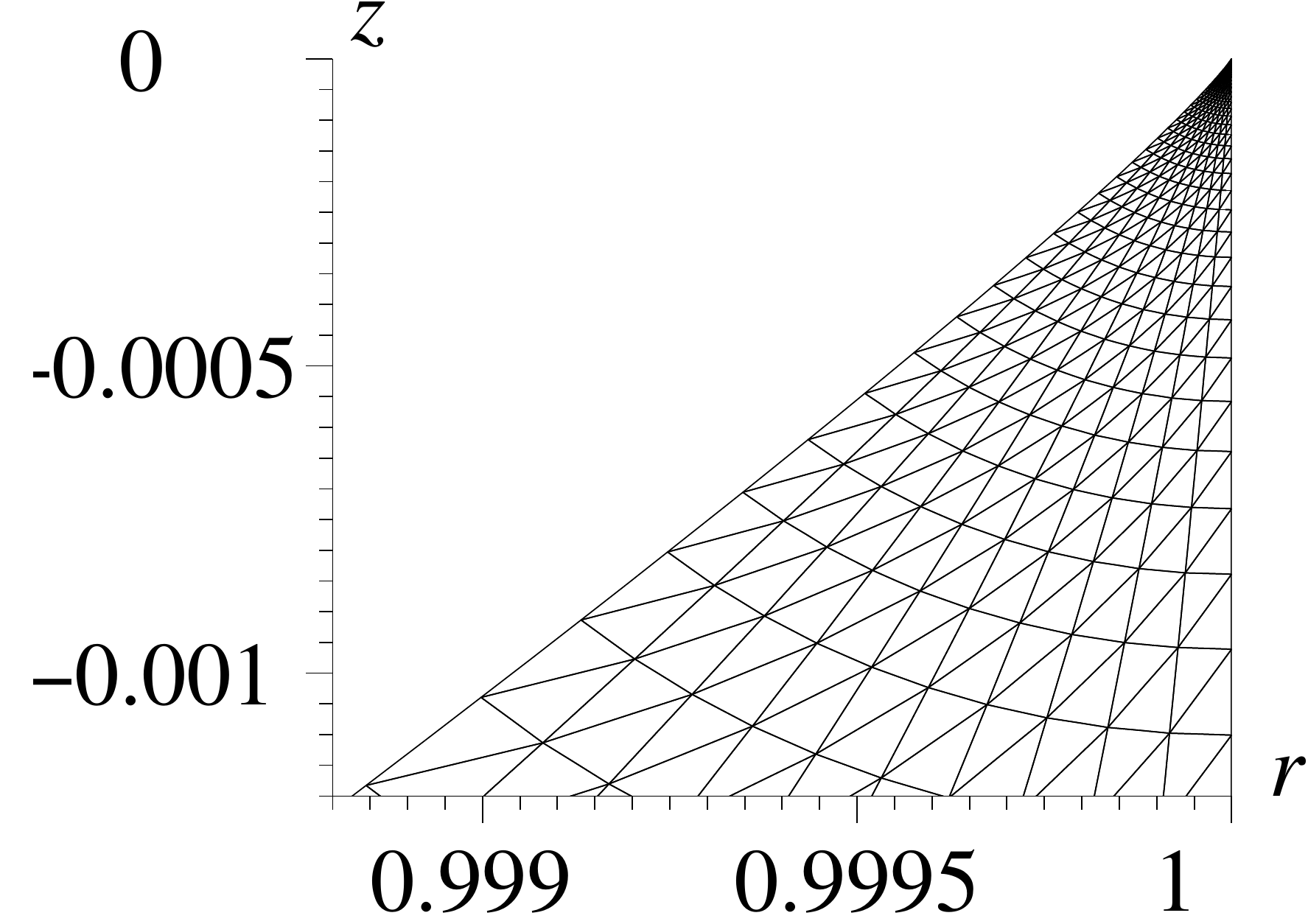}
\caption{Left: streamlines near the contact line in increments of $\psi=-2\times 10^{-5}$ computed
for $Ca=0.01,\bar{\beta}=10^{5},\theta=30^{\circ}$.  Right: corresponding mesh used in this
region showing a high degree of spatial resolution.} \label{F:cap_zoom}
\end{figure}

For the same parameter values, in Figure~\ref{F:u_asymp} a comparison of the computed velocity $u_s$ tangential to the liquid-solid and liquid-gas free surface with the local asymptotic result (\ref{u_asymp}) is shown. Although the computed free surface is curved, it is almost flat on the scale considered and hence $u_\rho$, the radial velocity calculated in the asymptotic analysis, should agree well with $u_s$. In Figure~\ref{F:u_asymp}, one can observe that the agreement between numerical and asymptotic results is good, with both velocities observed to be linear as the contact line is approached. Agreement between the computed and the asymptotic results is seen to occur on both surfaces for a distance from the contact line of $s<\bar{\beta}^{-1}=10^{-5}$, i.e.\ inside the region of slip where the velocity starts to increase from its value at the contact line $u_s=0$ up to its far field value.
\begin{figure}
\centering
\includegraphics[scale=0.40]{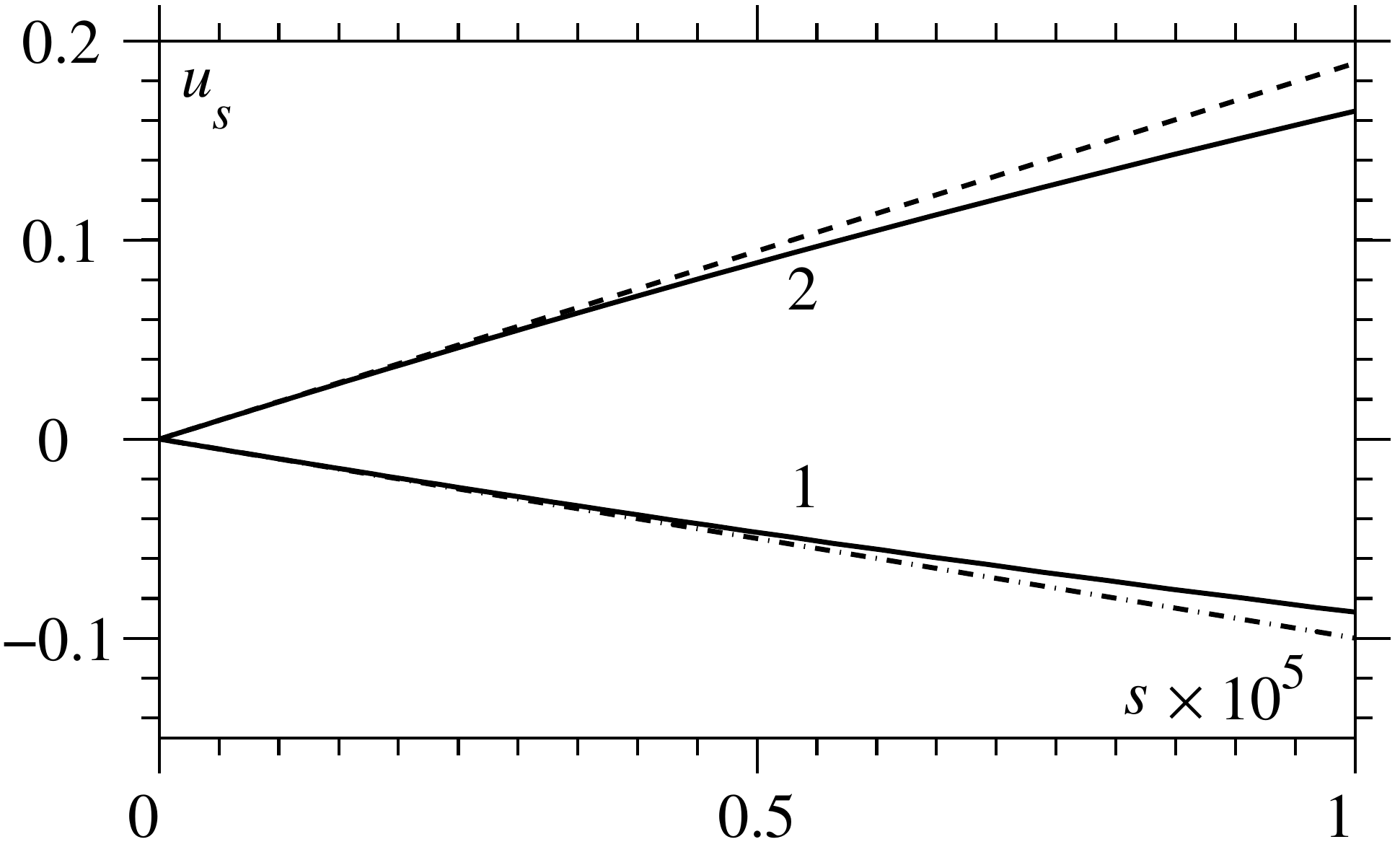}\hspace{5 mm}
\includegraphics[scale=0.40]{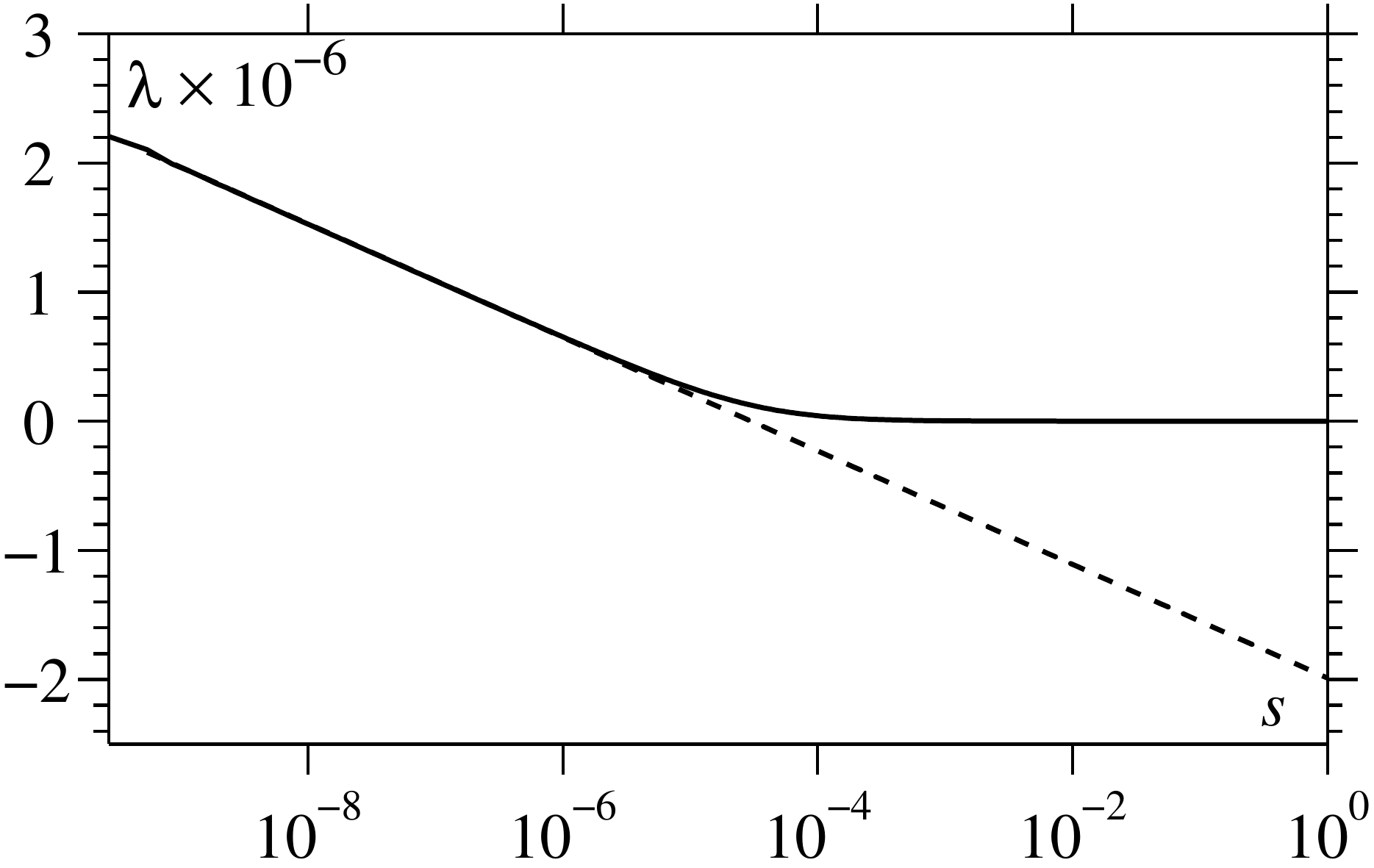}
\caption{Left: comparison of the velocity $u_s$ tangential to the free surface, curve~$1$, and solid surface, curve~$2$, with the local asymptotic result (dashed lines) from  (\ref{u_asymp}). The arclength along the interface from the contact line is $s$. Right: comparison of the normal stress along the liquid-solid interface $\lambda$ with the local asymptotic result (dashed line) from (\ref{u_asymp}). The pressure constant $p_0$ in (\ref{u_asymp}) has been chosen so that the numerical and asymptotic result coincide at the node
adjacent to the contact line.}\label{F:u_asymp}
\end{figure}

The computed variable representing the normal stress $\lambda$ on the liquid-solid interface is seen in Figure~\ref{F:u_asymp} to agree very well with the asymptotic result (\ref{u_asymp}); in particular, the normal stress is logarithmically singular as the contact line is approached.  The agreement confirms that introducing this variable and incorporating it naturally into the weak form of the momentum equations has the desired effect.  In particular, the level of approximation of this variable is sufficient.

Thus, a comparison of our numerical results with local asymptotic analysis has shown that our code provides an accurate description of the flow near the contact line. Next, we study the convergence of our scheme as the mesh is refined.

\subsection{Convergence of the new implementation}\label{convergence}

We now consider whether the new implementation, in which the contact angle is imposed naturally into the weak form of the equations, ensures that $\theta_c$ is equal to $\theta$. In particular, high capillary number flow is considered as this is cited as the most problematic parameter regime in previous publications.

In Figure~\ref{F:ca_com}, the influence that the capillary number has on the free surface shape is shown.  For $Ca \le 10^{-4}$ the free surface shapes are graphically indistinguishable and represent the spherical cap one would expect a static meniscus to form.  As one increases the capillary number, the apex of the free surface moves from below ($z_a<0$) the contact line height to above it ($z_a>0$). On the scale of the whole capillary radius the free surface does not seem heavily deformed; however, zooming in on the contact line region (Figure~\ref{F:ca_com}) reveals that the free surface actually makes an angle $\theta=30^{\circ}$ in all of these cases.  Therefore, at larger capillary numbers there is a small region of large curvature near the contact line.
\begin{figure}
\centering
\includegraphics[scale=0.420]{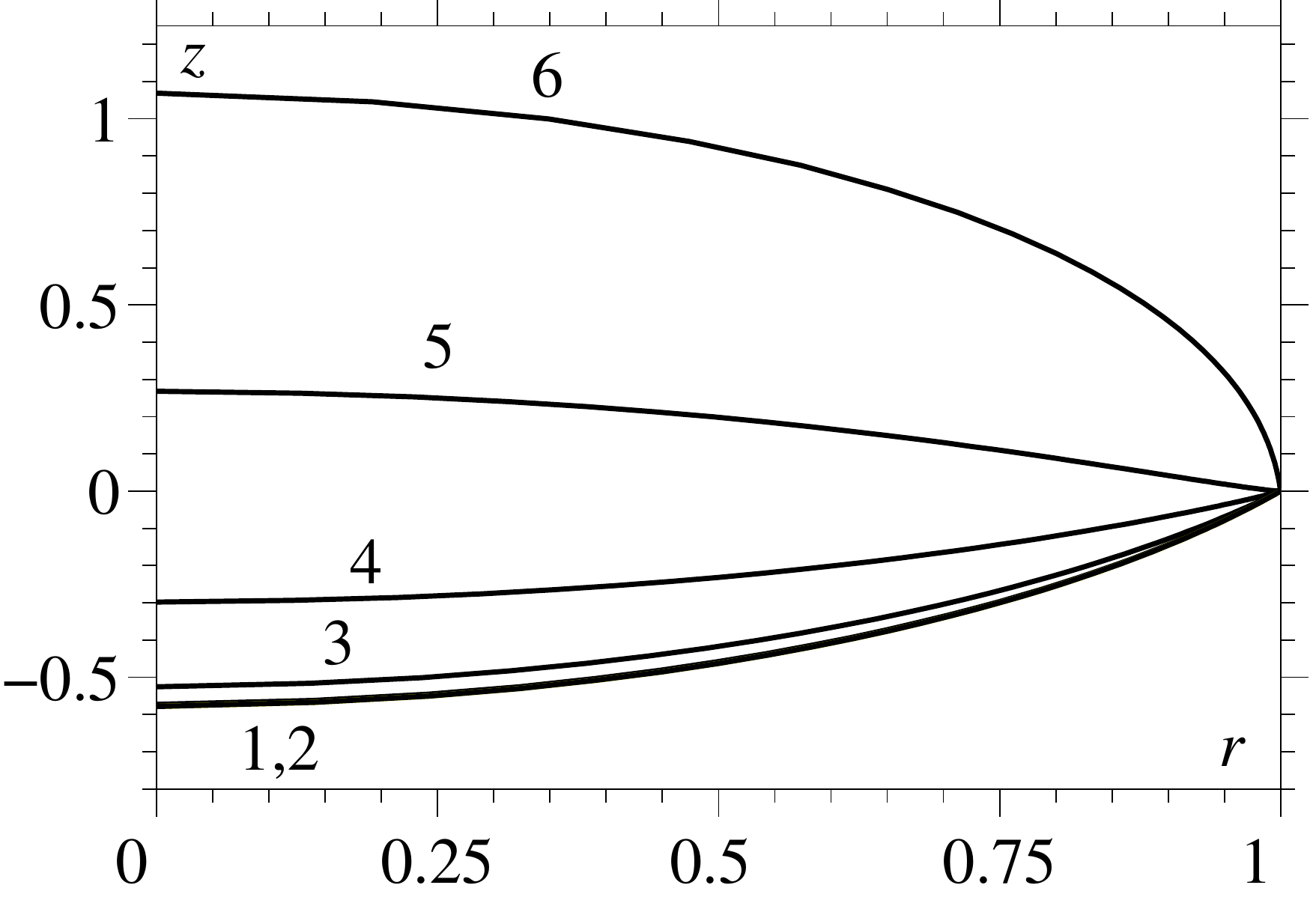}\hspace{5 mm}
\includegraphics[scale=0.370]{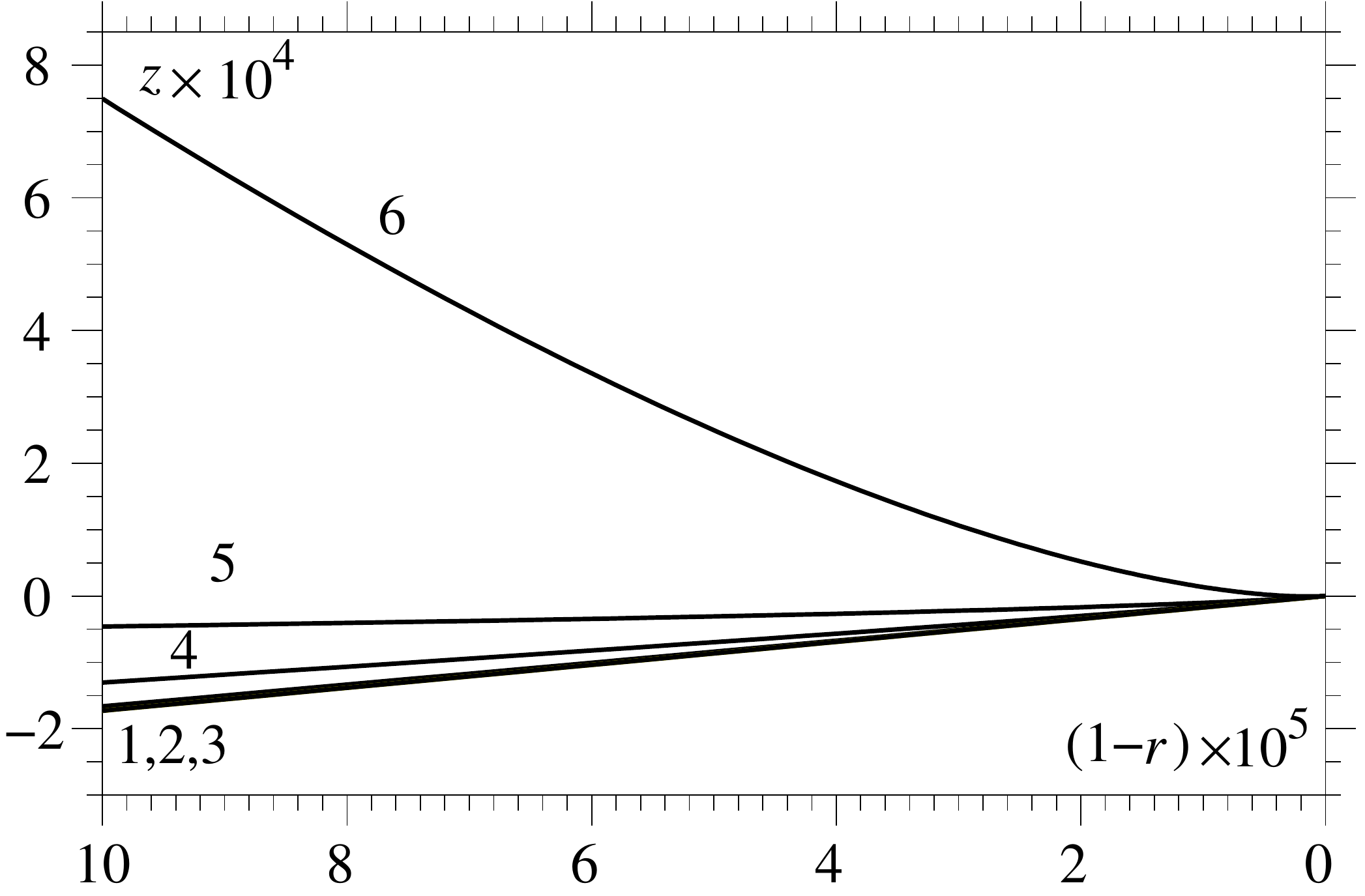}
\caption{Computed free surface shape on the scale of the capillary (left) and close to the contact line (right) as the capillary number is varied with
$Re=10,~\theta=30^\circ,~ \bar{\beta}=10^5$ fixed. Curves $1-6$ are for capillary numbers
$Ca=10^{-5},10^{-4},10^{-3},10^{-2},10^{-1},1$, respectively.} \label{F:ca_com}
\end{figure}

Each curve in Figure~\ref{F:ca_com} was obtained using the new implementation and in each case the computed contact angle $\theta_c$ was within $0.01^{\circ}$ of $\theta$ that was imposed.  It appears that this can always be achieved when the mesh has sufficiently small elements near the contact line.  To quantify this claim, we study the computed contact angle and, as a measure of the entire free surface shape, the difference in height $H$ between the contact line $z=0$ and the apex $z=z_a$ for a range of meshes of increasing resolution, characterized by the smallest element size $l_{min}$, for $Ca=0.1$. Note that the spatial resolution in the bulk of the mesh is the same in all of these simulations, it is only the resolution near the contact line which is altered.  The results are presented in the table below.
\begin{center}
\begin{tabular}[c]{|c|c|c|c|c|}
  \hline
  $l_{min}\times 10^{8}$ & Computed Angle ($\theta_c$)  & $H=z_a-z_c$ & Error in H (\%) \\
  \hline
  32000   & 66.666 & 0.2318781   &  4.9 \\
  11000   & 32.567 & 0.2738979   &  2.6$\times 10^{-3}$ \\
  400     & 30.060 & 0.2764014   &  1.6$\times 10^{-5}$ \\
  1       & 30.001 & 0.2764592   &  0       \\
  \hline
\end{tabular}
\end{center}    


With increasing spatial resolution, the computed angle $\theta_c$ converges to the applied one $\theta=30^{\circ}$ imposed via the weak formulation.  For coarser meshes than those considered above, $\theta_c$ exceeded $180^\circ$ so that the free surface passes (unphysically) through the solid.  It seems likely that such results are the reason that the contact angle is so often imposed as an essential condition, as in implementation (A-b). Critically, we see that the vertical distance $H$ between the contact line and apex position is influenced by the resolution close to the contact line. So, errors generated by a poor approximation of the dynamics \emph{near} the contact line greatly influence a measure of the \emph{entire} free surface shape.

We have shown that our results converge as the mesh size is reduced: using a specially designed mesh (i.e.\ high resolution with the appropriate increment of the element size) the computed contact angle $\theta_c$ can be made arbitrarily close to the applied one $\theta$, i.e. to the one featuring in the mathematical formulation of the problem.

In the next section, we will examine the predictions of the numerical schemes previously used in the literature and will see that the conclusions from above also apply for implementation (A-a).  What remains to be checked is whether by using implementation (A-b), in which the contact angle is imposed as an essential condition, it is possible to obtain accurate results at a lower spatial resolution, i.e.\ at less computational cost, than considered above.  This is the implicit assumption of numerical schemes which do not attempt to resolve the dynamics of slip and use (A-b) to ensure that the computed contact angle is always equal to the applied value.

\subsection{Examination of implementations (A-a) and (A-b)}\label{exam}

The same problem is now solved using the implementations previously considered in the literature. We found that implementation (A-a), with the one momentum equation applied at the contact line which is parallel to the solid, gives results which, for the accuracy considered in this paper, are indistinguishable from those obtained using the new implementation (B).  This means that for the particular flow configuration considered, the effect of reinstating the momentum equation normal to the solid surface is only to determine the function $\lambda$, i.e.\ no additional accuracy is gained. Implementation (B) remains advantageous over (A-a), since there is no room for user `input' at the contact line and the surfaces that are not aligned to coordinate axes require no additional work.  It may be the case that the agreement between the results of the two implementations is due to the particular type of flow considered here with, for example, no component of velocity normal to the solid.  More complex flows in which there is a flux of mass into the solid, such as in the wetting of a porous medium, or in which the solid surface is curved may distinguish between the accuracy of the two methods.  However, for the purposes of the present paper, the results from  (A-a) and (B) may be presented together.

Our focus is now on comparing the results from implementations (A-a) and (B) with those obtained using (A-b), where the contact angle is imposed as an essential condition, which is the most commonly used implementation in the literature. As shown in the table below, there is a significant difference in the results which these two methods produce.
\begin{center}
\begin{tabular}{|c|c|c|c|c|}
  \hline
 \multirow{2}{*}{$l_{min}\times 10^{8}$} &    \multicolumn{2}{|c|}{$H=z_a-z_c$} &   \multicolumn{2}{|c|}{Error in H (\%)} \\
 \cline{2-5}
                         &   (A-b) &   (A-a) \& (B)             & (A-b) &  (A-a) \& (B)  \\
  \hline
  940000    & 0.0945618 & N/A         & 65.8   & N/A \\
  32000     & 0.2318781 & 0.2901297  & 16.1   & 4.9 \\
  11000     & 0.2738979 & 0.2764526  & 0.93   & $2.6\times 10^{-3}$ \\
  400       & 0.2764014 & 0.2764598  & 0.02   & $1.6\times 10^{-5}$ \\
  1         & 0.2764592 & 0.2764597  & $1.9\times 10^{-4}$ & 0       \\
  \hline
\end{tabular}
\end{center}

By design, in implementation (A-b) the computed angle is equal to $\theta$ for all meshes and, as hoped, this allows a solution to be found for coarser meshes.  Furthermore, the value of $H$ converges to the same value in all implementations, i.e.\ for a sufficiently refined mesh all implementations considered give the same result. However, imposing the contact angle in the weak form, as in approaches (A-a) and (B), is seen to create a formulation which converges to the correct solution significantly faster than approach (A-b).
For example, for the third mesh, the percentage error in $H$ using implementation (A-b) is more than a hundred times greater than that obtained from (A-a) and (B). This is despite the computed angle for (A-a) and (B) being more than $2^{\circ}$ away from the imposed one, as shown in the first table, whereas (A-b) ensures equality between $\theta$ and $\theta_c$.

For the coarsest mesh considered, implementation (A-b) allows a solution to be obtained whose free surface is wildly different from the converged solution. This is despite the smallest element still being over one hundred times smaller than the radius of the capillary, i.e.\ being what one could easily interpret as `relatively small'.  This serves as a representative example of the dangers of computing a solution on a mesh which is not specially designed to simulate dynamic wetting phenomena and then employing implementation (A-b) to `compensate' for this and ensure a `solution' is produced.

As previously discussed, when the computed contact angle does not equal the one imposed in the weak form, it is tempting to impose the contact angle as an essential condition. This is because it is very easy to observe that the computed angle does not equal the applied one, more easy than, say, observing that the stress conditions on the free surface are not well satisfied near the contact line.  However, what we have found is that the computed angle not equalling the imposed one should be interpreted by the user as a warning sign: the mesh is under-resolved. As seen from our results, attempting to bypass this problem by imposing the contact angle as an essential condition actually \emph{increases} the global error of the scheme.

Qualitatively, our conclusions are similar to the general remark made in the famous paper of Gresho \& Lee \cite{gresho81}, ``Don't suppress the wiggles - They're telling you something'', where it is warned that artificially suppressing spurious numerical artifacts merely shifts the generated error to another, less observable, part of the computed solution. We see, that when the contact angle is imposed naturally via the weak formulation, the deviation of $\theta_c$ from $\theta$ actually provides a simple quantitative error control on the numerical scheme.  The error should not be suppressed by enforcing the angle as an essential condition, as this only exasperates errors in less easily observable parts of the scheme. One could insist that the computed angle is, say, within $0.1^\circ$ of the applied value and, from the results obtained above, in the parameter range considered, this would provide a very accurate numerical scheme. This is an idea which we now examine in order to provide practical quantitative recommendations for the spatial resolution required at given values of $Ca$ and $\bar{\beta}$.





%


\subsection{A practical guide to mesh design}

Using the aforementioned criterion that $\theta_c$ is within a chosen distance from $\theta$, we will now determine the required spatial resolution for given parameter values.  At lower capillary numbers, roughly $Ca<10^{-2}$, it is observed that the main constraint on a mesh-independent solution comes from the need to resolve the velocity field which varies on the (non-dimensional) length $\bar{\beta}^{-1}$ near the contact line.  The computed angle is seen to be within $0.1^{\circ}$ of the applied value before this spatial resolution is achieved.

At higher capillary numbers, where the interfacial curvature is high close to the contact line, the required spatial resolution must be determined.  To quantify the results of the numerical experiments, we determine the size of the smallest element $l_{min}$ that will ensure the computed contact angle is within $0.1^\circ$ of the imposed one, in this case $\theta=30^\circ$.  The results are summarized in the table below.
\begin{center}
\begin{tabular}{|c|c||c|c|c|}
  \cline{3-5}
     \multicolumn{1}{c}{} & \multicolumn{1}{c}{} &  \multicolumn{3}{|c|}{ $\bar{\beta}$ } \\
   \cline{3-5}
      \multicolumn{1}{c}{}   & \multicolumn{1}{c|}{} & $10^{4}$ & $10^{5}$ & $10^{6}$ \\
  \cline{1-2}
  \hhline{|~|~||=|=|=|}
  \multirow{3}{*}{$Ca$} &  $10^{-2}$ & $6\times 10^{-5}$ & $6\times 10^{-6}$ & $4\times 10^{-7}$ \\
  \cline{2-5}
  & $10^{-1}$ & $6\times 10^{-6}$ & $5\times 10^{-7}$ & $5\times 10^{-8}$ \\
  \cline{2-5}
  & $1$       & $8\times 10^{-7}$ & $7\times 10^{-8}$ & $8\times 10^{-9}$ \\
  \cline{1-5}
\end{tabular}
\end{center}

A general trend can be seen that for higher $Ca$ and $\bar{\beta}$ it is required that $l_{min}$ is smaller, i.e.\ more spatial resolution is needed. Surprisingly, in this range of
values, a simple empirical formula provides a good estimate for $l_{min}$: $l_{min} =5\times 10^{-3}Ca^{-1}\bar{\beta}^{-1}$. It is important to remember that this is only applicable for large values of $Ca$, whereas for small $Ca$ we still need to resolve the velocity field and this guarantees the closeness of $\theta_c$ to $\theta$.   In practice, to ensure that the velocity field is resolved, we require that at least $l_{min}<\bar{\beta}^{-1}$ so that, finally, one needs
\begin{equation}
l_{min} \le \bar{\beta}^{-1}\min\left(5\times 10^{-3}Ca^{-1},1\right).
\end{equation}

The linear dependence of $l_{min}$ on $Ca^{-1}\bar{\beta}^{-1}$ can be rationalized by recalling that the local asymptotic result (\ref{fs_asymp}) predicted that the deviation of the
free-surface from wedge-shaped ($\Theta=\theta$) near the contact line is proportional to $Ca\bar{\beta}$.  Thus, it is this product that determines the length scale on which the curvature of the free surface becomes significant and consequently on what scale the smallest element needs to be to ensure accurate results. We also considered the smallest element required for a $1^\circ$ error in the computed contact angle and found that increasing the coefficient of proportionality by a factor of ten to $5\times 10^{-2}$ gave a good estimate for $l_{min}$.

The above results can be used as a practical guide to the spatial resolution required to ensure accurate simulation of dynamic wetting flows. It is apparent that very small elements are required near the contact line which can only be obtained in a computationally tractable way by using a specially designed mesh. As we have seen, without this special attention to the contact line region the error in the computed solution can be huge. This will apply not only to the FEM but also to other numerical methods which, with a few exceptions (e.g.\ \cite{zhou90}), rarely incorporate the spatial resolution shown to be imperative for the recovery of meaningful results.

In this section, we have fully resolved all issues surrounding the numerical implementation of a contact line at which an arbitrary angle is imposed.  We now consider if the contact angle can be specified on the basis of physical considerations and how the resulting equations fit most naturally into the FEM framework which has been developed.

\section{FEM implementation of interfacial physics}\label{surf}

The contact angle is a boundary condition for the normal stress equation on the free surface, and, thus far, it has been considered as a prescribed input to the model.  Now we will look into the `extra' physics which actually determines its value, and examine the resulting effects on the flow.   The contact angle is introduced into macroscopic fluid mechanics by the Young equation \cite{young05}, which expresses the balance of the surface tension forces on the contact line acting tangentially to the solid surface:
\begin{equation}\label{young}
\sigma_1\cos\theta_d + \sigma_2 = 0.
\end{equation}
Here, $\sigma_1$ and $\sigma_2$ are the surface tensions acting on the contact line from the free surface and the liquid-solid interface, respectively, with the surface tension on the gas-solid interface being zero.  Thus, the only way in which the contact angle $\theta_d$ can vary, i.e.\ be dynamic in the model and not as an artifact of numerics, is if at least one of the surface tensions varies from its equilibrium value $\sigma_{1e}$ and $\sigma_{2e}$, respectively. If the surface tensions become variables of the problem, this must be accounted for in the boundary conditions on the interfaces. Here, we allow for a dynamic surface tension and, being interested in the numerical implementation of the boundary conditions and for the sake of  generality, do not concern ourselves with the mechanism for its variation (see \cite{shik07} for such a mechanism). As it turns out, the equations on the free surface, and their implementation into the FEM scheme, already allow for gradients in surface tension. It is the equations on the liquid-solid interface where the influence of gradients in surface tension is not yet clear.

We will now derive a generalized Navier condition by considering a balance of forces on an interface and an equation relating the torque acting on the interface to the velocity difference, i.e.\ the slip, across it. A balance of stresses acting on the liquid-solid interface is
\begin{equation}\label{lag_stress}
Ca~\mathbf{n}_2\cdot \left(\mathbf{P}^{+}-\mathbf{P}^{-}\right) +
\nabla^s\cdot\left(\sigma^s_{2}(\mathbf{I}-\mathbf{n}_2\mathbf{n}_2)\right)=\mathbf{0},
\end{equation}
where superscript $+$ and $-$ now denote, respectively, the value of a variable on the liquid-facing side of an interface, which is the value involved in the boundary to the Navier-Stokes equations (these variables have been referred to throughout the paper), and on the other side, in this case the solid-facing side of the interface\footnote{The same equation applies on the liquid-gas interface which, when combined with a condition on the gas-facing side of the interface $\mathbf{n}\cdot\mathbf{P}^{-}=\mathbf{0}$, yielded (\ref{fs}).}. The stress on the solid-facing side of the liquid-solid interface is unknown and, if needed, is determined as part of the solution.  Instead, the velocity on the solid-facing side of the liquid-solid interface $\mathbf{U}$ is known.  A link between the two is obtained by relating the velocity difference across the interface to the difference in tangential stress acting on the interface, i.e.\ the torque on the interface, which gives
\begin{equation}\label{torque}
\frac{1}{2}\mathbf{n}_2\cdot\left(\mathbf{P}^{+}+\mathbf{P}^{-}\right)\cdot\left(\mathbf{I}-\mathbf{n}_2\mathbf{n}_2\right)
= \bar{\beta}\left(\mathbf{u}^{+}_{||} - \mathbf{U}_{||}\right).
\end{equation}
This condition is considered, for example, in \cite{barenblatt63} and later derived thermodynamically in \cite{shik07}. Upon elimination of the stress on the solid-facing side of the interface $\mathbf{n}_2\cdot\mathbf{P}^{-}$ from (\ref{torque}) using (\ref{lag_stress}), we arrive at the so-called generalized Navier condition \cite[p.198]{shik07}, which has the form
\begin{equation}\label{gennav}
\mathbf{n}_2\cdot\mathbf{P}^{+}\cdot\left(\mathbf{I}-\mathbf{n}_2\mathbf{n}_2\right)
+\frac{1}{2Ca}\nabla^s\sigma_2= \bar{\beta}\left(\mathbf{u}^{+}_{||} - \mathbf{U}_{||}\right),
\end{equation}
and can see that, when the surface tension is a constant, this generalized condition reduces to the standard Navier condition (\ref{ss}).

Using (\ref{gennav}), we now have the contribution to the momentum residuals (\ref{stress}) on the liquid-solid interface given by
\begin{equation}\label{ns555}
\left(R^{M,\alpha}_{\textrm{i}}\right)_{A_{2}} =\int_{A_2}
\phi_{2,\textrm{i}}\left[\lambda~(\mathbf{e}_\alpha\cdot\mathbf{n}) +
\bar{\beta}~\mathbf{e}_\alpha\cdot\left(\mathbf{u}_{||}-\mathbf{U}_{||}\right) - \frac{1}{2Ca}\mathbf{e}_\alpha\cdot\nabla^s\sigma_2   \right]~dA_2.
\end{equation}
This new formulation has been implemented into our numerical scheme.


\subsection{Verification of the method}

The numerical scheme containing the additional physics has been thoroughly tested and gives excellent results.  To demonstrate this, let us examine the same axisymmetric flow in a capillary as considered in \S\ref{comp_loc}, taking $Ca=0.1,~Re=10,~\bar{\beta}=10^5$ and prescribing the surface tension on the interfaces to be
\begin{equation}\label{st_ss}
\sigma_1 = \sigma_{1e}=1,\qquad \sigma_2 = \sigma_{2e} + \frac{1}{2}\exp(\bar{\beta}z).
\end{equation}
Here, $\sigma_{2e}=-\sqrt{3}/2$ is taken to be the value of the equilibrium surface tension on the liquid-solid interface, which is the asymptotic limit of $\sigma_2$ in the far field (as $z\rightarrow-\infty$) and ensures that in equilibrium the Young equation (\ref{young}) gives that $\cos\theta_e = -\sigma_{2e}/\sigma_{1e}=\sqrt{3}/2$, so that $\theta_e=30^\circ$.  In the dynamic case, at the contact line $z=0$, from (\ref{st_ss}) one has $\sigma_2 = (1-\sqrt{3})/2$ so that the dynamic contact angle $\theta_d = \arccos(-\sigma_2) = 68.53^\circ$.

Although the formulated problem contains artificially devised surface tension distributions, it is still of interest to see the influence which these have on the flow field. In Figure~\ref{F:surfac}, we show streamlines for this flow in the region where both the tangential stress and the surface tension gradients contribute to slip at the liquid-solid interface.  It is seen that the surface tension gradients can be strong enough to reverse the flow close to the contact line.
\begin{figure}
\centering
\includegraphics[scale=0.320]{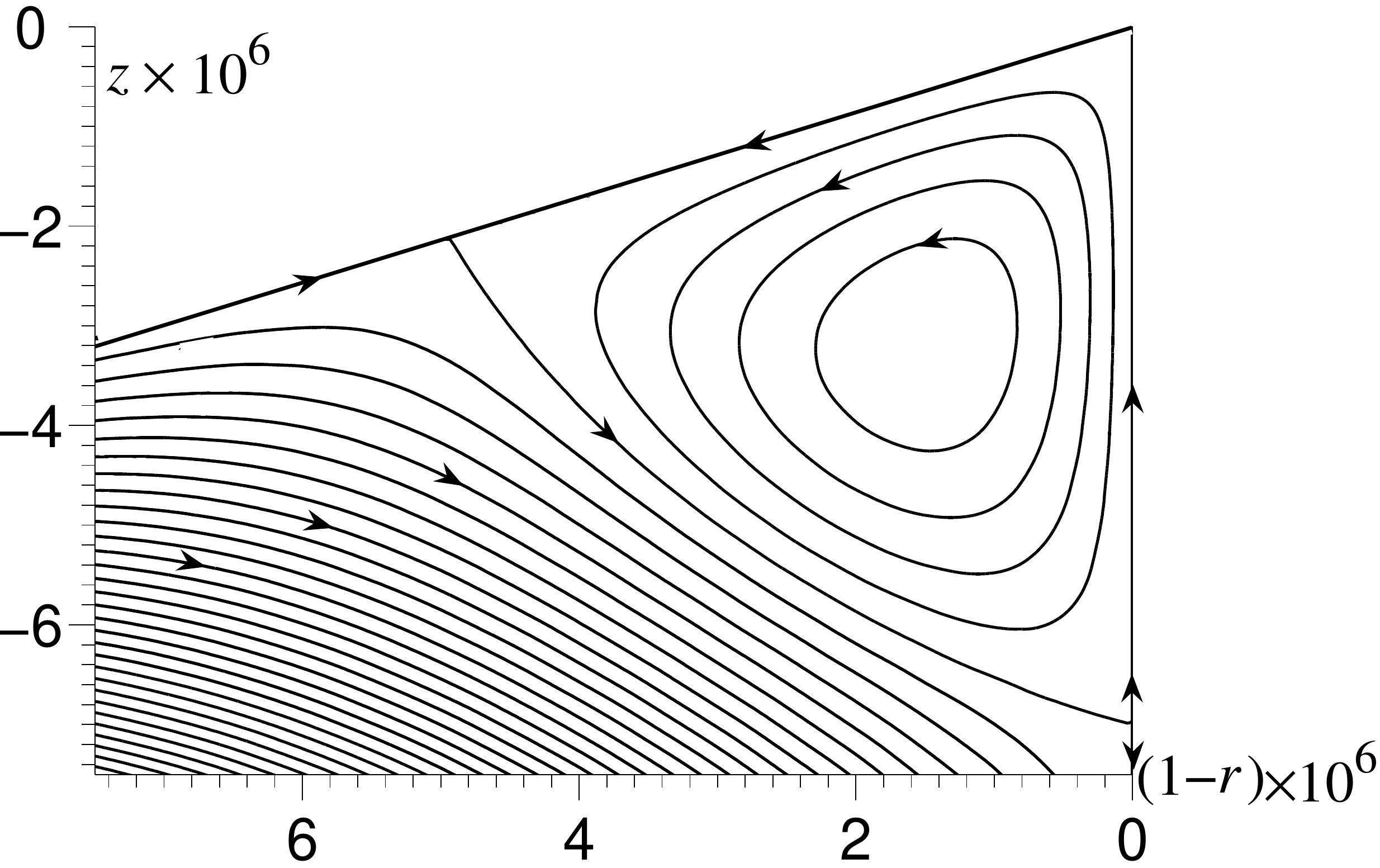}
\includegraphics[scale=0.350]{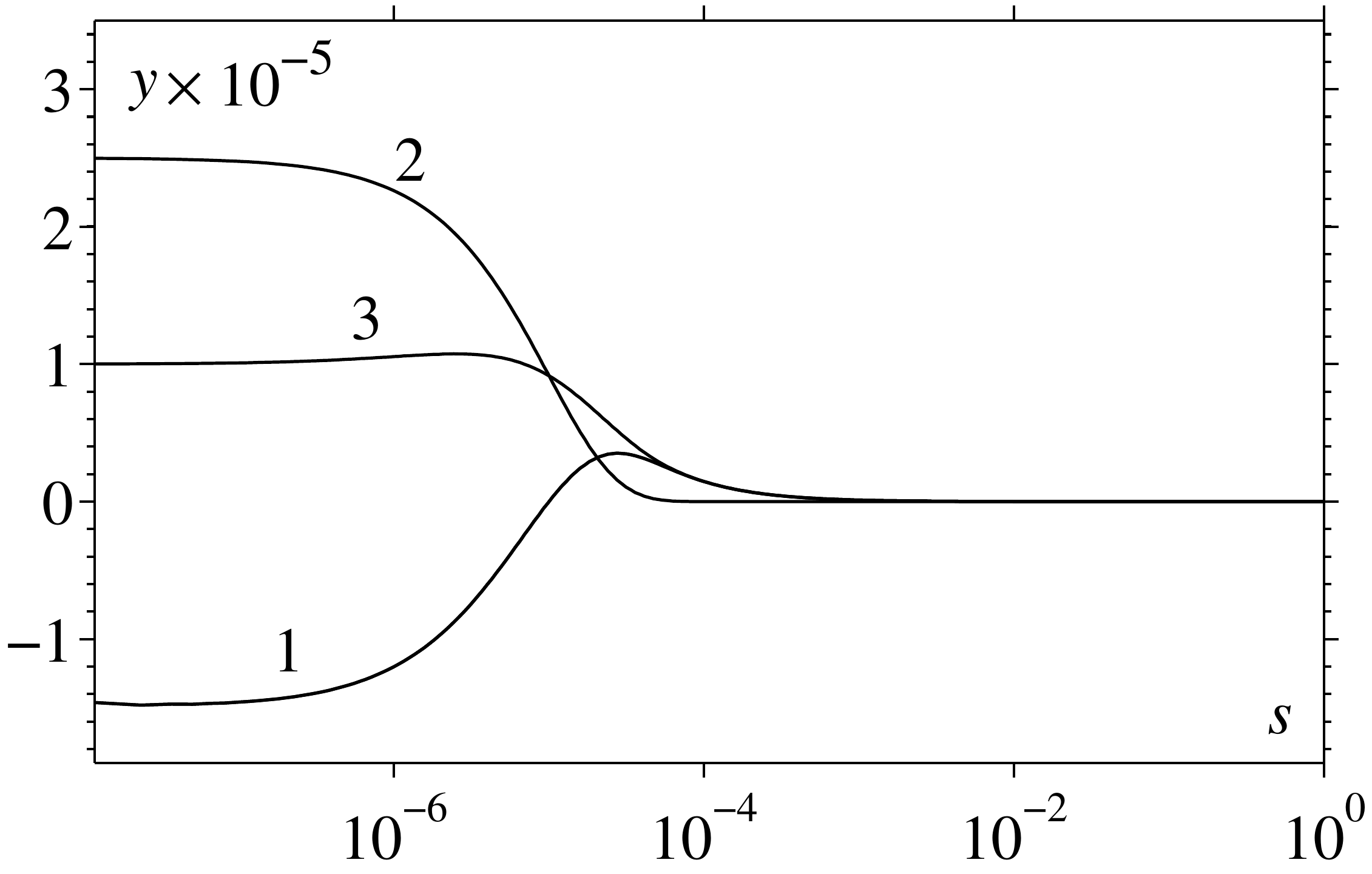}
\caption{Left: streamlines near the contact line in increments of $\psi=10^{-8}$ computed for
$Ca=0.1,\bar{\beta}=10^{5},\theta_d=68.53^{\circ}$. Right: components of equation (\ref{gennav_scalar}) plotted against distance from the contact line
along the solid surface $s$. Curves 1: $y=\pdiff{w}{r}$, 2: $y=\frac{1}{2Ca}\pdiff{\sigma_2}{z}$,
3: $y=\bar{\beta}\left(w+1\right)$.} \label{F:surfac}
\end{figure}
%

For axisymmetric flow, equation (\ref{gennav}) can be written down in scalar form as
\begin{equation}\label{gennav_scalar}
\pdiff{w}{r}+\frac{1}{2Ca}\pdiff{\sigma_2}{z}= \bar{\beta}\left(w + 1\right),
\end{equation}
and in Figure~\ref{F:surfac} we plot the behaviour of the individual terms from  this equation.  The equation is well satisfied with curve $1$ and curve $2$ adding to give curve $3$.  From this figure the relative strengths of the factors which cause slip can be seen: it is the surface tension gradient which is largest for $s<10^{-5}$ and hence it is able to drive a flow reversal.

In summary, it has been shown that our new framework is easily generalizable and, in particular, allows interfacial physics associated with dynamic wetting to be naturally incorporated into it.

\section{Discussion}\label{conclusions}

As has been discussed, a number of different FEM implementations have been used to simulate dynamic wetting phenomena.  We have shown that to guarantee accurate results, particularly at
higher capillary numbers, sufficient spatial resolution of the computational mesh must be used near the contact line. When this is the case, many of the seemingly different implementations, including our own, converge to give the same result.  Despite these similarities, approaches in which the weak formulation is used to enforce the contact angle are seen to converge to the correct solution faster.  An apparent `window' for a numerical error imposing the contact angle using the weak formulation is seen to actually be an advantage: the difference between the applied angle and the computed one provides a simple quantitative test on the accuracy of the scheme. This constraint is lost if the angle is imposed as an essential condition whilst the error is shifted elsewhere in the code but not removed.  In particular, imposing the angle as an essential condition when using a mesh which does not incorporate sufficient extra spatial resolution near the contact line can lead to huge global errors in the free surface shape.

In the new implementation, the momentum equations are applied everywhere and any surface which is curved or does not align with coordinate axes causes no additional complexity.  A key advantage of our framework is that, since there are no `loose ends' at the contact line, there are no choices left for a user in how to implement the contact line region.  This, together with the obtained quantitative criterion for the mesh, ensures that the implementation of dynamic wetting phenomena into the FEM satisfies the mantra in \cite[p.~35]{gresho1} that, once the weak form is derived and the choice of mesh design, element type, etc are taken, ``the rest of the recipe is well defined - just turn the crank to generate the complete spatial approximation".


It has been shown that when the weak form is used to impose the contact angle, the computed value on an under-resolved mesh can vary significantly as the mesh resolution is altered near the contact line, and to interpret the computed value, a mesh-dependent quantity, as the dynamic contact angle would be fundamentally incorrect.  If one wishes to define an `apparent' angle, that is some measure of the free-surface shape away from the contact line, then this should be a well-defined variable, see e.g.\ \cite{wilson06}, which converges to a specific value as the mesh is refined.  In phenomena where the dynamic contact angle starts from an out-of-equilibrium value, e.g.\ in drop impact and spreading phenomena, where the angle starts at $180^\circ$, it is a fundamental \emph{modelling mistake} to prescribe the contact angle to be equal to its equilibrium value.  In this case, the applied angle cannot be reconciled with the initial free-surface shape in principle, and the code will be unable to provide mesh-independent solutions for early time.

One may hope that the restrictions on mesh design described in this paper could be relaxed by removing the region near the contact line and, if an appropriate limit is available, incorporating some local asymptotic results there. This approach has been considered, for example, in \cite{somalinga00,afkhami09} and is discussed in \cite{sikalo05}. The idea is to incorporate the small capillary number asymptotic results of \cite{cox86}, which relate the contact angle to the free surface shape at a given distance from the contact line. In both investigations, it is seen that such an approach fails for $Ca>0.1$. Therefore, there is a limited range of validity to these numerical schemes which prevents them from being used to simulate flows where the capillary number is of order one or higher regularly encountered in processes of technological interest, e.g.\ the impact and spreading of liquid drops ejected from ink-jet printers. Furthermore, the approach taken in these articles relies on the assumption that the flow field has no influence on the dynamic contact angle.  This has been shown by experiments to be incorrect, e.g.\ in \cite{blake99,clarke06}.  The model aimed at describing this effect has to be based on the Young equation that actually introduces the contact angle as a macroscopic fluid-mechanical concept and hence its numerical implementation has to rely on incorporating gradients of the surface tension along the contacting interfaces \cite{shik07}.  We have shown that the developed FEM framework allows such a generalization to be made in a naturally way without any deviation from the FEM methodology.




\section{Appendix. FEM implementation for 2D and 3D axisymmetric flow}\label{residuals}

The Appendix provides a user-friendly ready-to-use guide to the specific implementation of our FEM framework for a dynamic wetting process. This guide enables one to easily reproduce all the results presented in the paper. For a more detailed exposition about using the FEM to solve fluid flows the reader is referred to \cite{gresho1,gresho2}, whilst specific information on free surface flows, including alternative mesh designs, can be found in \cite{kistler84,tezduyar92,christodoulou92,ramanan96,cairncross00}.  For a beginner's guide to developing and structuring finite element code we suggest \cite{smith88}.

Two-dimensional and three-dimensional axisymmetric flows are considered simultaneously by using a variable $n$ which takes the values $n=0$ in the former case and $n=1$ in the latter. Two-dimensional flow occurs in a Cartesian coordinate system $(r,z)$ whilst axisymmetric flow is in a cylindrical polar coordinate system $(r,z,\vartheta)$, where $r$ is now the radial coordinate and $\vartheta$ is the azimuthal angular coordinate around the $z$-axis about which symmetry is assumed (Figure~\ref{F:sketch}). For both coordinate systems, the governing equations are solved in the $(r,z)$-plane.

The Appendix is organized as follows. First, in Section~\ref{scalar}, the tensorial expressions in the main body of the text are specified for the coordinate systems considered.  Section~\ref{spine_design} describes the arbitrary Lagrangian-Eulerian mesh design, based on the method of spines, which tessellates the domain into elements whose position becomes dependent upon the free surface shape. In Section~\ref{procedure}, the interpolating functions are constructed piecewise over each element so that both the functions and the residuals are calculated element-by-element. Expressions for the residuals on an arbitrary element are determined. Section~\ref{mapping} shows that by mapping each element in the computational domain onto a \emph{master element} the integral expressions from Section~\ref{procedure} can be calculated in a systematic way. In Section~\ref{solution}, a method for solving the resulting set of algebraic equations using a Newton-Raphson method is outlined. Specific attention is given to the construction of the Jacobian which is required in the iteration process. Finally, in Section~\ref{additional}, additional details of the numerical scheme are given.

The Appendix provides information of our own specific implementation of the framework described in the main body of the paper.  Many alternatives exist (e.g.\ different element types) which may produce equally accurate results but here we present what we have used and tested, which for a `practitioner' gives a ready-to-use algorithm.

\subsection{Scalar expressions}\label{scalar}

We will first give coordinate-specific expressions for both two-dimensional flow ($n=0$) and three-dimensional axisymmetric flow ($n=1$) in the integral form and then consider the differential terms in these integrals.

\subsubsection{Integral expressions}

The integral of a function $f$ over a volume $V$, corresponding to an appropriate portion of the $(r,z)$-plane, is
\begin{equation}
\int_V f~dV = \int_z\int_r f~ (2\pi r)^n~dr~dz,
\end{equation}
where integration over the $\vartheta$ coordinate for the axisymmetric case has yielded the $(2\pi r)^n$ term.  Surfaces are parameterized in terms of the arclength $s=s$ in the ($r,z$)-plane and, in the axisymmetric case, the angular coordinate $t=\vartheta$.  The integral of a function $f$ over the surface $A$, which is a line in the $(r,z)$-plane, will then be
\begin{equation}
\int_A f~dA = \int_s f~ (2\pi r)^n~ds,
\end{equation}
for appropriate values of $s$.  Finally, integration over the contact line contour $C_{cl}$,  which in the ($r,z$)-plane is a point $(r_c,z_c)$, will give
\begin{equation}
\int_{C_{cl}} f~dA = (2\pi r_c)^n f(r_c,z_c).
\end{equation}
Note, the factor $(2\pi)^n$ will appear from all integrals and is thus cancelled throughout the equations given in Section~\ref{procedure}.

\subsubsection{Bulk expressions}

Using unit basis vectors of the coordinate system $\mathbf{e}_r$ and $\mathbf{e}_z$, we decompose the bulk velocity into components along the coordinate axes as $\mathbf{u} = u\mathbf{e}_r + w\mathbf{e}_z$ and the components of the solid substrate speed as $\mathbf{U}=U\mathbf{e}_r + W\mathbf{e}_z$. We can now derive the coordinate-specific expressions for the differential terms that form the integrands.

The bulk gradient operator is
\begin{equation}\label{grad}
\nabla = \mathbf{e}_r\pdiff{}{r} + \mathbf{e}_\vartheta\frac{n}{r}\pdiff{}{\vartheta} +
\mathbf{e}_z\pdiff{}{z}.
\end{equation}
Then, the gradient of a vector $\mathbf{A}=a_r(r,z)\mathbf{e}_r + a_z(r,z)\mathbf{e}_z$, which will be a second order tensor, will be required and is given by
\begin{equation}\label{app_diff}
\nabla \mathbf{A} =  \pdiff{a_r}{r}\mathbf{e}_r\mathbf{e}_r + \pdiff{a_r}{z}\mathbf{e}_r\mathbf{e}_z  +\frac{n a_r}{r}\mathbf{e}_\vartheta\mathbf{e}_\vartheta + \pdiff{a_z}{r}\mathbf{e}_z\mathbf{e}_r+ \pdiff{a_z}{z}\mathbf{e}_z\mathbf{e}_z,
\end{equation}
whilst the divergence of the vector $\mathbf{A}$, which is the trace of its gradient, is
\begin{equation}\label{app_diff1}
\nabla\cdot\mathbf{A} = \pdiff{a_r}{r} + \frac{n a_r}{r} + \pdiff{a_z}{z}.
\end{equation}
The stress tensor is given by
\begin{equation}\label{app_stress}
\mathbf{P} = P_{rr}\mathbf{e}_r\mathbf{e}_r + P_{rz}\mathbf{e}_r\mathbf{e}_z + P_{\vartheta\vartheta}\mathbf{e}_\vartheta\mathbf{e}_\vartheta
+P_{zr}\mathbf{e}_z\mathbf{e}_r + P_{zz}\mathbf{e}_z\mathbf{e}_z,
\end{equation}
with components
\begin{equation}
P_{rr} = -p + 2\pdiff{u}{r},\qquad P_{rz} = P_{zr} = \pdiff{w}{r}+\pdiff{u}{z},\qquad P_{\vartheta\vartheta} =
-p+\frac{2 n u}{r},\qquad P_{zz} =
-p+2\pdiff{w}{z}.
\end{equation}

Using (\ref{app_diff}) with $\mathbf{A}=\phi_i\mathbf{e}_r$ and then $\mathbf{A}=\phi_i\mathbf{e}_z$ combined with (\ref{app_stress}), an expression for $\nabla(\phi_i\mathbf{e}_\alpha):\mathbf{P}$, which is required in the bulk momentum residuals, is obtained:
\begin{equation}
\nabla(\phi_i\mathbf{e}_r):\mathbf{P} = \pdiff{\phi_i}{r} P_{rr} + \pdiff{\phi_i}{z}P_{rz}+\frac{n\phi_i}{r}P_{\vartheta\vartheta}, \qquad \nabla(\phi_i\mathbf{e}_z):\mathbf{P} = \pdiff{\phi_i}{r} P_{zr} + \pdiff{\phi_i}{z}P_{zz}.
\end{equation}

\subsubsection{Boundary expressions}

In the $(r,z)$-plane, surfaces are parameterized in terms of the arclength $s$, which increases away from the contact line, so that the coordinates of a point on a surface are
$(r(s),z(s))$ or in the case of the axisymmetric system $(r(s),z(s),\vartheta)$. The tangent vector in the $(r,z)$-plane, which points in the direction of increasing $s$, is $\mathbf{t}=t_r\mathbf{e}_r+t_z\mathbf{e}_z$ and the vector normal to the surface, which points into the liquid, is $\mathbf{n}=n_r\mathbf{e}_r+n_z\mathbf{e}_z$.

The vectors normal and tangent to the surface are then
\begin{equation}
(t_r,t_z) = \frac{(r',z')}{(r'^2 +
z'^2)^{1/2}}, \qquad (n_r,n_z)=\pm(-t_z,t_r),
\end{equation}
where the prime denotes partial differentiation with respect to $s$ and the sign is chosen so that the normal vector will point into the liquid. In axisymmetric coordinates, the second tangent vector, perpendicular to the one in the $(r,z)$-plane, is the unit basis vector in the azimuthal direction $\mathbf{e}_\vartheta$.

Now, the tensor $\left(\mathbf{I}-\mathbf{n}\mathbf{n}\right)$, which is used to extract the tangential component of vectors, has the form
\begin{equation}
\left(\mathbf{I}-\mathbf{n}\mathbf{n}\right) = \mathbf{t}\mathbf{t} +
n\mathbf{e}_\vartheta\mathbf{e}_\vartheta,
\end{equation}
and hence we can also find the surface gradient operator
\begin{equation}
\nabla^s = \left(\mathbf{I}-\mathbf{n}\mathbf{n}\right)\cdot\nabla =
\mathbf{t}\left(\mathbf{t}\cdot\nabla\right) +
n\mathbf{e}_\vartheta\left(\mathbf{e}_\vartheta\cdot\nabla\right) = \mathbf{t}\pdiff{}{s} + \frac{n
\mathbf{e}_\vartheta}{r}\pdiff{}{\vartheta}.
\end{equation}
Therefore, in particular, expressions for the terms associated with the stress on the free surface are
\begin{align}
\nabla^s\cdot\left(\phi_i\mathbf{e}_r\right) = \left(\mathbf{t}\cdot\mathbf{e}_r\right)\pdiff{\phi_i}{s} + \frac{n\phi_i}{r},\\
\nabla^s\cdot\left(\phi_i\mathbf{e}_z\right) = \left(\mathbf{t}\cdot\mathbf{e}_z\right)\pdiff{\phi_i}{s},
\end{align}
where we have used $\pdiff{\mathbf{e}_r}{\vartheta} =\mathbf{e}_\vartheta,~\pdiff{\mathbf{e}_z}{\vartheta} =0$.

All of the components which appear in the residual equations have now been derived.  Next we consider how the finite element procedure begins by tessellating the domain into elements.

\subsection{Method of spines mesh design}\label{spine_design}

We use a well known arbitrary Lagrangian-Eulerian approach, known as the method of spines, which has been successfully applied to many free surface flows, e.g.\ \cite{heil04,wilson06}. This method, which is described in detail in \cite{kistler84}, builds on the boundary location method proposed in \cite{ruschak80}. In the method, nodes which define the free surface are located at the end of a line on which bulk nodes are attached. These are the so called spines which form the skeleton of the mesh and usually run between a solid surface and the free surface, with nodes often spaced equally along each spine. As an iteration proceeds, all the free surface nodes evolve to new positions and hence the spine, which is attached to the free surface node, moves and drags the bulk nodes accordingly.  Elements are tesselated around the nodes and therefore their position is dependent on the free surface shape.

In the $(r,z)$ computational domain the free surface is a line parameterized by the arclength $s$.  Then, the position of the free surface can be defined by a function of one variable $h=h(s)$ which depends on the mesh design chosen. Finding $h$ at each node $h_\textrm{i}$ on the free surface ($\textrm{i}=1\hbox{--}N_1$) gives the free surface shape. An example of the free-surface parameterization is given in \S\ref{c_ex}, and more specific schemes are presented below.

The ideal spines to capture the dynamics near the contact line, where the geometry is wedge-like, are arcs of circles whose centre is the contact line (Figure~\ref{F:cap}) and whose arc angles are the $h$'s.  We will show that this can be easily achieved by introducing a polar coordinate system centred at the contact line.  The drawback of this polar method is that it can be difficult to match the mesh design near the contact line with the geometry of the computational domain, which can often involve straight boundaries, further away.  To overcome this problem, we also present a method based on the bipolar coordinate system which creates  circular spines near the contact line and straight spines further away

To ensure the computational problem is tractable and that the smallest scales in the model are well resolved, a graded mesh is used with the smallest element near the contact line.  This is achieved by defining each spine to be a given distance from the contact line.  In particular, if we choose to have $N_k$ spines, with $k=1$ corresponding to the contact-line point and $k=2$ the first proper spine, i.e.\ an arc, and increase the distance between adjacent spines by a ratio $r$, then the distance $R_k$ of spine $k$ from the contact line is
\begin{equation}\label{weight}
R_k = R_{max}\frac{r^{k-1}-1}{r^{N_{k}-1}-1}\qquad k=1\hbox{--}N_k,
\end{equation}
where $R_{max}$ is the distance of the final spine ($k=N_k$) from the contact line.  In this paper, we used $r=1.07$, with $N_k$ varied to create meshes of differing spatial resolutions.  Expression (\ref{weight}) is used for each of the two methods presented.  For simplicity, when discussing the mesh design we shall consider, as is the case for the capillary, a free surface meeting a flat solid ($r=1$) at a contact line $(1,z_c)$.  We note that in the actual code the contact line is at $z=z_c$ and is free to move.  Only once we have obtained a converged solution do we map $z\rightarrow z - z_c$ so that the contact line is at $(1,0)$ as in the presented results.

\subsubsection{Polar spines}

A polar coordinate system $(R,\varphi)$ is created around the contact line, which is related to the $(r,z)$ coordinates by
\begin{equation}
\left(r,z\right)= \left(1 - R\sin\varphi, z_c-R\cos\varphi\right).
\end{equation}
Therefore, given the polar coordinates of a node, we can obtain its actual position in the computational domain.  The free surface unknown at the contact line $k=1$ is the contact line's height $h_1=z_c$. Spines $k=2\hbox{--}N_k$ are arcs of a circle of radius $R_k$ and arc angle $\varphi_k$ from the solid surface to the free surface.  The angle of the arc is the free surface unknown which is to be determined as part of the solution, i.e.\ $h_k=\varphi_k$. Nodes are then placed uniformly across each spine, so that node $m$ out of $M$ nodes on a given spine $k$ is at angle
\begin{equation}
\varphi_{m,k} = \varphi_k\left(\frac{m-1}{M-1}\right),
\end{equation}
and the subscript $m,k$ refers to the fact that this is the $m$--th node of spine $k$.   Then, for given $m,k$ a node in the domain is dependent upon the vertical coordinate of the contact line $h_1=z_c$, which determines the (known) distance of spine $k$ from the contact line calculated from (\ref{weight}), and the angular coordinate $h_k=\varphi_k$ which determines how nodes are distributed across that arc (Figure~\ref{F:spines}).
\begin{figure}
\centering
\includegraphics[scale=0.60]{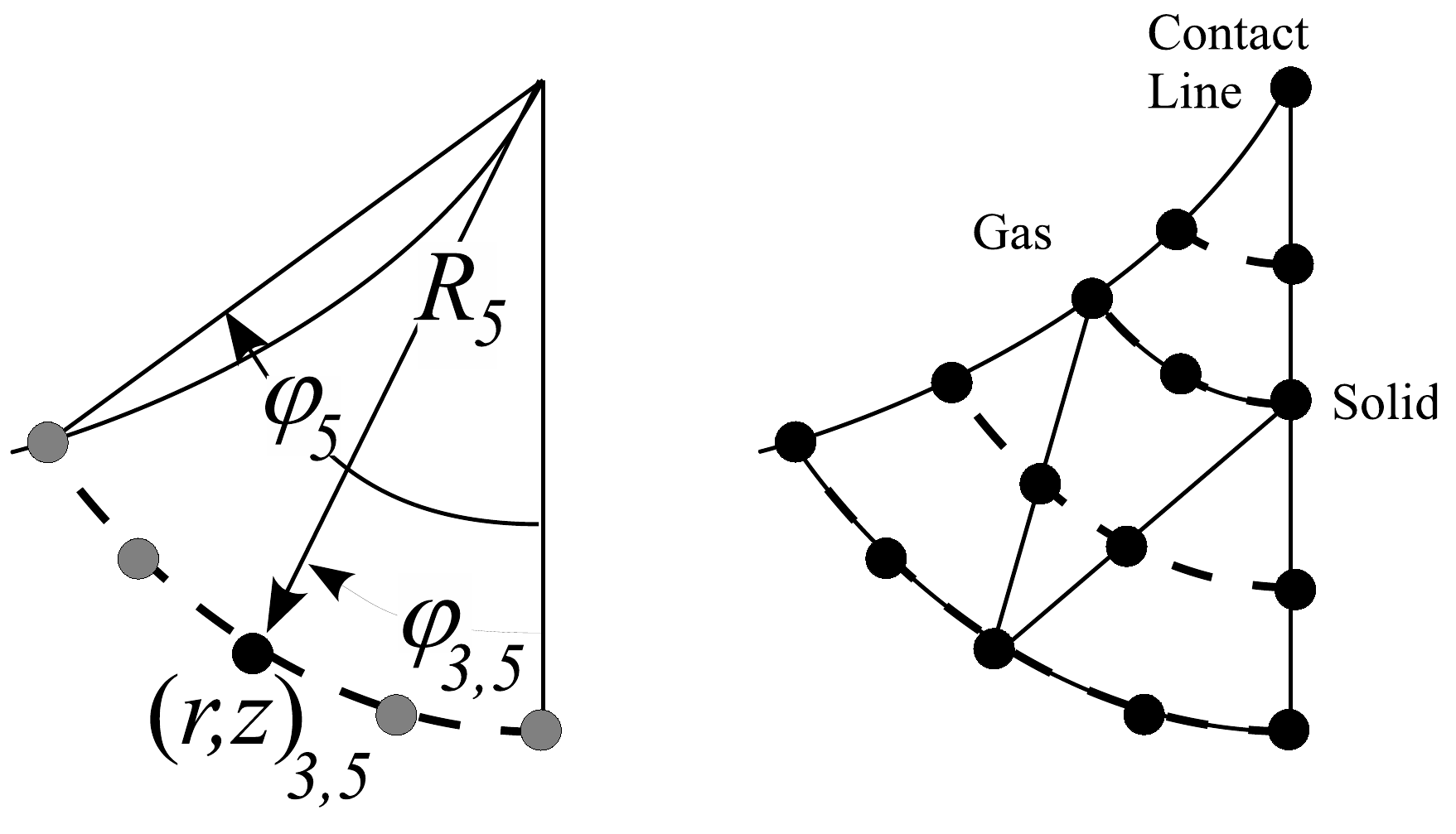}
\caption{Left: illustrative example of how the position of the third node of the fifth spine would be determined if that spine contained five nodes.  Right: how the elements are tesselated around these spines.} \label{F:spines}
\end{figure}

\subsubsection{Bipolar spines}

Far away from the contact line, in the region of the capillary between the solid surface and the axis of symmetry, we would like to use straight spines to match with the capillary's geometry.  Such spines cannot be matched to circular spines emanating from the contact line region.  However, by using the bipolar coordinate system, the spines near the contact line can remain circular whilst those further away, at a distance chosen by the user, are straight.  To achieve this, the bipolar coordinates are first related to arbitrary Cartesian coordinates.  Then, an appropriate mapping is used to embed the mesh in the computational domain.

The bipolar coordinate system $(\chi,\zeta)$ is related to Cartesian coordinates $(x,y)$ by
\begin{equation}\label{FEM_torr}
x=f\frac{\sinh{\chi}}{\cosh\chi+\cos\zeta},\quad y=f\frac{\sin{\zeta}}{\cosh\chi+\cos\zeta},
\end{equation}
where $(x,y)=(f,0)$ is the focus of the bipolar coordinate system.  The bipolar coordinate $\chi$ is used to define spines, as $R$ was previously used,  with spine $k$ having $\chi = \chi_k$ and being a circle with radius and centre given, respectively, by
\begin{equation}\label{circles}
\frac{f}{\sinh\chi_k}, \qquad (x,y) = \left(\frac{f}{\tanh\chi_k},0\right),
\end{equation}
so that as $\chi_k\rightarrow  \infty$ the spines' centres approach the focus and their radius tends to zero whilst as $\chi_k\rightarrow 0$ the radius of the circles tends to infinity, i.e.\ the radius of curvature tends to zero, and we approach straight lines at $x=0$ (Figure~\ref{F:bipolar}).
\begin{figure}
\centering
\includegraphics[scale=0.60]{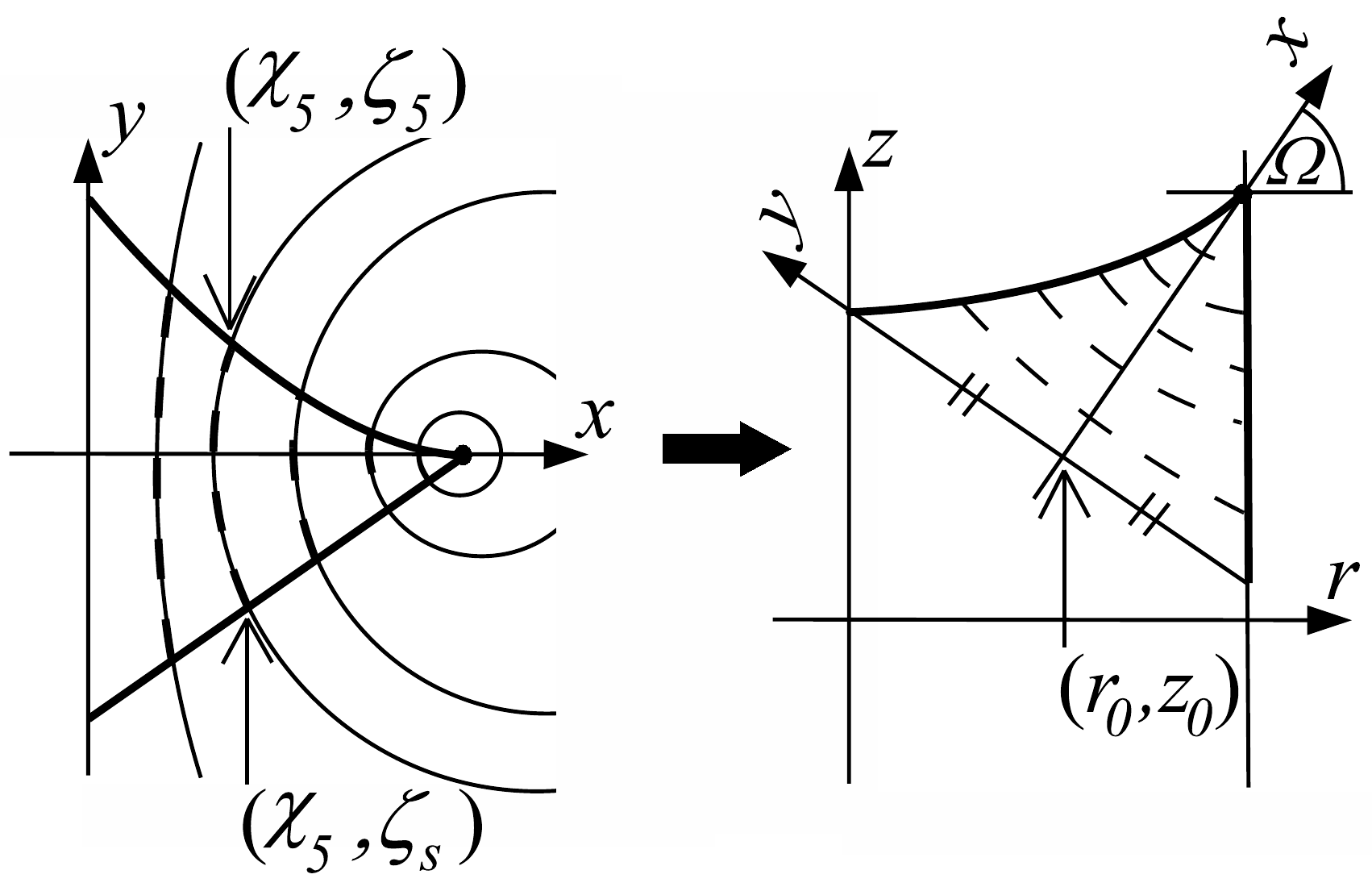}
\caption{Left: bipolar spines in a Cartesian frame highlighting the coordinates at the start and end of the fifth spine.  Right: embedding of the bipolar mesh system into the computational domain.} \label{F:bipolar}
\end{figure}

We want to map the bipolar coordinate system in $(x,y)$ with focus at $(f,0)$ and straight spines at $x=0$  onto the computational domain $(r,z)$ with the focus at the contact line $(1,z_c)$, where spines of a small radius are required, and straight spines at a chosen distance $R_{max}$ from the contact line so that $f=R_{max}$.  Then, on a given spine $k$ with $\chi=\chi_k$, varying $\zeta$ will take us from the (fixed) solid surface at $\zeta = \zeta_s$, with $\zeta_s$ to be determined, up to the free surface at $\zeta = \zeta_k$, i.e.\ $\zeta$ takes the role which $\varphi$ previously played for the polar method.  Notably, here we have not assumed that the solid surface occurs at $y=0$ which would correspond to $\zeta = 0$. Choosing $\Omega$ to be the angle at which the $x$-axis meets the $r$-coordinate this can be achieved by the transformation
\begin{equation}\label{transf}
\left( \begin{array}{c} r - r_0  \\ z - z_0\end{array}\right) =
  \left(
  \begin{array}{cc}
    \cos(\Omega)  & -\sin(\Omega)   \\
    \sin(\Omega)  &  \cos(\Omega)
  \end{array}\right)
\left( \begin{array}{c} x \\ y \end{array}\right),
\end{equation}
where the origin of the $(x,y)$ coordinate system is placed at $(r_0,z_0) = (1-R_{max}\cos\Omega,z_c-R_{max}\sin\Omega)$.  Therefore, given values of ($\chi,\zeta$), we can determine $(x,y)$ from (\ref{FEM_torr}) and hence a position in the $(r,z)$-plane from (\ref{transf}).  One can still use (\ref{weight}) to define the distance along the $x$-axis ($\zeta=0$) of a given spine from the contact line, except one must now use (\ref{FEM_torr}) and (\ref{transf}) to determine the appropriate value of $\chi_k$ from $R_k$.  For our particular geometry we take $R_{max}$ such that the $y$-axis passes through the apex of the free surface at $z=z_a$ and choose the angle $\Omega$ such that the distance along $y$ to the apex is the same as the distance along $y$ to the solid (Figure~\ref{F:bipolar}).

Outside the bipolar region, one is left with a straight-sided domain whose upper boundary is dependent on the position of the final, straight, bipolar spine. The spines in this region are chosen to run from the solid surface to the axis of symmetry, both of which are fixed, so their positions may be easily defined in terms of the final bipolar spine. What has been described is only one possible mesh, which is shown in Figure~\ref{F:cap}, that can be designed by mapping the bipolar coordinate system onto the computational domain and is seen to work well for the range of parameters considered in this paper.

\subsubsection{Remarks on mesh design}

The aforementioned procedures create elements throughout the domain which have curved sides.  However, only the element sides which form the free surface need to be curved and so, if required, to straighten all the other element sides one can loop through every node and define every midpoint node to lie exactly half way between its two adjacent vertex nodes.  The drawback of this approach is that the midpoint nodes become dependent on the position of the adjacent spines.  This means that the nodal positions become dependent on, at most: the position of the contact line; the apex position, which is used to define $\Omega$ in (\ref{transf}); and on the free surface unknowns of the adjacent spines.

The polar and the bipolar meshes can be used local to the contact line with a different mesh design, or even a different numerical method \cite{lukyanov08}, taking over further away.  In our case, we used straight spines for the rest of the capillary as the geometry is simple; however, it would be possible to use unstructured meshing away from the contact line which would allow a greater degree of flexibility.

\subsection{Finite element procedure}\label{procedure}

Having derived the expressions which will be required in the residuals and described a particular mesh design, we proceed to outline the finite element method.  The method is not restricted to the capillary geometry and our only assumption about the free surface is that in the $(r,z)$-plane, i.e.\ the computational domain, it is a line parameterized by arclength $s$ with position defined by a function of one variable $h=h(s)$ which depends on the particular mesh design. In other words, to determine the free surface shape we must determine the value of $h_\textrm{i}$ at every node on the free surface $\textrm{i}=1\hbox{--}N_1$.

The domain is tesselated into $e=1\hbox{--}N_e$ non-overlapping subdomains of area $E_e$, the \emph{elements}, which in our case will be two-dimensional curved-sided triangles whose positions are defined by the spines of the mesh and hence by the values of free surface unknowns $h_\textrm{i}$.  The boundary of the computational domain is composed of one-dimensional elements of arclength $s_e$ formed from the sides of the bulk elements which are adjacent to the boundary.  We refer to element level quantities as \emph{local} and those defined over the entire domain, which were used in \S\ref{FEM}, as \emph{global}.   Each element contains a set number of local nodes and it is the set of all local nodes which form the global nodes referred to in \S\ref{FEM}.  We have used Roman letters $(\textrm{i},\textrm{j})$ for global quantities and italicized letters $(i,j,k,l)$ for local ones. In the FEM, one must store a function $I$ which relates each local node $i$ in each element $e$ to its global node number $\textrm{i}$ so that ${\textrm{i}}=I(e,i)$.  Each global node often belongs to more than one element, see Figure~\ref{F:numbering} where, for example $7 = I(1,3) = I(2,1)$.
\begin{figure}
\centering
\includegraphics[scale=0.50]{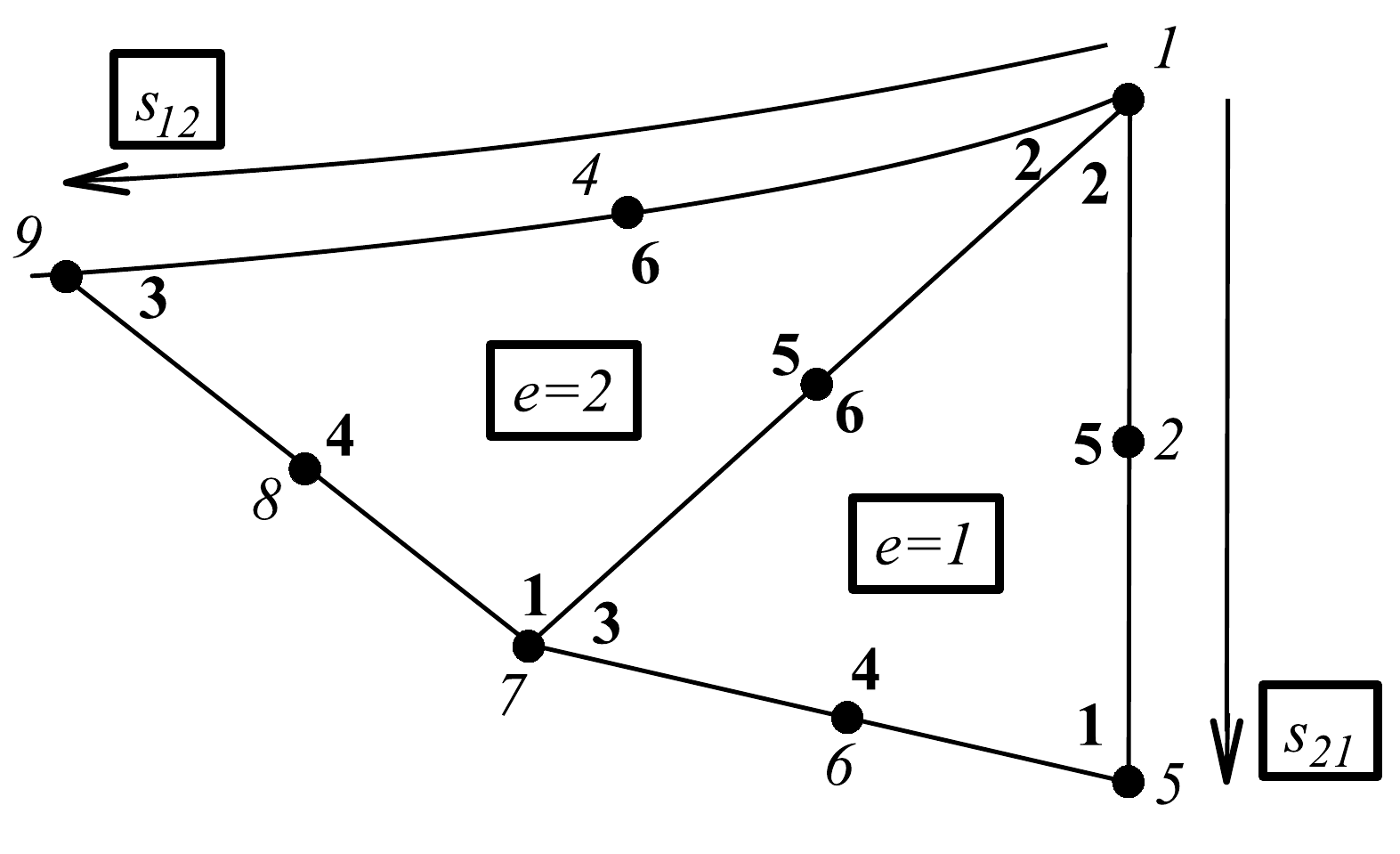}
\caption{Two elements adjacent to the contact line showing the relationship between the element numbers $e=1,2$, the global nodes (which are given on the outside of the elements) and the local node numbers (on the inside of the elements).  Also shown are the first line element on the free surface $s_{1e}$ which is a side of element $e=2$ and the first line element on the solid surface $s_{2e}$ which is a side of element $e=1$.} \label{F:numbering}
\end{figure}

In this section, it will be shown how the global functions and residuals are constructed in a piecewise manner from local quantities, by ensuring that, as required in the FEM, the
interpolating function associated with a given node is constructed to be zero outside the elements to which that node belongs.  This allows us to derive expressions for the residuals in an arbitrary element which, after summing up contributions from every element in the domain, will form a set of algebraic equations whose coefficients are integrals to be calculated using the methods described in Section~\ref{mapping}.

\subsubsection{Construction of the interpolating functions}

The global interpolating functions are constructed piecewise from local interpolating functions which are defined across elements as low order polynomial functions.  A global interpolating function, say $\phi_{\textrm{i}}$, is zero in all elements which do not contain global node $\textrm{i}$.  In an element $e$ whose local node $i$ is global node $\textrm{i}$, the global function is equal to the local one $\phi_i$.  For example, the global velocity interpolating function in an arbitrary element $e$ is:
\begin{equation}
\phi_{\textrm{i}}(r,z) =
\left\{
  \begin{array}{ll}
    \phi_{i}(r,z), & {\textrm{i}}=I(e,i) ; \\
    0, & \hbox{otherwise.}
  \end{array}
\right.
\end{equation}

To avoid having to construct each local interpolating function in an element-specific manner, they are constructed on a master element with coordinates ($\xi,\eta$), see Figure~\ref{F:master}, so that $\phi_{i} = \phi_{i}(\xi,\eta),~\psi_{i} = \psi_{i}(\xi,\eta)$.  Mixed interpolation is achieved, to ensure the Ladyzhenskaya-Babu\u{s}ka-Brezzi \cite{babuska72} condition is satisfied, by using the V6P3 Taylor-Hood triangular element which approximates velocity by means of bi-quadratic local interpolating functions $\phi_{i}(\xi,\eta)~~(i=1\hbox{--}6)$ and pressure using bi-linear ones $\psi_{i}(\xi,\eta)~~(i=1\hbox{--}3)$.  As discussed, the interpolating functions are constructed to take the value one at local node $i$ and zero at all others associated with that function, so that
\begin{align}\label{interp1}
&\psi_{1}=\frac{1+\eta}{2},\quad \psi_{2}=-\frac{\xi+\eta}{2}, \quad
\psi_{3}=\frac{1+\xi}{2},
\end{align}
and
\begin{equation}\label{interp2}
 \begin{array}{rclrclrclrcl}
\phi_{1}& = &\psi_{1}(2\psi_{1}-1),& \phi_{2}&=&\psi_{2}(2\psi_{2}-1) , &  \phi_{3}&=&\psi_{3}(2\psi_{3}-1) , & \\
\phi_{4}& = &4\psi_{1}\psi_{3} ,&       \phi_{5}&=&4\psi_{2}\psi_{1}  ,&      \phi_{6}&=&4\psi_{3}\psi_{2} . &
\end{array}
\end{equation}

In a given element $e$, whose nodes in the global domain are located at $(r_j,z_j),~(j=1\hbox{--}6)$, the relationship between a position in the global domain $(r,z)$ and a position in the master element $(\xi,\eta)$ is given by
\begin{equation}\label{FEM_disc_isopara}
r = \sum^{6}_{j=1} r_j(h)\phi_{j}(\xi,\eta) \qquad z = \sum^{6}_{j=1} z_j(h)\phi_{j}(\xi,\eta),
\end{equation}
where the dependence of \emph{all} of the nodal positions on $h$, i.e.\ the free surface unknowns, is shown here explicitly but henceforth assumed.  Being able to find the global coordinates from only the nodal positions and the interpolating functions gives the mapping the property of being \emph{isoparametric}.

Next, we form the one-dimensional local interpolating functions which are required for the approximation of boundary terms.   For example, on the side of the master element (Figure~\ref{F:master}) with nodes $j=2,6,3$ a one-dimensional master element $\eta=-1$ and $\xi\in[-1,1]$ can be defined on which a one-dimensional quadratic (as opposed to bi-quadratic) interpolating function $\phi_{1,j}(\xi) = \phi_{j}(\xi,\eta=-1)$ is obtained from (\ref{interp2}):
\begin{align*}
&\psi_{1,2}(\xi)=\frac{1-\xi}{2}, \quad
\psi_{1,3}(\xi)=\frac{1+\xi}{2},\\
&\phi_{1,2}(\xi)=\psi_{1,2}(2\psi_{1,2}-1),\quad
\phi_{1,3}(\xi)=\psi_{1,3}(2\psi_{1,3}-1),\quad \phi_{1,6}(\xi)=4\psi_{1,3}\psi_{1,2}.
\end{align*}
Then, the coordinates in the global domain on this side of the element are
\begin{equation}\label{FEM_disc_isopara1}
r_1 = \sum_{j=2,6,3} r_{j}\phi_{1,j}(\xi) \qquad z_1 = \sum_{j=2,6,3} z_j\phi_{1,j}(\xi).
\end{equation}
By ensuring that elements whose sides form the free surface always have local nodes $2,6,3$ associated with them, we can use (\ref{FEM_disc_isopara1}) to define the free surface shape.  Similarly, by ensuring that the local nodes on the solid surface are always $j=1,5,2$, we can define $\phi_{2,j}(\eta)=\phi_j(\xi=-1,\eta)$.
\begin{figure}
\centering
\includegraphics[scale=0.80]{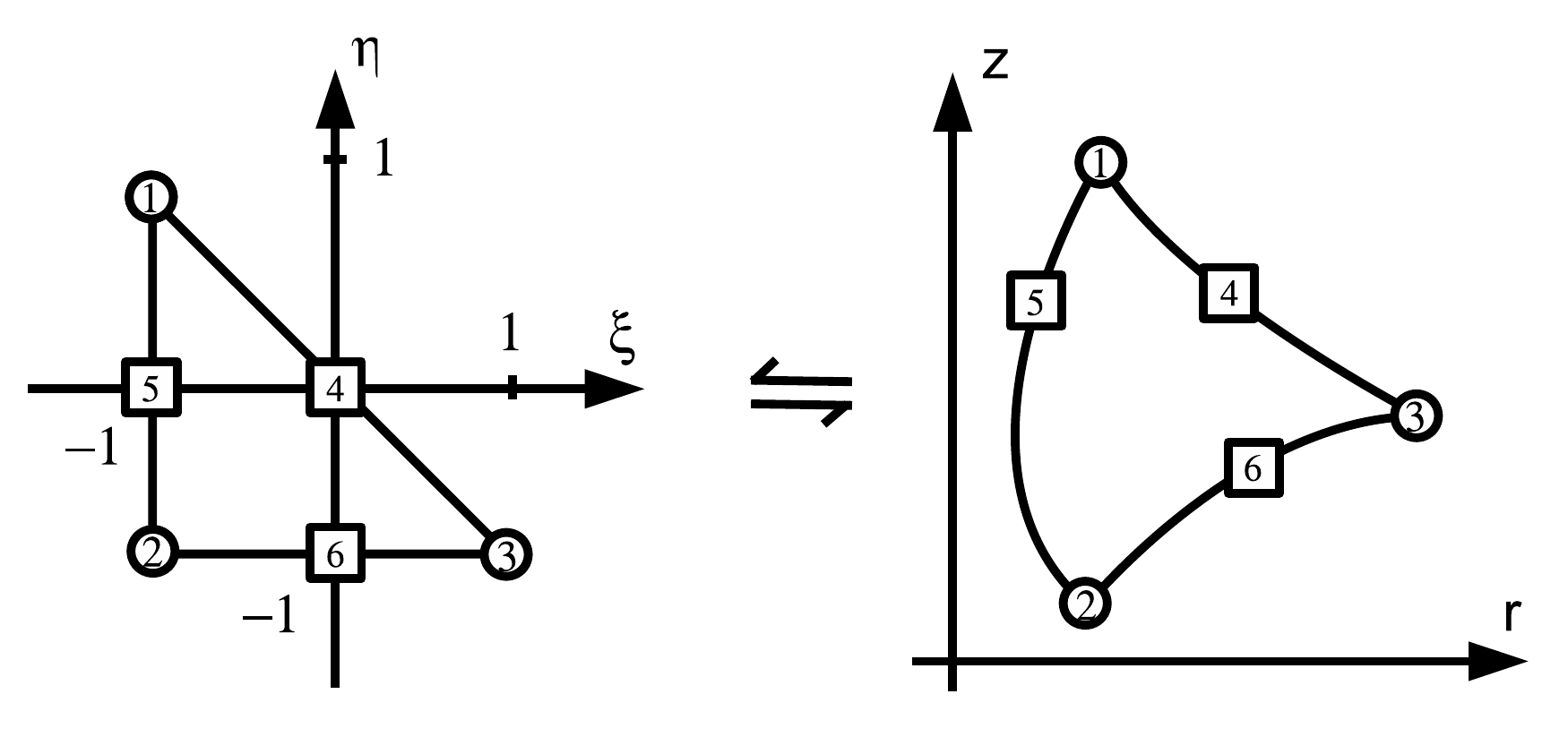}
\caption{The V6P3 master element with local coordinates $(\xi,\eta)$ which is mapped into the
physical $(r,z)$ space via the isoparametric mapping (\ref{FEM_disc_isopara}).  The numbered
circles represent nodes at which the velocity and pressure are to be found whilst the squares are
velocity only nodes.} \label{F:master}
\end{figure}

Having constructed the required interpolating functions, over an element $e$ the following functional forms are obtained
\begin{equation}
(u,w) = \sum^{6}_{j=1} (u_{j},w_{j})\phi_{j}(\xi,\eta),\qquad p
= \sum^{3}_{j=1} p_{j}\psi_{j}(\xi,\eta),\qquad (r,z)
= \sum_{j=1}^6 (r_j,z_j)\phi_{j}(\xi,\eta).
\end{equation}
The surface tension is also approximated on the surfaces quadratically as
\begin{equation}
\sigma_1 = \sum_{j=2,6,3} \sigma_{1,j}\phi_{1,j}(\xi),\qquad \sigma_2 = \sum_{j=1,5,2} \sigma_{2,j}\phi_{2,j}(\eta).
\end{equation}
Note that, in this article the nodal values of the surface tension were prescribed.

The global residuals (the $R_{\textrm{i}}$'s) presented in \S\ref{FEM}, which involve integrals over the entire domain, are formed by summing up local residuals (the $R_{e,i}$'s), denoted with a superscript $e$ to indicate which element the local residual is calculated in, obtained by integrating over each of the $N_e$ elements in the global domain. Let $\bar{N}_1$ be the set of numbers of those elements that form part of the free surface and $\bar{N}_2$ be the numbers of elements forming a part of the liquid-solid interface.  Then, the continuity of mass, momentum equations, kinematic equation on the free surface and impermeability equation at the liquid-solid interface are:
\begin{align}\label{summ_resid}
R^{C}_{{\textrm{i}}} = \mathop{\sum_{e=1}^{N_e} \sum_{i=1}^{6}}_{\textrm{i}=I(e,i)} R^{C}_{e,i},\quad R^{M,\alpha}_{{\textrm{i}}} = \mathop{\sum_{e=1}^{N_e} \sum_{i=1}^{3}}_{\textrm{i}=I(e,i)} R^{M,\alpha}_{e,i}, \quad  R^{K}_{{\textrm{i}}} = \mathop{\sum_{e=1}^{N_e} \sum_{i=2,6,3}}_{\textrm{i}=I(e,i),~e\in \bar{N}_1} R^{K}_{e,i}, \quad R^{I}_{\textrm{i}} = \mathop{\sum_{e=1}^{N_e} \sum_{i=1,5,2}}_{\textrm{i}=I(e,i),~e\in \bar{N}_2} R^{I}_{e,i},
\end{align}
where the constraint under the summation symbols ensures that, through $\textrm{i}=I(e,i)$, the local residuals are assigned to the correct global ones and, via $e\in \bar{N}$, that the surface equations are only evaluated when an element is adjacent to the boundary of the domain.

The local residuals in (\ref{summ_resid}) are simply the equations of \S\ref{weakform} consider over a subdomain of $V$.  Notably, in \S\ref{weakform} it was shown that the manipulation of the momentum residuals resulted in terms $\left(R^{M,\alpha}_{{\textrm{i}}}\right)_A$ which represented the stress acting on the bounding surface of the domain. One may expect that these terms will contribute to the momentum residuals on the boundaries of each element.  However, at nodes on inter-element boundaries, such as global node $3$ in Figure~\ref{F:numbering}, there will be a boundary contribution to the global residual from two adjacent elements, with normal vectors of opposite signs, both representing the stress on the edge of the element. By cancelling these terms at inter-element boundaries continuity of stress is ensured across the interior of the domain. Then, surface contributions to the local residuals only occur when one of the sides of an element $e$ forms part of the free surface $e\in \bar{N}_1$ or the solid surface $e\in \bar{N}_2$.  In the same way, by doing nothing, we can ensure continuity of surface stress across all free surface elements: this actually amounts to continuity of the tangent when the surface tension is continuous.  Contributions from the edge of a free surface element only occur at the contact line.

Taking the equations of \S\ref{FEM} over an arbitrary element $e$, with area $E_e$ and boundary $s_e$, and using the expressions in Section~\ref{scalar} we now derive the local residuals required in (\ref{summ_resid}).  The expressions appearing in these equations, which are usually integrals of products of the interpolating functions and their derivatives over the element or its boundary, are given in Section~\ref{thecomponents}.

\subsubsection{Element-level residuals}\label{theequations}

For an arbitrary element $e$, with local nodes $i=1\hbox{--}6$, contributions from the bulk of the element $E_e$ to the momentum residuals are
\begin{align}\label{sum_ns1}
R^{M,r}_{e,i} = & \left(Re~A_{ij}(u,w,h)+K^{11}_{ij}(h)\right) u_j   +K^{12}_{ij}(h)w_j+ C^{1}_{ik}(h)p_k\qquad j=1\hbox{--}6 \quad k=1\hbox{--}3,  \\ \label{sum_ns2}
R^{M,z}_{e,i} =& K^{21}_{ij}(h)u_j+\left(Re~A_{ij}(u,w,h)+K^{22}_{ij}(h)\right) w_j + C^{2}_{ik}(h)p_k   \qquad j=1\hbox{--}6 \quad k=1\hbox{--}3,
\end{align}
where summation over repeated indices is henceforth assumed.  The $K$ terms are associated with viscous forces, the $A$ terms are from the nonlinear convective terms and the $C$ terms are with pressure forces. Expressions inside brackets indicate the dependencies of each term on other functions in the problem, i.e.\ they show the nonlinearity of the problem.

For $i=1\hbox{--}3$ we have the incompressibility residuals
\begin{equation}\label{sum_cont}
R^{C}_{e,i} = C^{1}_{ji}(h) u_j + C^{2}_{ji} (h) w_j \qquad j=1\hbox{--}6  .
\end{equation}

If an element $e$ forms a part of the free surface, $e\in \bar{N}_1$, then for $i=2,6,3$, i.e.\ the free surface nodes, there are additional contributions to the momentum equations from capillary stress terms so that
\begin{align}\label{sum_fs1}
R^{M,r}_{e,i} = &R^{M,r}_{e,i} + F^{1}_{ij}(h)\sigma_{1,j} \qquad j=2,6,3,  \\ \label{sum_fs2}
R^{M,z}_{e,i} = &R^{M,z}_{e,i} + F^{2}_{ij}(h)\sigma_{1,j} \qquad j=2,6,3.
\end{align}
Additionally, for $i=2,6,3$ there is kinematic equation of impermeability
\begin{equation}\label{sum_kin}
R^{K}_{e,i}  = I^{1}_{ij}(h)u_j +  I^{2}_{ij}(h)w_j \qquad j=2,6,3.
\end{equation}

If an element $e$ forms a part of the liquid-solid interface, $e\in\bar{N}_2$, then for $i=1,5,2$ there are additional contributions to the momentum equations of
\begin{align}\label{sum_ss1}
R^{M,r}_{e,i} = &R^{M,r}_{e,i} + N^{11}_{ij}(h) u_j   +  N^{12}_{ij}(h) w_j + L^{1}(h)\lambda_j + S^{1}_{ij}(h) \sigma_{2,j}\qquad j=1,5,2,  \\ \label{sum_ss2}
R^{M,z}_{e,i} = &R^{M,z}_{e,i} + N^{21}_{ij}(h) u_j   +  N^{22}_{ij}(h) u_j + L^{2}(h)\lambda_j + S^{2}_{ij}(h) \sigma_{2,j}\qquad j=1,5,2,
\end{align}
where the $N$ terms come from either the Navier condition (\ref{ss}) or its generalization (\ref{gennav}), whilst the $S$ terms are associated with gradients in surface tension which are only accounted for in the generalized Navier condition (\ref{gennav}). The $L$ terms are associated with the normal stress on the interface.  Additionally, for $i=1,5,2$ one has the impermeability equation (\ref{ss}) in the form
\begin{equation}\label{sum_imp}
R^{I}_{e,i}  = I^{1}_{ij}(h)u_j +  I^{2}_{ij}(h)w_j\qquad j=1,5,2.
\end{equation}

If an element $e$ forms the part of the free surface $e\in\bar{N}_1$ which is adjacent to the contact line, so that local node $i=2$ is the contact line node, then there is an additional contribution
\begin{align}\label{sum_theta1}
R^{M,r}_{e,2} = &R^{M,r}_{e,2} + T^{1}(h,\theta), \\ \label{sum_theta2}
R^{M,z}_{e,2} = &R^{M,z}_{e,2} + T^{2}(h,\theta),
\end{align}
which is where the contact angle is imposed into the weak formulation.

\subsubsection{Required matrices}\label{thecomponents}

The viscous terms, see Section~\ref{scalar}, are
\begin{equation}\label{a_k}
K^{11}_{ij} = 2k_{ij11} + k_{ij22} + 2 n~k^r_{ij},\quad K^{12}_{ij} = k_{ij21},\quad K^{21}_{ij} = k_{ij12},\quad K^{22}_{ij} = k_{ij11}  + 2k_{ij22},
\end{equation}
where
\begin{eqnarray}\label{a_kk}
k_{ijkl} = \int_{E_e} \pdiff{\phi_{i}}{y_k} \pdiff{\phi_{j}}{y_l} \,r^n\,dE_e, \qquad k^r_{ij} =  \int_{E_e} \frac{\phi_{i}\phi_{j}}{r^2} \,r^n\,dE_e,
\end{eqnarray}
with  $y_1 = r,~y_2=z$.   Nonlinear convective terms are
\begin{equation}\label{a_a}
A_{ij}(u,w) = a_{ijk1}u_{k}+a_{ijk2}w_{k}\qquad k=1\hbox{--}6,
\end{equation}
where
\begin{equation}\label{a_aa}
a_{ijkl} = \int_{E_e}\phi_{i} \phi_{k} \pdiff{\phi_{j}}{y_l}  \,r^n\,dE_e.
\end{equation}
The $C$ matrices represent pressure terms whilst their transpose is required in the continuity of mass equation:
\begin{equation}\label{a_c}
C^{1}_{ij}= -\int_{E_e}\psi_{j}\left(\pdiff{\phi_{i}}{r} + \frac{n\phi_{i}}{r}\right)\,r^n\,dE,\qquad
C^{2}_{ij}=-\int_{E_e}\psi_{j}\pdiff{\phi_{i}}{z} \,r^n\,dE_e.
\end{equation}

On a free surface (one-dimensional) element $s_{1e}$, the capillary stress terms are
\begin{equation}\label{a_f1}
F^{1}_{ij}=\frac{1}{Ca}\int_{s_{1e}} t_{1r}\left(\pdiff{\phi_{1,i}}{s}+\frac{n\phi_{1,i}}{r}\right)\phi_{1,j}
\,r^n\,ds_{1e},\qquad
F^{2}_{ij}=\frac{1}{Ca}\int_{s_{1e}} t_{1z}\pdiff{\phi_{1,i}}{s}\phi_{1,j} \,r^n\,ds_{1e}.
\end{equation}
The components of the vectors tangential $\mathbf{t}= (t_{r},t_{z})$ and normal $\mathbf{n} = (n_{r},n_{z})$ to a surface have been derived in Section~\ref{scalar}.

On a (one-dimensional) element $s_{2e}$ on the liquid-solid interface, the tangential stress terms from the Navier-slip boundary condition are
\begin{equation}\label{a_ss3}
N^{kl}_{ij}=\bar{\beta}\int_{s_{2e}} \phi_{2,i}\phi_{2,j} t_{2k}t_{2l}   \,r^n\,ds_{2e},
\end{equation}
with $t_{21}=t_{2r}$ and $t_{22}=t_{2z}$. In the generalized Navier condition (\ref{gennav}), additional terms associated with gradients in surface tension are
\begin{equation}\label{a_f1}
S^{1}_{ij}=-\frac{1}{2Ca}\int_{s_{2e}}\phi_{2,i}\pdiff{\phi_{2,j}}{s} t_{2r} \,r^n\,ds_{2e},\qquad
S^{2}_{ij}=-\frac{1}{2Ca}\int_{s_{2e}}\phi_{2,i}\pdiff{\phi_{2,j}}{s} t_{2z} \,r^n\,ds_{2e},
\end{equation}
and the normal stress terms give
\begin{equation}\label{a_f1}
L^{1}_{ij}=\int_{s_{2e}}\phi_{2,i}\phi_{2,j} n_{2r} \,r^n\,ds_{2e},\qquad
L^{2}_{ij}=\int_{s_{2e}}\phi_{2,i}\phi_{2,j} n_{2z} \,r^n\,ds_{2e}.
\end{equation}

At the contact line ($r_c,z_c$) contributions will occur at local node $i=2$ of the first free surface element so that
\begin{equation}\label{a_t1}
T^{1} = \frac{\sigma_{1,i=2}}{Ca}\left[\left( t_{2r}\cos\theta + n_{2r} \sin\theta \right)
r_c^n\right],\qquad T^{2} = \frac{\sigma_{1,i=2}}{Ca}\left[\left(t_{2z}\cos\theta  + n_{2z}\sin\theta
\right)r_c^n\right],
\end{equation}
where $\theta$ is either prescribed or determined from the Young equation $\theta = \arccos(-\sigma_{2,i=2}/\sigma_{1,i=2})$, with the surface tension values evaluated at the contact line.

On both the free surface ($\gamma=1$) and the liquid-solid interface ($\gamma=2$), expressions for the impermeability equation in surface element $s_{\gamma e}$ are
\begin{equation}\label{a_k1}
I^{1}_{ij} = \int_{s_{\gamma e}} \phi_{\gamma,i}\phi_{\gamma,j} n_{\gamma r} \,r^n\,ds_{\gamma e},\qquad I^{2}_{ij} = \int_{s_{\gamma e}}
\phi_{\gamma,i}\phi_{\gamma,j} n_{\gamma z} \,r^n \,ds_{\gamma e}.
\end{equation}

We have now derived all the expressions required.  In the next section we present a systematic way of determining these integrals and hence forming a set of nonlinear algebraic equations with numerical coefficients.

\subsection{Mapping local integrals onto the master element for evaluation}\label{mapping}

We have a set of algebraic equations, that is the momentum equations (\ref{sum_ns1})--(\ref{sum_ns2}) with contributions from the free surface (\ref{sum_fs1})--(\ref{sum_fs2}), solid surface (\ref{sum_ss1})--(\ref{sum_ss2}) and contact line (\ref{sum_theta1})--(\ref{sum_theta2}); the continuity equation (\ref{sum_cont}); the kinematic condition on the free surface (\ref{sum_kin}); and the impermeability condition on the solid surface (\ref{sum_imp}).  The coefficients of the functions' nodal unknowns in these equations are the integrals over deformed elements in the global domain given in Section~\ref{thecomponents}.  These integrals are most easily determined in a systematic way by mapping them onto the master element (Figure~\ref{F:master}), over which we already have constructed the interpolating functions, for evaluation. Therefore, each integral will be over the same geometrically simple domain and, as described in Section~\ref{quad}, quadrature methods may be used to accurately approximate them.

From the mapping (\ref{FEM_disc_isopara}) we have a relationship of the form $r=r(\xi,\eta),z=z(\xi,\eta)$ which will be required to transform our integrals.  Derivatives of the global coordinates with respect to the master element's coordinates will also be required and are easily expressed as
\begin{equation}\label{derivs}
\pdiff{r}{\xi} =  r_j\pdiff{\phi_{j}}{\xi}, \quad \pdiff{r}{\eta} = r_j\pdiff{\phi_{j}}{\eta},\quad \pdiff{z}{\xi} =  z_j\pdiff{\phi_{j}}{\xi}, \quad \pdiff{z}{\eta} = z_j\pdiff{\phi_{j}}{\eta}\qquad j=1\hbox{--}6.
\end{equation}
Therefore, we would like all integrands to contain the interpolating functions and their derivatives with respect to the master element's coordinates, as these are quantities are easily calculable.

\subsubsection{Evaluating local `volume' integrals}

The integral of a function $f$ over a (two-dimensional) element $E_e$ in the computational domain is transformed into an integral over the master element as follows
\begin{equation}\label{interv}
\int_{E_e} f~dE_e =  \int_{\eta=-1}^{\eta=1}\int_{\xi=-1}^{\xi=-\eta} f\,\det{J_e}~d\xi~d\eta,
\end{equation}
where  the Jacobian of the isoparametric mapping is
\begin{equation}\label{jacobian1}
J_e =\left(\begin{array}{cc}
                            \pdiff{r}{\xi} & \pdiff{r}{\eta} \\
                            \pdiff{z}{\xi} & \pdiff{z}{\eta}
                          \end{array}\right),
\end{equation}
so that, using (\ref{derivs}), its determinant is
\begin{equation}\label{jacobian2}
\det{J_e} =\pdiff{r}{\xi}\pdiff{z}{\eta} -  \pdiff{r}{\eta}\pdiff{z}{\xi} = T_{ij} r_i z_j\qquad i,j=1\hbox{--}6,
\end{equation}
where we have introduced the useful matrix
\begin{equation}\label{jacobian3}
T_{ij} = \pdiff{\phi_{i}}{\xi}\pdiff{\phi_{j}}{\eta} - \pdiff{\phi_{i}}{\eta}\pdiff{\phi_{j}}{\xi}.
\end{equation}
We can ensure that the determinant is positive, and therefore do not require its modulus, by making sure that the local numbering of our master element and of our elements in the global domain are both, say, anti-clockwise (Figure~\ref{F:master}).

The derivatives of the master element's coordinates with respect to the global ones may be found by noting that the
inverse of the Jacobian may be obtained either by considering $\xi=\xi(r,z)$, $\eta = \eta(r,z)$ or by directly inverting (\ref{jacobian1}), so that
\begin{equation}\label{inv_j}
J_e^{-1} =\left(\begin{array}{cc}
                            \pdiff{\xi}{r} & \pdiff{\xi}{z} \\
                            \pdiff{\eta}{r} & \pdiff{\eta}{z}
                          \end{array}\right)
                          =\frac{1}{\det{J_e}}\left(\begin{array}{cc}
                            \pdiff{z}{\eta} & -\pdiff{r}{\eta} \\
                            -\pdiff{z}{\xi} & \pdiff{r}{\xi}
                          \end{array}\right).
\end{equation}
Then, for example, $\pdiff{\xi}{r} = \pdiff{z}{\eta}/\det{J_e}$.

To determine the derivatives of a variable with respect to global coordinates, in terms of the (known) element's global coordinates and (known) derivatives of the local interpolating functions (\ref{interp1}) and (\ref{interp2}) with respect to local coordinates, we use the chain rule and the expressions in (\ref{inv_j}) to obtain
\begin{align}\label{tmatrix1}
\pdiff{\phi_{i}}{r} = \pdiff{\phi_{i}}{\xi}\pdiff{\xi}{r} + \pdiff{\phi_{i}}{\eta}\pdiff{\eta}{r} =  \frac{T_{ik}z_k}{\det{J_e}}\qquad k=1\hbox{--}6,\\ \label{tmatrix2}
\pdiff{\phi_{i}}{z} = \pdiff{\phi_{i}}{\xi}\pdiff{\xi}{z} + \pdiff{\phi_{i}}{\eta}\pdiff{\eta}{z} = -\frac{T_{ik}r_k}{\det{J_e}}\qquad k=1\hbox{--}6.
\end{align}

We have now obtained all the required expressions to allow us to evaluate integrals over a standard master element instead of over the deformed global elements. Next, we consider the same procedure for the `surface' integrals.

\subsubsection{Evaluating local `surface' integrals}

In the computational domain, the `surface' is divided into line elements $s_e$ which are the curved sides of bulk elements.  In this section, it is shown how the integrals over the global domain may be mapped onto a one-dimensional master element for evaluation. In particular, we consider an element $e$ whose side, a line element $s_{1e}$, with local nodes $2,6,3$ forms part of the free surface $e\in\bar{N}_1$ and the extension to other surfaces is straightforward.

Integrals over a line element $s_{1e}$ of the liquid-gas free surface are transformed onto a one-dimensional master element $\xi\in[-1,1]$ using
\begin{equation}\label{incom}
\int_{s_{1e}} f~ds_{1e} = \int_{\xi=-1}^{\xi=1} f \, \left|\diff{s_{1e}}{\xi}\right|\, d\xi, \qquad \diff{s_{1e}}{\xi}  = \sqrt{r_{1}'^{2}+z_{1}'^{2}},
\end{equation}
where the position in the computational domain $(r_1(\xi),z_1(\xi))$ may be calculated from (\ref{FEM_disc_isopara1}}) and derivatives with respect to $\xi$, denoted with a prime, are easily found from (\ref{derivs}).  Derivatives of functions with respect to $s_{1e}$ may then be transformed using
\begin{equation}\label{incom}
\diff{}{s_{1e}} = \diff{\xi}{s_{1e}}\diff{}{\xi} = \frac{1}{\sqrt{r_{1}'^2 + z_{1}'^2}}\diff{}{\xi}.
\end{equation}

The tangent and normal vectors in a free surface element will be
\begin{equation}
\mathbf{t}_1 = (t_{1r},t_{1z}) = \frac{(r_{1}',z_{1}')}{(r_{1}'^2 + z_{1}'^2)^{1/2}},\qquad \mathbf{n}_1 =(n_{1r},n_{1z}) = \pm(-t_{1z},t_{1r}),
\end{equation}
where the normal vector is defined to point into the liquid.

Given this information, the tensor $\left(\mathbf{I}-\mathbf{n}_1\mathbf{n}_1\right)$ on the free surface can be determined in terms of the master element's coordinates
and hence the surface gradient operator can also be obtained:
\begin{equation}
\left(\mathbf{I}-\mathbf{n}\mathbf{n}\right) = \mathbf{t}_1\mathbf{t}_1 +
n\mathbf{e}_\vartheta\mathbf{e}_\vartheta,\qquad \nabla^s = \left(\mathbf{I}-\mathbf{n}_1\mathbf{n}_1\right)\cdot\nabla =
\frac{\mathbf{t}_1}{\sqrt{r_{1}'^{2}+z_{1}'^{2}}}\pdiff{}{\xi} +
\frac{n\mathbf{e}_\vartheta}{r_{1}}\pdiff{}{\vartheta}.
\end{equation}

In a completely analogous way, the expressions required for the evaluation of integrals on the liquid-solid interface may be formed.

\subsubsection{Determining local `line' integrals}

In our computational domain, the contact line corresponds to one point.   The vector $\mathbf{m}$, which is defined to be normal to the contact line and inwardly tangent to a given surface, will then coincide with the tangent vector $\mathbf{t}$ of the surface in the $(r,z)$-plane.  The value of the normal and tangent vectors at the contact line may easily be determined by setting the one-dimensional master element's coordinate to the value which it will take at the node corresponding to the contact line, i.e.\ at local node $2$ we have $\xi=\eta=-1$.

\subsubsection{Evaluating the integrals over the master element}\label{quad}

We have arrived at a set of algebraic equations whose coefficients are integrals over the master element which may be easily calculated using numerical integration. The integrands contain products of interpolating functions, which are low order polynomials, and hence Gaussian quadrature is well-suited to the task.  Gaussian quadrature allows the integral to be approximated by a weighted sum of the integrand evaluated at a finite number of points, called the Gauss points.  Using enough points, one can evaluate polynomial expressions of any degree exactly.

The rules for integration over a one-dimensional master element, say $\xi\in[-1,1]$, are easiest and hence presented first.  Using $N_{1G}$ Gauss points one can exactly integrate a polynomial of degree $2N_{1G}-1$ using
\begin{equation}\label{FEM_gps0}
\int_{\xi=-1}^{1} f(\xi) d\xi=\sum^{N_{1G}}_{i=1} f(\xi_{i})W_{i},
\end{equation}
where $\xi_i$ are the specially chosen Gauss points and $W_i$ are the appropriate weights.  Taking $N_{1G}=4$ with
\[ \begin{array}{rclrclrclrcl}
\xi_1 & = &-0.86113~63115~94052 , &\xi_3&=& -0.33998~10435~84856,  &\xi_2&=&-\xi_1,&\xi_4&=&-\xi_3, \\
W_1&=&~~0.34785~48451~37453,      &W_3  &=&~~0.65214~51548~62546,  &W_2  &=&  W_1, &W_4        &=& W_3,
\end{array} \]
is sufficient for exact integration of all surface integrals.

For integration over the master element's area we have
\begin{equation}\label{FEM_gps1}
\int_{\eta=-1}^{1}\int_{\xi=-1}^{-\eta} f(r(\xi,\eta),z(\xi,\eta)) d\xi d\eta=\\ \sum^{N_G}_{i=1} f(\xi_{i},\eta_{i})W_{i}.
\end{equation}
%
One can use tensor products of the one-dimensional case, i.e.\ integrate over one coordinate and then the other, and hence ensure that with $N_G^2$ Gauss points we can exactly integrate a polynomial of degree $2N_G-1$.  However, these expressions are not optimal: one can usually integrate a polynomial of degree $2N_G-1$ with less than $N_G^2$ points \cite{cowper73}.  The disadvantage of the non-tensor expressions is that the determination of the points and the weights becomes more complex.  Here, we simply list the $N_G=9$ points and weights which will exactly integrate polynomials of order $5$ and were found to be sufficient for a range of parameter values. They are:
\[ \begin{array}{rclrclrcl}
\xi_1 &=&~~\,0.00000~00000~00000 ,     &\xi_4   &=&  ~~\,0.77459~66692~41483            , \\
\eta_1&=& -0.88729~83346~20741,         &\eta_2  &=& -0.50000~00000~00000,           &\eta_3&=&-0.11270~16653~79258, \\
\eta_4&=& -0.97459~66692~41483,        &\eta_6  &=& -0.80000~00000~00000,           &\eta_9&=& ~~\,0.57459~66692~41483, \\
W_1   &=&~~\,0.24691~35802~46913,        &W_2     &=&  ~~\,0.39506~17283~95061 ,          &W_4   &=& ~~\,0.03478~44646~23227, \\
W_5   &=&~~\,0.05565~51433~97164,        &W_7     &=& ~~ \,0.27385~75106~85414 ,          &W_8   &=& ~~\,0.43817~20170~96662 , \\
\xi_{2,3}   &=& \xi_1,~\xi_{5,6}=-\xi_{7,8,9}=\xi_4   &\eta_{5}&=&\eta_1,~\eta_7=\eta_6,~\eta_8=\eta_3,         &W_3   &=& W_1,~W_6=W_4,~W_9=W_7 . \\
\end{array} \]
%
%
Polynomials of higher order than $5$ may exist in the integrands of the bulk integrals, but for a range of parameter values the order of quadrature suggested here has been sufficient.  It is recommended that a user tests different numbers of Gauss points to ensure optimal speed without loss of accuracy.

When very small elements are used, roughly for element sizes of $l_{min}<10^{-8}$, extreme care has to be taken to ensure that the finite precision arithmetic does
not create significant errors in the values of the integrals. In particular, it was found that for the
calculation of the determinant of the Jacobian $J_e$ in (\ref{jacobian2}), which is of order $l_{min}^2$, and only depends on the relative positions of the nodes in the global domain, the coordinate system should be shifted so that the contact line, near which the smallest elements reside, is at $(0,0)$.

\subsection{A Newton-Raphson solution method}\label{solution}

At each global node we have a number of algebraic equations whose unknowns are the functions' nodal values.  These equations and unknowns must each be given a unique position and assembled into a set of (nonlinear) algebraic equations to which standard solution methods can be applied.  A solution vector $\bd{\alpha}$, which contains all of the functions' nodal unknowns, can be easily formed by looping through every global node and assigning the functions' unknowns there consecutive positions in $\bd{\alpha}$, i.e.\ each unknown in the problem is assigned a unique position in $\bd{\alpha}$.  Then, the residual which determines a given unknown is given the same position in a vector containing all of the residual values $\bd{R}$.

To solve the assembled set of (nonlinear) algebraic equations the Newton-Raphson method is used.  This method takes an approximate solution $\bd{\alpha}^m$ with associated residuals $\bd{R}(\bd{\alpha}^m)$ and then provides an updated solution $\bd{\alpha}^{m+1}$ for each entry in the solution vector  by solving
\begin{equation}\label{FEM_nr}
\mathbf{J}\left[\bd{\alpha}^{m+1}-\bd{\alpha}^{m}\right]=-\bd{R}(\bd{\alpha}^{m}), \qquad
J_{{\textrm{i}} \textrm{j}}=\pdiff{R_{{\textrm{i}}}(\bd{\alpha}^{m})}{ \alpha^m_{ \textrm{j} } }.
\end{equation}
The method has quadratic convergence and may be repeated until a sufficient degree of accuracy is achieved. In our computations, the resulting linearized equations were solved using the MA41 solver provided by the Harwell Subroutine Library. The main complexity of the Newton-Raphson method is in the formation of the Jacobian $J_{{\textrm{i}} \textrm{j}}$ in (\ref{FEM_nr}), and this is now described. It may not be necessary to do this at every iteration: methods which do not reform the entire Jacobian after each iteration are known as quasi-Newton methods, see \cite{engelman81}.

\subsubsection{Formation of the Jacobian}

Entries of the Jacobian may be calculated numerically, using a difference formula, or determined analytically.   Analytic calculation is computationally significantly faster and also improves the convergence of the code, particularly when using small elements where the combination of truncation and round-off errors in the
calculation of the Jacobian can become significant if calculated numerically. Analytic calculation is easily achieved for the dependence of the residuals on everything except the free surface unknowns which will be considered later.

Entries of the Jacobian are built, as the residuals are, element-by-element.  Below, as a representative example, the derivative of the nonlinear local term  $A_{ik}u_k = (a_{ikl1}u_l+a_{ikl2}w_l)u_k$, appearing in the $r$-momentum residual (\ref{a_aa}), with respect to velocity $u_j$ is determined:
\begin{equation}\label{a_n}
\pdiff{\left(A_{ik}(u,w)u_k\right)}{u_j} = A_{ik}\pdiff{u_k}{u_j} + u_k\pdiff{A_{ik}}{u_j} = A_{ik}\delta_{jk} + u_k a_{ikl1}\delta_{jl} = A_{ij} + u_k a_{ikj1} \qquad k=1\hbox{--}6.
\end{equation}
As with the residuals, the local nodes $(i,j)$ will need to be related to the global ones in (\ref{FEM_nr}) using the function $I(e,i)$.

The only non-analytic calculation in our code is the dependence of the bulk nodal positions on the free surface nodal unknowns, which can become extremely complex when using the bipolar coordinate system described in Section~\ref{spine_design}.  As the residuals and the Jacobian are evaluated in an element-by-element manner, it is most useful to know the dependency of the positions of an element's nodes $(r_i,z_i)~~(i=1\hbox{--}6)$ on the global unknowns $h_{\textrm{j}}~~(\textrm{j}=1\hbox{--}N_1)$.  A quick way of calculating them numerically is by noting the unknowns which the nodal positions depend on, in our case the contact line position, apex height and nodal position at the end of the spine on which they lie and possibly the adjacent ones. The original nodal positions $(r_i, z_i)$ are updated by changing the free surface unknown $h_{\textrm{j}}$, on which the dependency is required, to $h_{\textrm{j}}\rightarrow h_{\textrm{j}} + \Delta h_{\textrm{j}}$, where $\Delta h_{\textrm{j}}/h_{\textrm{j}} \ll 1$, and obtaining a new mesh with nodal positions $(\tilde{r}_i, \tilde{z}_i)$.  Repeating this procedure for each global nodal unknown $(\textrm{j}=1\hbox{--}N_1)$ gives the required numerically calculated dependencies
\begin{equation}\label{globald}
\pdiff{r_i}{h_{\textrm{j}}} = \frac{\tilde{r}_i-r_i}{\Delta h_{\textrm{j}}},\qquad \pdiff{z_i}{h_{\textrm{j}}} = \frac{\tilde{z}_i-z_i}{\Delta h_{\textrm{j}}},
\end{equation}
and it is noted that $(r_i,z_i)$ is a local position whilst $h_{\textrm{j}}$ is a global unknown.

As a representative example, we will show how to calculate the dependence of the pressure terms in the $z$-momentum residual (\ref{a_c}), $C^2_{ik}p_k$, on a global free surface unknown $h_{\textrm{j}}$. A possible issue is that the domain for the integral depends on the position of the nodes themselves; however, by mapping our integral to the fixed master element this problem is resolved.  The calculation required is
\begin{equation}
\pdiff{\left(C^2_{ik}p_k\right)}{h_{\textrm{j}}} = -p_k\int_{\eta=-1}^{\eta=1}\int_{\xi=-1}^{\xi=-\eta} \pdiff{f_{ik}}{h_{\textrm{j}}} \,d\xi~d\eta\qquad k=1\hbox{--}3,
\end{equation}
where
\begin{equation}
f_{ik}=\psi_{k}\pdiff{\phi_{i}}{z} \,r^n\,\det{J_e}.
\end{equation}
Using (\ref{tmatrix2}), gives
\begin{equation}
\pdiff{\phi_{i}}{z}\,r^n\,\det{J_e}= -T_{im}r_m\,\left(r_l\phi_{l}\right)^n, \qquad l,m=1\hbox{--}6
\end{equation}
so that
\begin{equation}
\pdiff{f_{ik}}{h_{\textrm{j}}}= -\psi_{k}\left[T_{im}\,\left(r_l\phi_{l}\right)^n + n~r_l T_{il}\phi_{m}\right]\pdiff{r_{m}}{h_{\textrm{j}}}, \qquad l,m=1\hbox{--}6.
\end{equation}

It is clear that the derivation and analysis of these integrals is rather cumbersome and prone to error.  However, one can easily check the correctness of the semi-analytically determined Jacobian by comparing it to an approximate one calculated using a forward difference for all entries.  This procedure will allow errors to be quickly found and rectified. In the FEM, many equations are only dependent on unknowns in adjacent elements, so that most entries in the Jacobian are zero. Then, if the equations and unknowns are ordered in a sensible manner, the matrix is sparse and banded with the exception of the dependence of all nodes on the contact line position and apex height.

In some cases, to ensure convergence of the solution, the code was started with the solid moving slowly, i.e.\ $|\mathbf{U}\cdot\mathbf{e}_z| \ll 1$.  Then, using the previous solution as an initial guess, the procedure was repeated, with a higher speed, until $|\mathbf{U}\cdot\mathbf{e}_z| = 1$.  This was the case at higher capillary numbers where the final free surface shape differs considerably from the initial guess of the static shape.

\subsection{Additional details}\label{additional}

From the local asymptotics in \S\ref{comp_loc} it can be seen that both the pressure and normal stress on the solid are logarithmically singular as the contact line is approached, and so, strictly speaking, special singular elements \cite{georgiou89,georgiou90} should be used to capture this behaviour near the contact line. Such an approach has previously been considered for this class of flows in \cite{wilson06}. The singular elements have been incorporated into our code, see \cite{sprittles10}; however, to simplify our exposition we have opted against using them here and, as can be seen from Figure~\ref{F:u_asymp}, the agreement of the (integrably singular) asymptotic result for $\lambda$ with the computed one is still excellent right up to the node adjacent to the contact line.


\section*{Acknowledgements}
The authors would like to thank Dr Mark Wilson, Dr Paul Suckling and Dr Alex Lukyanov for many stimulating discussions about the FEM implementation of dynamic wetting phenomena and JES kindly acknowledges the financial support of EPSRC via a Postdoctoral Fellowship in Mathematics.



\end{document}